\documentclass[11pt]{article}
\usepackage{cite}
\usepackage{amsmath,amsfonts,amssymb}
\usepackage[small,bf,hang]{caption}
\usepackage{slashed}

\def\hybrid{
        \topmargin -20pt
        \oddsidemargin 0pt
        \headheight 0pt \headsep 0pt
        \textwidth 6.25in 
        \textheight 9.5in 
        \marginparwidth .875in
        \parskip 5pt plus 1pt \jot = 1.5ex}

\hybrid

 \csname
@addtoreset\endcsname{equation}{section}

\def\moth{\mathsurround=0pt}
\newdimen\zo \zo=0pt

\def\tick{\leaders\hrule height 0.5ex depth 0pt \hskip 0.5pt}
\def\upboxfill{$\moth \setbox\zo\hbox{\tick}%
  \hskip 3pt\hbox to 0pt{$\tick$\hss}\hrulefill \hbox to 7.5pt{$\tick$\hss}$}

\def\dtick{\leaders\hrule height .34pt depth 0.5ex \hskip 0.5pt}
\def\downboxfill{$\moth \setbox\zo\hbox{\dtick}%
  \hskip 2pt\hbox to 0pt{$\dtick$\hss}\hrulefill \hbox to 2pt{$\dtick$\hss}$}

\def\fancys{\mathbb{S}}

\def\bec{\begin{center}}
\def\ec{\end{center}}

 \def\det{{\rm det\,}}
\def\be{\begin{equation}}
\def\ee{\end{equation}}
\def\bea{\begin{eqnarray}}
\def\eea{\end{eqnarray}}
\def\ba{\begin{array}}
\def\ea{\end{array}}

\def\ket#1{|#1\rangle}
\def\bra#1{\langle#1|}

\begin{document}

\begin{titlepage}
\rightline{}
\rightline{\tt MIT-CTP-4278}
\rightline{June 2011}
\begin{center}
\vskip 2.5cm
{\Large \bf {
Double Field Theory of Type II Strings
}}\\
\vskip 2.5cm
{\large {Olaf Hohm, Seung Ki Kwak, and Barton Zwiebach}}
\vskip 1cm
{\it {Center for Theoretical Physics}}\\
{\it {Massachusetts Institute of Technology}}\\
{\it {Cambridge, MA 02139, USA}}\\
ohohm@mit.edu, sk\_kwak@mit.edu, zwiebach@mit.edu
\vskip 0.7cm

\vskip 1.5cm
{\bf Abstract}
\end{center}

\vskip 0.5cm

\noindent
\begin{narrower}
We use double field theory to give a unified description
of the low energy limits of type IIA and type IIB superstrings.
The Ramond-Ramond potentials fit into spinor representations of the duality group $O(D,D)$ and field-strengths are obtained by acting with the Dirac operator on the potentials.
The action, supplemented by a Spin${}^+(D,D)$-covariant
self-duality condition on field strengths, reduces to the IIA and IIB theories in different frames.  As usual, the NS-NS gravitational variables are
described through the generalized metric.  Our work suggests that the fundamental gravitational variable is 
a hermitian element
of the group Spin$(D,D)$ whose natural projection to $O(D,D)$ gives
the generalized metric.

\end{narrower}

\vspace{4cm}

\end{titlepage}

\newpage

\tableofcontents



\section{Introduction and summary}

T-duality transformations along circles of compactified type II superstrings show that type IIA and type IIB superstrings are, in fact, the same theory for toroidal backgrounds of odd dimension (see~\cite{joep} and references therein).  This naturally leads to the question of whether there
exists a formulation of type II theories that makes this feature manifest. In this paper we will
address this question, reporting on results that have recently been announced in \cite{Hohm:2011zr}.

The understanding of the T-duality transformation rules for the Ramond-Ramond (RR) 
fields has been the subject of many works in a number of
formalisms \cite{Dine:1989vu,Hull:1994ys,Bergshoeff:1995as,Hassan:1999mm,Cvetic:1999zs,Benichou:2008it}.
Experience with bosonic strings, or with the NS-NS sector of type II superstrings, has shown that the duality group is $O(d,d)$, where $d$
is the number of toroidal dimensions~\cite{Meissner:1991zj,Maharana:1992my}.
In double field theory ---
an approach to make T-duality manifest for the massless sector of string theory by
doubling the coordinates~\cite{Hull:2009mi,Hull:2009zb,Hohm:2010jy,Hohm:2010pp} --- it has been useful to work with the group $O(D,D)$,
where $D$ is the total number of {\em spacetime} dimensions.
(See
\cite{Siegel:1993th} for earlier work by Siegel and \cite{Tseytlin:1990nb,Kwak:2010ew,Hohm:2010xe,Hohm:2011gs,Hohm:2011dz,Hohm:2011ex,Berman:2010is,West:2010ev,Jeon:2010rw,Copland:2011yh,Andriot:2011uh,Thompson:2011uw}
for related papers.)
Conservatively, one can focus on the elements of $O(D,D)$ that
act only on $d$ compact space dimensions. In bosonic double field theory, however, the full $O(D,D)$
is a symmetry when all spacetime coordinates
are non-compact and doubled.  The 
symmetry is manifest, acting both on the fields and on the coordinates.

In an important work,  Fukuma, Oota and Tanaka~\cite{Fukuma:1999jt}
discussed the IIA and IIB supergravity limits of superstrings
compactified on a torus
$T^d$.  The authors verified that 
the dimensionally reduced theory
arising from the RR 
sector contains scalars, one-forms, and
higher forms, each of which fit into the {\em spinor} representation of $O(d,d)$.  The kinetic operator was shown to use the spin representative   
of the familiar $O(d,d)$ matrix 
of scalar fields that arise from the metric and $b$-field components along the compact directions. The required spin representatives of $O(d,d)$ elements were discussed in the earlier work of Brace, Morariu, and Zumino~\cite{Brace:1998xz} in their study of RR backgrounds in the matrix model. The relevance of $O(d,d)$ spinors for dimensionally reduced RR fields was first noted by
Hull and Townsend~\cite{Hull:1994ys}. 

In this paper we construct the double field theory of the RR
massless sector of superstring theory.  The NS-NS massless sector
is described by the same theory that describes 
the massless sector of the bosonic string \cite{Hull:2009mi,Hull:2009zb,Hohm:2010jy,Hohm:2010pp}. 
The fields are a duality invariant dilaton $d$
and  the generalized metric ${\cal H}_{MN}$, that encodes
the metric and $b$-fields in a matrix called ${\cal H}$:
 \be\label{firstH}
  {\cal H}_{MN} \ = \  \begin{pmatrix}    g^{ij} & -g^{ik}b_{kj}\\[0.5ex]
  b_{ik}g^{kj} & g_{ij}-b_{ik}g^{kl}b_{lj}\end{pmatrix} \, \equiv \, 
  {\cal H} \;.   
 \ee
Here $M,N,\ldots =1,\ldots,2D$ denote 
fundamental $O(D,D)$ indices.
The double field theory action then takes the Einstein-Hilbert-like form
  \be\label{actR}
   S \ = \  \int dx \, d\tilde x  \, e^{-2d} \, {\cal R}({\cal H},d)\;,
 \ee
where ${\cal R}({\cal H},d)$ is an $O(D,D)$ invariant scalar.
In here all fields depend on the doubled coordinates $X^{M}=(\tilde{x}_{i},x^{i})$,
and after setting $\tilde{\partial}^{i}=0$ the action (\ref{actR}) reduces to the conventional
low-energy action for the massless NS-NS fields.
The action also features
a gauge symmetry 
with an $O(D,D)$ vector    
 parameter $\xi^{M}=(\tilde{\xi}_{i},\xi^{i})$ that combines the
diffeomorphism parameter $\xi^{i}$ and the Kalb-Ramond gauge parameter $\tilde{\xi}_{i}$:
 \be\label{manifestH}
 \begin{split}
  \delta_{\xi}{\cal H}_{MN} \ &= \  \widehat{\cal L}_{\xi} {\cal H}_{MN} \ \equiv  \ \xi^{P}\partial_{P}{\cal H}_{MN}
  +\big(\partial_{M}\xi^{P} -\partial^{P}\xi_{M}\big)\,{\cal H}_{PN}
  +
  \big(\partial_{N}\xi^{P} -\partial^{P}\xi_{N}\big)\,{\cal H}_{MP}\;, \\
  \delta d ~\ &= \ \xi^M \partial_M d - {1\over 2}  \partial_M \xi^M \,.  
 \end{split}
 \ee
Here $\widehat{\cal L}_{\xi}$ defines a generalized Lie derivative and
$\partial_{M}=(\tilde{\partial}^{i},\partial_{i})$.
The gauge invariance
of the action requires the $O(D,D)$ covariant
constraints
  \be\label{ODDconstr}
   \partial^{M}\partial_{M}A \ = \ \eta^{MN}\partial_{M}\partial_{N}A \ = \ 0\;, \qquad
   \partial^{M}A\,\partial_{M}B \ = \ 0\;, \qquad
   \eta^{MN} \ = \  \begin{pmatrix}
    0&1 \\1&0 \end{pmatrix}\,,
 \ee
for all fields and parameters $A$ and $B$, where $\eta$ is the $O(D,D)$ invariant metric.
This constraint implies that locally one can always
find an $O(D,D)$ transformation that rotates into a frame in which the fields depend only
on half of the coordinates, e.g., only on the $x^{i}$ or the $\tilde{x}_{i}$.

Let us now turn to
the RR sector, which
requires some new ingredients.  The first one is that the RR gauge fields fit naturally into the spinor representation of $O(D,D)$. In the case of interest, the physical dimension is $D=10$ and we have
a spinor of $O(10,10)$.
The spinor representation of $O(D,D)$ of 
dimension $2^D$ 
is real (or Majorana) and reducible.  This dimension equals  the sum of the number of components of all the forms in a $D$-dimensional spacetime.   An additional Weyl condition yields two
spinor representations of opposite chirality, each of dimension $2^{D-1}$,
that  can be matched with even and odd forms and therefore with
the RR fields in the type II theories.  The RR potentials of the IIA and IIB
theories do not include all odd and all even forms, but duality relations
can be naturally imposed on the field strengths to reduce the spectrum
to the desired one.  This `democratic' formulation of the type II supergravities uses field strengths of degrees $2,4,6,$ and $8$ for
type IIA and field strengths of degrees $1,3,5,7,$ and $9$ for type IIB \cite{Fukuma:1999jt}.

The type II theories are formulated in a ten-dimensional spacetime
with Lorentzian signature.  
In fact, the requisite self-duality condition of type IIB is consistent only with
this signature.  A number of features arise from this choice of signature that require a careful discussion of the relevant duality groups, in particular of the `spin' groups that provide the double covers of the orthogonal duality groups.  The RR fields, as mentioned above, fit into a spinor of
$O(D,D)$, but the so-called `spinor' representation of $O(D,D)$
is only defined up to signs.  A true representation exists for the group Pin$(D,D)$, which provides a double cover of $O(D,D)$, or for the group Spin$(D,D)$, which provides a double cover of
$SO(D,D)$.  Just like $SO(D,D)$ is a subgroup of $O(D,D)$, Spin$(D,D)$ is a subgroup of Pin$(D,D)$. Because of the double covering, each element in $O(D,D)$ has two lifts to Pin$(D,D)$
and similarly each element in $SO(D,D)$ has two lifts to Spin$(D,D)$.
Moreover, there is a group homomorphism $\rho$\;: Pin$(D,D)\to$ $O(D,D)$
that also takes Spin$(D,D)$ to $SO(D,D)$. If $S$ is an element in Pin$(D,D)$, then $(-S)$ is also an 
element and both $S$ and $(-S)$ map to the same $O(D,D)$ element under $\rho$.

T-dualities about single circles are elements of $O(D,D)$ that are not
in $SO(D,D)$:  they are represented by matrices of determinant minus one. Their lifts are transformations in Pin$(D,D)$ that are not in Spin$(D,D)$ and have the effect of changing the Weyl condition of a spinor.  Since the chirality of the spinor that encodes the RR forms must be fixed in order to
write down the theory, the duality group is Spin$(D,D)$.  Calling $\chi$ the spinor that encodes the RR forms we have the duality transformations
\be
\label{duachi}
\hbox{Duality transformations:} ~~~ \chi  ~\to ~  S\, \chi \,, ~~~~ S\,\in\, \hbox{Spin} (D,D) \,.
\ee
In the doubled space it is natural to define a Dirac operator
\be
\slashed{\partial} \ \equiv \  {1\over \sqrt{2}} \Gamma^M \partial_M
\ = \ {1\over \sqrt{2}} \,(\Gamma^i \partial_i + \Gamma_i \tilde\partial^i) \,,
\ee
where $\Gamma^{M}$ are gamma matrices of $O(D,D)$.
Using the Clifford algebra and the constraint (\ref{ODDconstr}), we readily verify that $\slashed{\partial}\slashed{\partial}=0$.
We show that $\slashed{\partial}$ is duality invariant and as a result
(\ref{duachi}) implies
\be
\label{duachi99}
 ~~~\slashed{\partial} \chi  ~\to ~  S\, \slashed{\partial}\chi \;, ~~~~ S\,\in\, \hbox{Spin} (D,D) \,.
\ee
Since $\slashed{\partial}$ is first order in derivatives,
$\slashed{\partial}\chi$ is naturally interpreted as the field strength associated to
the RR potentials, to which it indeed reduces for $\tilde{\partial}^{i}=0$.

Following the insights of~\cite{Fukuma:1999jt} it is natural to
consider  the spin group
representative of ${\cal H}$
to discuss the coupling of the RR fields to the NS-NS fields. The generalized metric ${\cal H}$ is a symmetric matrix that is also an $O(D,D)$ element.  Since the determinant of ${\cal H}$ is
plus one, we actually have ${\cal H} \in\,SO(D,D)$.  The group
$SO(D,D)$ has two disconnected components:
the subgroup
$SO{}^+(D,D)$ that contains the identity and
a coset denoted by $SO{}^-(D,D)$.  One can check that in Lorentzian signature
${\cal H}$ is actually in $SO{}^-(D,D)$.
The associated spin representatives are in Spin${}^-(D,D)$;
they are elements
$S$ and $-S$, such that $\rho(\pm S) = {\cal H}$.
It turns out to be impossible to choose a spin
representative in a single-valued and continuous way
over the space of possible ${\cal H}$. We  
illustrate this
with an explicit example of a closed path in the space of ${\cal H}$ configurations ({\em i.e.}~a closed path in $SO{}^-(D,D)$)  for which forcing a continuous choice of representative results in an open path in Spin${}^-(D,D)$, a path in which the initial and final elements differ by a sign. 
We note  
that this phenomenon occurs whenever a \textit{timelike} T-duality is employed,  
and therefore does  not arise in Euclidean signature where  
${\cal H} \in SO{}^+(D,D)$ and a lift to
Spin${}^+(D,D)$ can be chosen continuously.

In light of the above topological subtlety we suggest that instead of viewing  ${\cal H}$ as the fundamental gravitational field,
from which a spin
representative needs to be constructed, we
view the spin element
itself as the dynamical field, denoted by $\fancys \,\in\;$Spin${}^-(D,D)$.
The generalized metric can then be defined uniquely
by the homomorphism: ${\cal H}= \rho(\fancys)$.
The condition that ${\cal H}$ is symmetric requires that $\fancys$ be
hermitian, $\fancys=\fancys^{\dagger}$. Under the duality transformation (\ref{duachi}) we declare that
\be
\label{duachiS}
\hbox{Duality transformations:} ~~~\quad \fancys  ~\to ~\fancys' \ = \   (S^{-1})^\dagger\, \fancys\, S^{-1}  ~~~~ S\in \hbox{Spin} (D,D) \,.
\ee
This transformation is consistent with that of the generalized metric, namely, $\rho(S)$ is an $SO(D,D)$ transformation that takes
${\cal H} = \rho(\fancys)$ to  ${\cal H}' =  \rho(\fancys')$.

We can now discuss the double field theory action for type II theories, whose independent fields
are $\fancys$, $\chi$ and $d$.
It is the sum of the action (\ref{actR}) for the NS-NS sector and a new action for the RR sector:
\be
\label{totaction}
\begin{split}
S \ &= \  \int dx d\tilde x\, \Bigl(  e^{-2d}\, {\cal R} ({\cal H} , d)  + \frac{1}{4}
 (\slashed{\partial}{\chi})^\dagger \;
 \fancys \; \slashed\partial\chi\,\Bigr)\, , \\
&{\cal H}\  =\  \rho (\fancys),  ~~~~  \fancys  \,\in\, \hbox{Spin}^-(D,D) \,, ~~~
\fancys^\dagger \ = \ \fancys\,.
\end{split}
\ee
The RR action is quadratic in the field strengths $\slashed{\partial} \chi$, and $\fancys$ is actually needed to produce the Hodge dual that then
leads to conventional kinetic terms.  The duality invariance of the
RR action is manifest on account of (\ref{duachi99}) and  (\ref{duachiS}).
The definition of the theory also requires
the field strength
$\slashed{\partial} \chi$ to satisfy
a self-duality constraint
that can be written in a manifestly duality covariant way,
\be\label{dualityintro}
 \slashed\partial\chi\  = \,
 -C^{-1} \fancys\,\,\slashed\partial\chi\, .
\ee
Here
the charge conjugation   matrix $C$ satisfies $C^{-1}
\Gamma^M C = (\Gamma^M)^\dagger$. While the action is invariant under
Spin$(D,D)$, the self-duality constraint breaks the duality symmetry down to Spin${}^+(D,D)$.
This is not unexpected since the epsilon tensor in the duality relations
is only left invariant
by the
orientation-preserving transformations
$GL^+(D)\subset SO^+(D,D)$.
It should be emphasized that the action is originally Pin$(D,D)$ invariant.
The Weyl condition on the spinor
reduces the duality symmetry
of the action to Spin$(D,D)$.
Finally, the self-duality constraint
reduces the symmetry of the theory to Spin${}^+(D,D)$.

The RR potentials have the usual abelian gauge symmetries in which the
form fields are shifted by exact forms. This symmetry also takes a manifestly duality covariant form,
\be
\delta_{\lambda} \chi \ = \ \slashed{\partial} \lambda\,,
\ee
and leaves (\ref{totaction}) invariant because 
$\slashed{\partial}^2 = 0$.  More nontrivially,
the invariance of the theory under the gauge symmetries parameterized by $\xi^{M}$ requires 
that $\chi$ transform as
\be \label{gtchi99}
\delta_\xi \chi \  = \ \widehat{\cal L}_\xi \chi \  \equiv \ \xi^M \partial_M  \chi
+  {1\over 2}   \partial_M \xi_N \Gamma^M
\Gamma^N \chi \,.
\ee
In here we defined the generalized Lie derivative $\widehat{\cal L}_\xi$ acting on a spinor.
To complete the analysis we require
a gauge transformation of the gravitational field $\fancys$
that satisfies two consistency conditions: (i) 
 together with (\ref{gtchi99}) it must leave the action invariant, and
(ii) it must imply the gauge transformation (\ref{manifestH}) for  ${\cal H}$ that is required for
gauge invariance of the NS-NS part of the action.
We find that these two conditions are satisfied by  
\be \label{gtS99}
\delta_{\xi}\, \fancys  \ = \ \xi^M \partial_M \,\fancys
+ {1 \over 2} C\big[ \Gamma^{PQ}  ,   C^{-1}\,\fancys  \big]  \partial_{P} \xi_{Q} \; .
\ee

In order to evaluate the action in different  
T-duality frames, i.e., for different solutions of the constraint (\ref{ODDconstr}), 
and to compare with the conventional formulation in terms of fields like 
$g$ and $b$, we need to choose a particular parametrization 
of the field $\fancys$. We start  
from the parametrization (\ref{firstH}) of the generalized metric 
${\cal H}= \rho (\fancys)$ implied by $\fancys$.
A spin representative $S_{\cal H}$ can then be defined locally, 
and we parametrize the field $\fancys$ by setting $\fancys = S_{\cal H}$.  
It turns out, however, that once a parametrization has been chosen 
in terms of $g$ and $b$, the original Spin$(D,D)$ symmetry of the action
cannot be fully realized as transformations of $g$ and $b$ since they
change the sign of the RR double field theory action for timelike T-dualities.
If the full Spin$(D,D)$  is to be a 
symmetry we must view $\fancys$ as the fundamental field. A manifestation
of the sign phenomenon is that
evaluating the action in  T-duality frames  related via timelike T-dualities results in RR actions that differ by 
an overall sign, a result that turns out to be consistent with proposals
in the literature. 
In order to explain this, let us  
discuss the evaluation of the action in different T-duality frames.

Suppose we have chosen a chirality of $\chi$ and a parametrization of $\fancys$ such that 
the theory reduces for $\tilde{\partial}^{i}=0$ to type IIA. All other
solutions of (\ref{ODDconstr}) can be obtained from this one by an $O(D,D)$ transformation.
For the bosonic double field theory, or for the NS-NS part of the type II theory,
it has been shown in \cite{Hohm:2010jy} that the action reduces in all frames to the
same theory: the conventional low-energy action of bosonic string theory,
but written in terms of different field variables, which are related by the corresponding T-duality
transformations. In type II theories, however, this changes, because generally
T-duality relates different type II theories to each other.
If, for instance, the theory reduces in one frame to type IIA, we will see that it reduces in any other
frame obtained by an \textit{odd} number of spacelike T-duality inversions to type IIB, and
vice versa. If, on the other hand, the frames are related by an even number of spacelike
T-duality inversions, the theory reduces in both frames to the same theory, either IIA or IIB.
We next consider the case of a frame that is obtained by a timelike T-duality transformation. 
First, let us review the status of timelike T-duality as discussed in the literature.

If one considers the reduction of the ten-dimensional low-energy type IIA or IIB theory
on a timelike circle, one finds that each RR $p$-form
gives rise to a form of the same degree in the nine-dimensional
Euclidean theory, together with a $(p-1)$-form, which originates from the timelike component.
The latter form enters  
with the wrong sign kinetic term. Consequently,
the timelike circle reductions of type IIA and type IIB do  
not give rise to the same theory
in nine dimensions; they give two theories  
that differ by an overall sign in the  RR kinetic terms. 
Therefore, the IIA and IIB theories cannot be T-dual on 
a timelike circle.  
It has been proposed  
by Hull that on a timelike circle the proper T-dual pairs are 
type IIA and type IIB$^{\star}$, or type IIB and type IIA$^\star$~\cite{Hull:1998vg}. 
In the low-energy description the type IIA$^\star$ and type IIB$^\star$
 differ from the IIA and IIB theories just by the overall sign of the 
RR kinetic terms, such that the timelike circle reductions of IIA and IIB$^{\star}$, and of
IIB and IIA$^{\star}$, give rise to the same theory.

If we start from a T-duality frame in which the double field theory reduces
to type IIA (IIB), we indeed find that the same theory reduces  to IIB$^{\star}$ (IIA$^\star$) in any frame obtained by a timelike T-duality
transformation.
In summary, the manifestly
T-duality invariant double field theory defined by (\ref{totaction}) and (\ref{dualityintro})
unifies these four different type II theories in that each of them arises in particular
T-duality frames.

This paper is organized as follows. In sec.~2 we review the properties of the
spinor representation of $O(D,D)$ and of its double covering group. 
Due to the aforementioned topological subtleties, we find it necessary to delve in some detail into  the mathematical issues.  In sec.~3
we discuss the field that is interpreted as the spinor representative of the
generalized metric. The duality covariant form of the action and duality relations is
introduced in sec.~4, while their evaluation in particular T-duality frames is done
in sec.~5 and 6. We conclude with a brief discussion in sec.~7. 
A number of technical proofs  as well as an example illustrating the topological obstructions in the 
construction of
 the spin representative of the generalized metric are given in an appendix.

\section{O(D,D) spinor representation}\setcounter{equation}{0}
In this section we review properties of the T-duality group $O(D,D)$
and its spinor representation or, more
precisely, the properties of its two-fold covering group Pin$(D,D)$ and its representations. Convenient references for this section are
\cite{Fukuma:1999jt}, \cite{Gualtieri}, and~\cite{Fulton}.

\subsection{O$(D,D)$, Clifford algebras, and Pin$(D,D)$}
In order to fix our conventions, we start by recalling some basic properties of $O(D,D)$.
This group is defined to be the group leaving the metric of signature $({\bf 1}_{D},-{\bf 1}_{D})$
invariant. We choose a basis where the metric takes the form
 \be
  \eta \ = \ \begin{pmatrix}  0 & 1 \\ 1 & 0 \end{pmatrix}\;,
 \ee
and we denote it by $\eta^{MN}$ or $\eta_{MN}$ which, viewed
as matrices, are equal. The indices $M, N$ run over the $2D$ values
$1, 2, \ldots , 2D$.
The preservation of $\eta$ implies that group elements $h\in O(D,D)$, viewed as matrices,
 satisfy
 \be
  \eta^{MN} \ = \ h^{M}{}_{P}\, h^{N}{}_{Q}\,\eta^{PQ} \; \quad \Leftrightarrow \,\quad
  \eta \ = \ h\,\eta\,h^{T}\;.
 \ee
This implies that $\det(h)=\pm 1$. The subgroup of $O(D,D)$ whose elements have determinant plus one
is denoted by $SO(D,D)$. While the group $O(D,D)$ has four connected components,  $SO(D,D)$ has two connected components.
In $SO(D,D)$ the component connected to the identity is
the subgroup denoted as $SO^{+}(D,D)$.  It can be shown that
in the basis where
 the metric  takes the diagonal form
 $\text{diag}({\bf 1}_{D},-{\bf 1}_{D})$, the
two $D\times D$ block-diagonal matrices of any
$SO^{+}(D,D)$ element have positive determinant.
The other component of  $SO(D,D)$ is denoted by
$SO^{-}(D,D)$.  It is not a subgroup of $SO(D,D)$ but rather a coset
of $SO^+(D,D)$.

The Lie algebra of $O(D,D)$ is spanned by generators
$T^{MN}=-T^{NM}$ satisfying
 \be\label{ODDalgebra}
   \big[ T^{MN},T^{KL}\big] \ = \ 
   \eta^{MK}\,T^{LN} - \eta^{NK}\,T^{LM}-\eta^{ML}\,T^{KN}+\eta^{NL}\,T^{KM}\;.
 \ee
Any group element connected to the identity can be written as an exponential of Lie algebra generators,
 \be
   h^{M}{}_{N} \ = \ \bigl[ \exp\big(\tfrac{1}{2}\Lambda_{PQ}T^{PQ}\big)\bigr]^{M}{}_{N}\;,
 \ee
where
 \be\label{fund}
  (T^{MN})^{K}{}_{L}  \ = \ 2\eta^{K[M}\delta^{N]}{}_{L}\;,  
 \ee
is the fundamental representation of the Lie algebra (\ref{ODDalgebra}).
We use the anti-symmetrization convention $X_{[MN]} \equiv {1\over 2} (X_{MN} - X_{NM})$.

\medskip
We turn now to the spinor representation of $O(D,D)$ and to
the groups Spin$(D,D)$ and Pin$(D,D)$, whose properties will be instrumental below.
The (reducible) spinor representation of $O(D,D)$ has dimension $2^D$ and can be chosen
to be real or Majorana.
Imposing an additional Weyl condition will yield two spinor
representations of opposite chirality, both of dimension $2^{D-1}$. These can be
identified with even and odd forms and thus with the RR fields in type II.

To begin with, we introduce the Clifford algebra $C(D,D)$ associated to the quadratic form
  $\eta\, (\, \cdot , \,\cdot )$ on  $\mathbb{R}^{2D}$.  With basis vectors
  $\Gamma_M$, $M= 1, \ldots , 2D$, we have
\be
\eta_{MN} = \eta \,(\Gamma_M, \Gamma_N) =  \begin{pmatrix}  0 & 1 \\ 1 & 0 \end{pmatrix} \,.
\ee
The main relation of the Clifford algebra states that for any $V \in \mathbb{R}^{2D}$
\be
\label{main-cliff}
V \cdot V = \eta ( V, V) \;{\bf 1}\,,
\ee
where ${\bf 1}$ is the unit element and the dot indicates the product in
the algebra.  This algebra is generated by the unit and basis vectors
$\Gamma_{M}$.  Writing $V = V^M \Gamma_M$,  substitution in (\ref{main-cliff}) gives
 \be \label{gammaClif}
  \big\{ \Gamma_{M},\Gamma_{N}\big\} \ \equiv \
   \Gamma_M \cdot \Gamma_N + \Gamma_N \cdot \Gamma_M
   =  \ 2\,\eta_{MN}\,.
 \ee
Using the quadratic form $\eta_{MN}$ and its inverse $\eta^{MN}$
to raise and lower indices, we can write arbitrary vectors as $V=
V^M \Gamma_M =  V_M \Gamma^M$,
which then allows to write (\ref{gammaClif}) with all indices raised.

An explicit representation of the Clifford algebra (and below of the Pin group)
can be conveniently constructed using fermionic oscillators
$\psi^i$ and $\psi_i$, $i=1,\ldots,D$, satisfying
 \be \label{deffermosc}
  \{ \psi_i ,  \psi^j\} \ = \ \delta_{i}{}^{j}\;, \qquad
  \{\psi_i,\psi_j\} \ = \  0\;, \qquad \{\psi^i,\psi^j\} \ = \ 0\;,
 \ee
where
 \be
  \left(\psi_i\right)^{\dagger}  \ = \ \psi^i\;.
 \ee
Defining
 \be \label{defgamma}
   \Gamma_{i} \ = \ \sqrt{2}\psi_{i}\;, \qquad
   \Gamma^{i} \ = \ \sqrt{2}\psi^{i}\;,
 \ee
the oscillators realize the algebra (\ref{gammaClif}).
Spinor states can be defined introducing a Clifford vacuum $\ket{0}$
annihilated by the $\psi_i$ for all $i$:
\be
\psi_i\ket{0}=0\,, ~~\forall \, i \,.
\ee
From this, we derive a convenient identity that will be useful below,
 \be\label{spinident}
  \psi_{j}\psi^{i_1}\cdots\psi^{i_p}\ket{0} \ = \
  p\,\delta_{j}{}^{[i_1}\psi^{i_2}\cdots\psi^{i_p]}\ket{0}\;.
 \ee
A spinor $\chi$ in the $2^D$-dimensional space can then be
identified with a general state
 \be \label{genstate}
  \ket{\chi} \ = \ \sum_{p=0}^{D}\frac{1}{p!}\,C_{i_1\ldots i_p}\,\psi^{i_1}\ldots\psi^{i_p}\ket{0}\;,
 \ee
where the coefficients are completely antisymmetric tensors. Thus, there is a natural identification
of the spinor representation with the $p$-forms on $\mathbb{R}^{D}$.
We define $\bra{0}$ to be the the `dagger' of the state $\ket{0}$ and
declare:
\be
\bra{0} 0 \rangle = 1 \,.
\ee
For more general states,
\be
(\psi^{i_1}\ldots\psi^{i_p}\ket{0})^\dagger =  \bra{0} \psi_{i_p}\ldots\psi_{i_i}\;.
\ee
We work on a real vector space, so the $\dagger$ operation does
not affect the numbers multiplying the vectors.  In the notation where
dagger takes  $\ket{a}$ to $\bra{a}$ and vice versa, we can quickly
show that $\bra{a}b\rangle = \bra{b} a\rangle$.
We see from these definitions that in the spinor 
representation 
$(\Gamma^i)^\dagger$ is indeed equal to $\Gamma_i$.  Since all 
matrix elements are real, the dagger operation is just transposition.

Let us now turn to the definition of the groups Spin$(D,D)$
and Pin$(D,D)$, which act on the spinor states.  These groups are, respectively, double covers of the
groups $SO(D,D)$ and $O(D,D)$.  To describe these groups we
need to introduce an anti-involution $\star$ of the Clifford algebra $C(D,D)$, which
is defined by
\be
(V_1\cdot V_2 \ldots  \cdot V_k )^\star
\equiv (-1)^k \,V_k\cdot \ldots   V_2\cdot V_1\,.
\ee
Note that for any vector $V$ in $\mathbb{R}^{2D}$, $V^\star = - V$.
For arbitrary elements $S, T$ of the Clifford algebra one has
$(S+ T)^\star =
S^\star + T^\star$
and $(S\cdot  T)^\star =  T^\star \cdot  S^\star$.
The group Pin$(D,D)$ is now defined as follows:
 \be\label{defpin}
  \text{Pin}(D,D) \ := \ \big\{ S\in C(D,D)\,|\,S\cdot S^\star=\pm{\bf 1}\,, \; V \in \mathbb{R}^{2D}
  \,\Rightarrow\,
  S\cdot V\cdot S^{-1}\in \mathbb{R}^{2D}\big\}\;.
 \ee
The first condition implies for all group elements that $S^\star$ is, up to a sign, the inverse of $S$.
The second condition indicates that acting by conjugation with $S$ on any vector $V\in \mathbb{R}^{2D}$
results in a vector in $\mathbb{R}^{2D}$.
One readily checks that $S \in$ Pin$(D,D)$ implies
$S^\star \in$ Pin$(D,D)$. In  what follows we will omit the dot indicating Clifford multiplication whenever no  confusion can arise. We finally note that the Lie algebras of $O(D,D)$ and Pin$(D,D)$ are
isomorphic, and in spinor representation the generators are given by
 \be\label{spinmat}
  T^{MN} \ = \ \frac{1}{2}\Gamma^{MN} \ \equiv \ \frac{1}{4}\big[\Gamma^{M},\Gamma^{N}\big]\;,
 \ee
which satisfy (\ref{ODDalgebra}).

Next, we define a group homomorphism
\be
\rho:~\hbox{Pin}(D,D)\to O(D,D)\,,
\ee
with kernel $\{ {\bf 1}, -{\bf 1}\}$, that encodes the two-fold covering of $O(D,D)$.
It is defined via its action on a vector $V\in \mathbb{R}^{2D}$ according to
 \be\label{Defrho}
  \rho(S)V \ = \ SVS^{-1} \;.
 \ee
It is easily seen that this is a homomorphism, i.e., for arbitrary $S_1, S_2\in \text{Pin}(D,D)$
 \be
  \rho(S_1 S_2)  \ = \ \rho(S_1)\rho(S_2)\;.
 \ee
Moreover, $\rho$ indeed maps into $O(D,D)$, for it preserves the quadratic form,
 \be
 \begin{split}
  \eta(\rho(S)V,\rho(S)V)\, {\bf 1}\ &= \ \eta(SVS^{-1},SVS^{-1}) \,{\bf 1}
  \ = \ SVS^{-1}\cdot SVS^{-1} \\
  \ & = \  S\cdot (V\cdot V) \cdot S^{-1}  =  S\cdot {\bf 1} \cdot S^{-1}
  \eta(V,V) =  \eta\,(V,V) \,{\bf 1}\,,
\end{split}
 \ee
where the Clifford algebra relation (\ref{main-cliff}) has been used. Finally,
$\rho$ is surjective, i.e., for any $h\in O(D,D)$ there is an $S_h\in \text{Pin}(D,D)$ such
that $\rho(S_h)=h$. More precisely, by the two-fold covering, both $S_{h}$
and $-S_h$ are mapped to $h$ under $\rho$.

The map $\rho$ can be written in a basis using $V= V^M \Gamma_M$
for the original vector
and $V' = {V'}^M \Gamma_M$,  with ${V'}^M = h^M{}_{N} V^N$, for the
rotated vector, where $h^M{}_{N}$ is
an $O(D,D)$ element.  With this, the map in (\ref{Defrho}) becomes
\be
 \ \rho(S)V \ = \ V' \ =  \ S V S^{-1} \quad \to \quad
 h^M{}_{N} V^N\Gamma_M \ =  \ S \,V^{M}\Gamma_{M}\, S^{-1}\;.
 \ee
 Relabeling and canceling out the vector components we find
  \be\label{invgamma2}
  S\,\Gamma_{M}\,S^{-1} \ = \ \Gamma_{N}\, h^{N}{}_{M}\;.
 \ee
Here $\rho(S) =h$,  and $h$ --- with matrix representative
$h^M{}_{N}$ --- is the $O(D,D)$ element associated with $S$.
We rewrite the above equation by raising the indices.  Using the invariance property  $\eta_{MN}(h^{-1})^{N}{}_{K}=\eta_{KN}h^{N}{}_{M}$, we find
 \be\label{invgamma}
  S\,\Gamma^{M}\,S^{-1} \ = \ (h^{-1})^{M}{}_{N}\Gamma^{N}\; .
 \ee
Rewritten as  $ h^M{}_{N}\, S\,\Gamma^{N}\,S^{-1}  =  \Gamma^M$,
this is the familiar statement that gamma matrices are invariant
under  the combined action of Pin$(D,D)$ on the spinor and vector indices.

\medskip
Let us now turn to the definition of the subgroup Spin$(D,D)$ of Pin$(D,D)$.
It is obtained if in (\ref{defpin}) we have
$S  \in  C(D,D)^{\rm even}$, which is the Clifford subalgebra
spanned by elements with an even number of products of basis vectors.
In this case the homomorphism $\rho$ above restricts to a
homomorphism
\be
\rho:~\hbox{Spin}(D,D)\to SO(D,D)\,,
\ee
with kernel $\{ {\bf 1}, -{\bf 1}\}$.
If, in addition to restricting to $C(D,D)^{\rm even}$, the normalization condition is changed to
$SS^\star = {\bf 1}$, the resulting group is  Spin${}^+(D,D)$ and $\rho$ would map
to $SO^+(D,D)$.

Let us consider a set of useful elements $S$ of Pin$(D,D)$.  We
write the elements using the oscillators $\psi_i$ and $\psi^i,$\footnote{Here
we are closely following \cite{Fukuma:1999jt} with a slightly different notation. }
\be
\label{spin-elem}
\begin{split}
S_b \ \equiv \  & e^{- \frac{1}{2}b_{ij}\psi^{i}\psi^{j}}  \; , \\
S_r  \ \equiv \ & \frac{1}{\sqrt{\det r}} \, e^{ \psi^i R_i{}^j \psi_j } \; , \quad
(r = (r_i{}^{j} ) = e^{R_{i}{}^{j}} \in GL^{+}(D))\,, \\
S_i  \ \equiv \ & \psi^i + \psi_i \; , \quad (i = 1, \ldots , D) \; ,
\end{split}
\ee
where $GL^{+}(D)$ is the group of $D\times D$ matrices with strictly positive determinant.
It is instructive and straightforward to verify that the first condition
in (\ref{defpin}) holds.  Noting that $(e^x)^\star = e^{x^\star}$
we have
\be
\label{sbinv}
(S_b)^\star  = e^{- \frac{1}{2}b_{ij}\psi^{j}\psi^{i}} = e^{ \frac{1}{2}b_{ij}\psi^{i}\psi^{j}}  = (S_b)^{-1} \,.
\ee
We note that $S_b\in$ Spin${}^+(D,D)$.
For $S_r$ we have
\be
\label{srinverse}
\begin{split}
(S_r)^\star &\ = \ \frac{1}{\sqrt{\det r}} \, e^{ \psi_j R_i{}^j \psi^i } \ = \
 \frac{1}{\sqrt{\det r}} \,\, e^{\, -\psi^i R_i{}^j \psi_j + R_k{}^k } \\
&\ =\ \frac{1}{\sqrt{\det r}} \,\, e^{\,- \psi^i R_i{}^j \psi_j} e^{\text{tr}R}
\ =\ {\det r \over\sqrt{\det r} } \,\, e^{\, -\psi^i R_i{}^j \psi_j} \\
&\  = \ \sqrt{\det r} e^{\,- \psi^i R_i{}^j \psi_j} \ = \
(S_r)^{-1}\;,
 \end{split}
\ee
which implies that $S_r $ is in Spin${}^+(D,D)$.
Since $S_i$ is linear in gamma matrices, $S_i^\star = -S_i$.  We thus
have \be
S_i S_i^\star = - S_i S_i = -(\psi^i + \psi_i)  (\psi^i + \psi_i) = -\psi^i \psi_i
- \psi_i\psi^i =- {\bf 1}\,.
\ee
It follows that $S_i \in$ Pin$(D,D)$, while \textit{even} powers of the $S_{i}$ are in Spin$(D,D)$.

Using the definition (\ref{Defrho}) we can
calculate the $O(D,D)$ elements associated with these Spin$(D,D)$ elements.
For this we expand (\ref{invgamma2}) to find 
\be
\begin{split}
S\, \Gamma_i\,  S^{-1}  & \ = \    \Gamma_k\, h^{k}{}_{i}   +  \Gamma^k\, h_{ki} \,, \\
S\, \Gamma^i \, S^{-1}  & \ = \    \Gamma_k\, h^{ki}   +  \Gamma^k\, h_{k}{}^{i} \,,
\end{split}
\ee
and we build the $h$ matrix as follows
\be
h^M{}_{N}  = \begin{pmatrix} ~ h_i{}^{k} & h_{ik}\, \\[0.8ex] ~ h^{ik} & h^i{}_{k}
\end{pmatrix}\,.
\ee
Applying the above to (\ref{spin-elem}) one finds the $O(D,D)$ matrices
associated to the Pin elements:
\bea\label{PinEl}
h_b \ \equiv \ \rho (S_b)  &=&  \begin{pmatrix} 1 &    -b \\ 0 & 1 \end{pmatrix} \ , \quad b^T = - b \; , \\  \label{Pin2}
h_r \ \equiv \ \rho (S_r) &=& \begin{pmatrix} r &   0 \\ 0 & (r^{-1})^T \end{pmatrix} \ , \quad r \in GL^{+}(D) \; ,
\\ \label{PinEl33}
h_i \ \equiv \ \rho (S_i)&=&  -\begin{pmatrix} 1 - e_i &   - e_i \\ - e_i & 1-e_i \end{pmatrix} \ , \quad (e_i)_{jk} \equiv  \delta_{ij} \delta_{ik} \; , \quad (i = 1, \ldots , D)\,.
\eea
The group elements $h_{b}$, $h_{r}$ and even powers of the $h_{i}$
generate the component $SO^{+}(D,D)$ connected to the identity.
It is convenient to also record that
\be\label{PinEl2}
 \rho\big(e^{ \frac{1}{2}b_{ij}\psi_{i}\psi_{j}}\big) \ = \
 \begin{pmatrix} \,1 &   0 \\ b & 1 \end{pmatrix}\;, \qquad
 \rho\big( \psi^i- \psi_i\big) \ = \  -\begin{pmatrix} \,1-e_i &   e_i \\ e_i & 1-e_i \end{pmatrix}\,.
\ee

We note that (\ref{Pin2}) provides an embedding
$r\rightarrow h_{r}$ of $GL^{+}(D)$ into $SO^{+}(D,D)$, preserving the group structure,
 \be
  h_{r}\,h_{s} \ = \ h_{rs}\;,
 \ee
and thereby, via (\ref{spin-elem}), an embedding into Spin$^{+}(D,D)$.
In order to represent $GL^{-}(D,D)$ in Spin$(D,D)$, we note that this group can be
identified with the coset $GL^{+}(D)\,h_{-}$, with an arbitrary $h_{-}\in GL^{-}(D)$.
An example for such an element $h_{-}$ is given by the transformation that changes
the orientation in one direction, and for this we
consider:
\be
\begin{split}
 \rho(\psi^{i}\psi_{i}-\psi_{i}\psi^{i}) \  &= \
  \rho\Bigl((\psi^{i} - \psi_{i}) (\psi^{i}+ \psi_{i})\Bigr)
  = \rho (\psi^{i} - \psi_{i})\rho (\psi^{i}+ \psi_{i})\\
   & = \  \begin{pmatrix} \,1-e_i &   e_i \\ e_i & 1-e_i \end{pmatrix}
   \begin{pmatrix} \,1-e_i &   -e_i \\ -e_i & 1-e_i \end{pmatrix}
   \ = \ \begin{pmatrix} \,1-2e_i &   0 \\ 0 & 1-2e_i \end{pmatrix}\,,
\end{split}
\ee
where we used (\ref{PinEl33}) and (\ref{PinEl2}) .
 This shows that
 \be\label{PinEl42}
  \rho(\psi^{i}\psi_{i}-\psi_{i}\psi^{i}) \ = \ h_{-} \ = \
  \text{diag}(k_{i},k_{i})\;,  ~~~~  i ~\hbox{not summed}\; ,
 \ee
with the diagonal $D\times D$ matrix  $k_{i}\equiv\text{diag}(1,\ldots,-1,\ldots,1)$
that has a $-1$ in the $i$-th diagonal entry.
We will use this result below to define a spinor
representative of a metric $g$ with Lorentzian signature.

\subsection{Conjugation in Pin$(D,D)$}\label{csection}

We turn next to the definition of the charge conjugation matrix.
The charge conjugation matrix $C$ can be viewed as an element of Pin$(D,D)$ in general and
as an element of Spin$(D,D)$ for even $D$.
It  is  defined in terms of the oscillators by
  \be\label{fform}
   C \ \equiv \  \left\{
  \begin{array}{l l}
    C_+ \ \equiv  \  (\psi^{1}+\psi_{1})(\psi^{2}+\psi_{2})\cdots (\psi^{D}+\psi_{D}) \,, & ~~~\text{if\; $D$\; odd}\;,\\
    C_{-} \ \equiv  \  (\psi^{1}-\psi_{1})(\psi^{2}-\psi_{2})\cdots (\psi^{D}-\psi_{D}) \,, & ~~~ \text{if\; $D$\; even}\;.\\
  \end{array} \right.
 \ee
Noticing that with $i$ not summed
$(\psi^i \pm  \psi_i) (\psi^i \pm \psi_i)  = \pm \{ \psi^i \,, \psi_i\} = \pm 1\,,$
simple calculations show that
\be
C_+ (C_+)^\star =  (-1)^D  \,, ~~~~ C_- (C_-)^\star =  1 \,.
\ee
It is useful to note that the charge conjugation matrix is proportional to its inverse,
 \be \label{Cinverse}
  C^{-1} \ = \ (-1)^{D(D-1)/2} \, C \;.
 \ee
Since $C$ and $C^{-1}$
just differ by a sign, all expressions of the form $C \ldots C^{-1}$ can
be rewritten as $C^{-1} \ldots C$.   It is straightforward to show that
\be
\begin{split}
C_+  \psi_i (C_+)^{-1} =  -(-1)^D \psi^i   \,, ~~~& ~~~
C_+  \psi^i (C_+)^{-1} =  -(-1)^D \psi_i  \,, \\
C_-  \psi_i (C_-)^{-1} =  \phantom{-} (-1)^D\psi^i   \,, ~~~& ~~~
C_-  \psi^i (C_-)^{-1} =   \phantom{-}(-1)^D\psi_i  \,.
\end{split}
\ee
It then follows from (\ref{fform}) that in {\em all} dimensions
 \be
 \label{conpsiaction}
  C\,\psi_{i}\,C^{-1} \ = \ \psi^{i}\;, \qquad C\,\psi^{i}\,C^{-1} \ = \ \psi_{i}\;.
 \ee
As $\psi^{i}=(\psi_{i})^{\dagger}$, these relations can be written in terms of gamma matrices
as follows
  \be
   \label{GammaunderC} C \, \Gamma^M \, C^{-1} =
   (\Gamma^M)^\dagger \; ,  ~~~\hbox{or} ~~~
   C \, \Gamma_M \, C^{-1} =
   (\Gamma_M)^\dagger\;.
  \ee
Introducing the $O(D,D)$ element
 \be
 \label{Jdef}
 J^\bullet{}_{\bullet}  \ = \  J \ \equiv \ \begin{pmatrix}  0 & 1 \\ 1 & 0 \end{pmatrix}  \;,
 \ee
we can use (\ref{invgamma2}) to write the second equation  in (\ref{GammaunderC}) as
 \be
\label{GammaunderC99} C \, \Gamma_M \, C^{-1} = \Gamma_N \,
(\rho(C))^N{}_{M}  \ = \
(\Gamma_M)^\dagger = \Gamma_N J^N{}_{M} \,  \; .
\ee
We thus learn that
\be
\rho (C ) \ = \  J \,.
\ee
Since $C$ and $C^{-1}$ just differ by a sign, $\rho (C^{-1} ) = J$
and equation (\ref{GammaunderC}) also implies that
 \be
\label{GammaunderC-inv} C^{-1} \, \Gamma^M \, C \ = \
(\Gamma^M)^\dagger \; .
\ee

\smallskip
More generally we define the action of dagger by stating that
${\bf 1}^\dagger = {\bf 1}$, and that on vectors $V$ dagger is
realized by $C$
conjugation:
\be
\label{daggeraction}
V^\dagger \equiv   C V C^{-1}  =  J\, V\,.
\ee
On general elements of  the Clifford algebra we define dagger using
\be
(V_1\cdot V_2 \cdot  \ldots \,\cdot V_n)^\dagger \equiv
V_n^\dagger \cdot \ldots \,\cdot V_2^\dagger \cdot   V_1^\dagger\,,
\ee
so that for general elements  $(S_1\cdot S_2)^\dagger = S_2^\dagger
\cdot S_1^\dagger$.   A short calculation gives
\be
\label{Cdagger=}
C^\dagger \ = \ C^{-1} \,.
\ee
 It is straightforward to verify that $S \in$ Pin$(D,D)$
implies $S^\dagger \in$ Pin$(D,D)$.  It is  then
 natural to ask how the
homomorphism $\rho$ behaves under the dagger conjugation.

To answer this and related questions it is convenient to describe
the dagger operation in $C(D,D)$ in terms of
$C$ conjugation and the anti-involution $\tau$ defined by
\be
\tau (V_1 \cdot V_2 \cdot \ldots \cdot V_n) = V_n\cdot \ldots \,\cdot V_2 \cdot V_1 \,,
\ee
which satisfies $\tau(S_1S_2)=\tau(S_2)\tau(S_1)$.
Indeed, it is clear that
\be
\label{dagger-nice}
S^\dagger = C\, \tau (S) \,C^{-1} \,.
\ee

We now prove that the action of $\tau$
in Pin$(D,D)$ maps under $\rho$
 to the inverse operation in $O(D,D)$:
 \be
 \label{tauinv}
 \rho(\tau (S)) =  \rho(S)^{-1} \,.
 \ee
We begin with the defining relation (\ref{Defrho}) applied to $S^{-1}$:
\be\label{Defrho-var}
 S^{-1} VS  \ =\  \rho (S^{-1}) V \;.
 \ee
Now take the $\tau$ action on both sides. Noticing that the right-hand side
is left unchanged we get, because for any vector $\tau (V) = V$,
\be
\begin{split}
 \tau (S)  V\tau (S^{-1})  \ &= \  \rho (S^{-1}) V
 \to ~~  \tau (S)  V\tau (S)^{-1}  \ = \  \rho (S^{-1}) V\\
\to ~~  \rho( \tau (S))  V  \ &= \  \rho (S)^{-1} V\,,
\end{split}
\ee
thus establishing (\ref{tauinv}).  It is now easy to calculate
$\rho(S^\dagger)$ using (\ref{dagger-nice}).  Indeed, taking $\rho$ of this equation gives
\be
\rho(S^\dagger)  =  J  \rho(\tau(S)) J =  J \rho(S)^{-1} J \,,
\ee
where we recognized that $\rho(C^{-1}) = J$
and used (\ref{tauinv}).
Recalling that $O(D,D)$ elements $h$ satisfy  $h J h^T = J$, we have
$h^T = J h^{-1} J$.  Thus the right-hand
side above is simply $\rho(S)^T$, showing that
\be
\label{dagg-identity}
\rho(S^\dagger) = \rho(S)^T \,.
\ee

\medskip
For elements $S$ of Spin$(D,D)$, $\tau(S)= S^\star$, thus
(\ref{dagger-nice})
becomes
\be
\label{dagger-spin}
S^\dagger = C\, S^\star \,C^{-1} \,, ~~~ S \in \hbox{Spin} (D,D) \,.
\ee
Using that $S^{\star}=\pm S^{-1}$ for $S\in {\rm Spin}^{\pm}(D,D)$, this implies
 \be\label{finalSDagger}
  \begin{split}
   S^{\dagger} \ &= \ C\,S^{-1}\,C^{-1}\qquad {\rm for} \qquad S\,\in\, {\rm Spin}^{+}(D,D)\;, \\
   S^{\dagger} \ &= \ -C\,S^{-1}\,C^{-1}\quad\, {\rm for} \qquad S\,\in\, {\rm Spin}^{-}(D,D)\;.
  \end{split}
 \ee
In particular, for the spin generators $S_b$ and $S_r$ we get
\be \label{SunderC}
\begin{split}
S_b^\dagger \ &=  \  C S_b^{-1} C^{-1}  \; , \\
S_r^\dagger  \  &=  \  C S_r^{-1} C^{-1}  \; . \\
\end{split}
\ee
Since $\tau(S_i) = S_i$, for the final generator we have
\be
S_i^\dagger \ = \ C S_i C^{-1}\,.
\ee

\subsection{Chiral spinors}\label{chiralspinors}
We close this section with a brief discussion of the chirality conditions to be imposed on
the spinors. To this end it is convenient to introduce a `fermion number operator' $N_{F}$, defined by
 \be
  N_{F} \ = \ \sum_{k}\,\psi^{k}\psi_{k}\;.
 \ee
It acts on a spinor state that is of degree $p$ in the oscillators as follows
 \be
 \begin{split}
  N_{F}\ket{\chi}_{p} \ &\equiv \ N_{F}\Big(\,\frac{1}{p!}C_{i_1\ldots i_p}\psi^{i_1}\cdots \psi^{i_p}\ket{0}\,\Big)\\
  \ &= \ \sum_{k} p\frac{1}{p!}C_{i_1\ldots i_p}\psi^{k} \delta_{k}{}^{[i_1}\psi^{i_2}\cdots \psi^{i_p]}\ket{0}
   \ = \ p\ket{\chi}_{p}\;,
 \end{split}
 \ee
where (\ref{spinident}) has been used.
Thus, acting with $(-1)^{N_{F}}$ on a general spinor state (\ref{genstate}), one obtains
 \be
  (-1)^{N_{F}}\chi \ = \ \sum_{p=0}^{D}\,(-1)^{p}\,\frac{1}{p!}\,C_{i_1\ldots i_{p}}\,\psi^{i_1}\cdots\psi^{i_p}\ket{0}\;.
 \ee
We conclude that the eigenstates of  $(-1)^{N_{F}}$ consist of a $\chi$ that
is a linear combination of only even forms, with eigenvalue $+1$, or of a $\chi$ that
is a linear combination of only odd forms, with eigenvalue $-1$. Given an arbitrary
spinor $\chi$, one can project onto the two respective chiralities,
  \be
  \chi_{\pm} \ \equiv \ \frac{1}{2}\big(1\pm(-1)^{N_{F}}\big)\chi \qquad \Rightarrow \qquad
  (-1)^{N_{F}}\chi_{\pm} \ = \ \pm\, \chi_{\pm}\;.
 \ee
Then $\chi_{+}$ has positive chirality, consisting only of even forms, and $\chi_{-}$
has negative chirality, consisting only of odd forms. The operator
$(-1)^{N_{F}}$ is the analogue of the $\gamma^5$ matrix in four dimensions.

Finally, we note that the chirality is preserved under an arbitrary
Spin$(D,D)$ transformation. In fact, since the group elements of Spin$(D,D)$
contain only an even number of fermionic oscillators, they map even forms into even forms
and odd forms into odd forms. In contrast, a general Pin$(D,D)$ transformation can act
with an odd number of oscillators and thereby map
spinors of positive chirality to spinors of
negative chirality and vice versa.  Thus, when fixing the chirality, as for the action
to be introduced below, we break the symmetry from Pin$(D,D)$ to Spin$(D,D)$.

\bigskip

\section{Spin representative of the generalized metric}
In this section we discuss the spin representative $S_{\cal H}$ of the
generalized metric ${\cal H}_{MN}$. We determine its transformation
behavior under gauge symmetries and T-duality. More fundamentally, we will adopt the
point of view that $S_{\cal H}$ is just a particular parametrization of the fundamental
field $\fancys$.

\subsection{The generalized metric in Spin$(D,D)$}
We take the fundamental field to be $\fancys$, satisfying
 \be\label{basicS}
   \fancys \ = \ \fancys^{\dagger}\;, \qquad \fancys  \,\in\, \text{Spin}^{-}(D,D)\;.
 \ee
The generalized metric ${\cal H}_{MN}$ will then be \textit{defined} as
 \be
  {\cal H} \ \equiv \  \rho(\fancys) \qquad \Rightarrow \qquad
  {\cal H}^{T} \ = \ \rho(\fancys^{\dagger}) \ = \  {\cal H}\;, \quad
  {\cal H}\,\in\,SO^{-}(D,D)\;.
 \ee
Moreover, we constrain ${\cal H}$ and thereby $\fancys$ by requiring that
the upper-left $D\times D$ block matrix encoding $g^{-1}$ has
Lorentzian signature. 
An immediate  consequence of (\ref{basicS})
follows with (\ref{finalSDagger})
\be
\label{dagger-spinH}
 \fancys  \ = \
 \fancys^\dagger
\ =\  - C \,\fancys^{-1}\, C^{-1} \,.
 \ee
Equivalently, recalling that $C=\pm C^{-1}$,
\be
\label{Ctrans}
\fancys \, C \, \fancys  \ = \ -C  \,.
 \ee

 It is also possible to adopt the opposite point of view, i.e., to take the
group element ${\cal H}$ as given and then determine a corresponding
spin group representative $S_{\cal H}$ as a derived object. However, as we will discuss in more
detail below and in the appendix, this cannot be done in a consistent way globally over the space of
${\cal H}$.
 In the
following we first determine a spin representative $S_{\cal H}$ locally from ${\cal H}$,
but we stress
that this should be viewed as just a particular parameterization of $\fancys$ --- in the same
sense that the explicit form of ${\cal H}_{MN}$ in terms of $g$ and $b$ is just
a particular parametrization of ${\cal H}$.

We start by writing the $O(D,D)$ matrix ${\cal H}_{MN}$  as a product
of simple group elements,\footnote{We note that our conventions differ slightly from those in
 \cite{Hohm:2010pp} in that what we denote by ${\cal H}$ has been denoted ${\cal H}^{-1}$ there.
All other conventions,  however, are the same.}
  \be
   {\cal H} \ = \  \begin{pmatrix}  g^{-1} & -g^{-1} b \\ bg^{-1} &
   g- b g^{-1} b \end{pmatrix} \ = \ \begin{pmatrix} 1 &   0 \\ b & 1 \end{pmatrix}
   \begin{pmatrix} g^{-1} &   0 \\ 0 & g \end{pmatrix}
   \begin{pmatrix} 1 &   -b \\ 0 & 1 \end{pmatrix} \ \equiv \
   h_b^T\,h_{g^{-1}}\,h_{b} \; .
 \ee
The matrices defined in the last equation are analogous to the matrices defined in
(\ref{PinEl}) and (\ref{Pin2}). More precisely, this is true for $h_{b}$ while for $h_{g}$
(or $h_{g^{-1}}=h_{g}^{-1}$) eq.~(\ref{Pin2}) is only valid if  $g$ has
euclidean signature, because then $g\in GL^{+}(D)$. Here,
however,  we assume that
$g$ has Lorentzian signature $(-+\cdots +)$. Accordingly, ${\cal H}$ is indeed an element
of $SO^{-}(D,D)$.

In order to find the corresponding spinor representative for $h_{g}$ and thereby for ${\cal H}$,
it is convenient to introduce vielbeins in the usual
way,
 \be
 \label{define-k-matrix}
   g_{ij} \ = \ e_{i}{}^{\alpha}\, e_{j}{}^{\beta}\,k_{\alpha\beta}\;, \qquad
   k_{\alpha\beta} \ = \ \text{diag}(-1,1,\ldots,1)\;,
 \ee
where $\alpha,\beta,\ldots =1,\ldots, D$ are flat Lorentz indices with invariant metric $k_{\alpha\beta}$.
In matrix notation, we also write
 \be
  g \ = \ e\,k \, e^{T}\;.
 \ee
We can choose $e$ to have positive determinant, and thus its spin representative
can be chosen to be $S_{e}$ as defined in (\ref{spin-elem}). Using (\ref{PinEl42}),
the spin representative of diag($k, k)$
can be taken to be
 \be\label{Ssubk}
  S_{k} \ = \ \psi^{1}\psi_{1}-\psi_{1}\psi^{1}\;,
 \ee
where the label one
denotes the timelike direction.  We note that
\be
S_k =  S_k^\dagger =  S_k^{-1} = - S_k^\star  \,.
\ee
Since $S_k S_k^\star = -1$,  we confirm that $S_k \in\,$Spin${}^-(D,D)$.

Thus, we can choose the spinor representative of $g$ to be
\smallskip
 \be
  S_{g} \ \equiv \ S_{e}\,S_{k}\,S_{e}^{\dagger} \ = \
  {1 \over \det (e)}\, e^{\psi^{i}E_{i}{}^{j}\psi_{j}}
  (\psi^{1}\psi_{1}-\psi_{1}\psi^{1})
  e^{\psi^{i}(E^{T})_{i}{}^{j}\psi_{j} }\;,
 \ee
\noindent
where $e_{i}{}^{\alpha}=\exp(E)_{i}{}^{\alpha}$, and we used   $(E^{T})_{i}{}^{j}=E_{j}{}^{i}$.
 From its definition it follows  that
\be\label{Sgdagger}
S_g^\dagger \ = \ S_g \,.
\ee
Similarly,
 \be
  S_{g}^{-1} \ \equiv \ (S_{e}^{-1})^{\dagger}\,S_{k}\,S_{e}^{-1}
  \ = \ \det{e}\, e^{-\psi^{i}(E^{T})_{i}{}^{j}\psi_{j}}(\psi^{1}\psi_{1}-\psi_{1}\psi^{1})
  e^{-\psi^{i}E_{i}{}^{j}\psi_{j} }\;.
 \ee
We note that $S_{g}$ is an element of Spin$^{-}(D,D)$
because it is the product of $S_k \in\,$Spin$^{-}(D,D)$ times elements
of Spin$^{+}(D,D)$.
From this and (\ref{dagger-spin})
we also infer that
 \be\label{SginverseC}
  S_{g}^{\dagger} \ = \ S_{g} \ = \ C\,S_{g}^{\star}\,C^{-1} \ = \ -C\,S_{g}^{-1}\,C^{-1}\;\,.
 \ee

We can finally
 define the element $S_{\cal H}$ of Spin$(D,D)$ as follows
\be
\label{definitionSH}
\phantom{\Biggl(}
S_{\cal H} \ \equiv  \ S_b^{\dagger}\,  S_g^{-1}\, S_{b} \ = \  e^{\frac{1}{2}b_{ij}\psi_{i}\psi_{j}}\, S_{g}^{-1}\,
e^{ -\frac{1}{2}b_{ij}\psi^{i}\psi^{j}}
\;.
\ee
Using (\ref{Sgdagger}) we infer that
\be
\label{sHdag}
S_{\cal H}^\dagger \ = \ S_{\cal H}\,.
\ee
By construction,  the image
of $S_{\cal H}$ under the group homomorphism $\rho$ is precisely ${\cal H}$:
\be
\rho (S_{\cal H}) = \rho(S_b )^{T}
\rho    (S_g^{-1})
\rho (S_{ b})\ = \  h_b^Th_g^{-1} h_{b} = {\cal H}\,.
\ee
Since $S_b, S_b^\dagger\in\,$Spin${}^+(D,D)$ and
$S_g^{-1}\in\,$Spin${}^-(D,D)$, we have
$S_{\cal H} \in$ Spin${}^-(D,D)$.
As a result,
$S_{\cal H}$
satisfies the identities (\ref{dagger-spinH})
and (\ref{Ctrans}) and therefore gives
a consistent
parametrization of $\fancys$.

The flat Minkowski background  $g= k$ with zero $b$-field gives a generalized metric that we denote as ${\cal H}_0\equiv  \hbox{diag}(k, k)$.
Since $S_g = S_k$ and $S_b = {\bf 1}$, we have
\be
\label{shzero}
S_{{\cal H}_0}  = S_k^{-1} = S_k  =  \psi^{1}\psi_{1}-\psi_{1}\psi^{1}\,.
\ee

\subsection{Duality transformations}\label{dualsubsection}

We discuss now
the transformation behavior of $\fancys$ under some
arbitrary element $S\in\,$Pin$(D,D)$.  Since we view $\fancys$
as an elementary field we can postulate such a transformation.
The transformation of $\fancys$, however, must
be consistent with the transformation of the associated ${\cal H} =
\rho(\fancys)$.  Writing also ${\cal H}' = \rho(\fancys')$,
we want to postulate a transformation for which
\be
 \fancys ~~  \xrightarrow[\hspace{15pt}]{S} ~~ \fancys'
  ~~\hbox{implies} ~~~   {\cal H} ~~  \xrightarrow[\hspace{18pt}]{\rho(S)} ~~ {\cal H} '\,.
\ee
In words, the $O(D,D)$ transformation $\rho(S)$ associated with
 $S\in$ Pin$(D,D)$
relates the corresponding generalized metrics.  The generalized metric
appears explicitly in the NS-NS action.

Recall that under an $O(D,D)$ transformation $h$ the generalized metric
transforms as
\be
{\cal H}'_{MN} \ = \ {\cal H}_{PQ}  (h^{-1})^P_{~~M} (h^{-1})^Q_{~~N}\,.
\ee
In matrix notation, we will write ${\cal H}$ transformations as follows:
 \be\label{Htransform}
 {\cal H}^{\prime} =  h\circ {\cal H} \ \equiv  \ (h^{-1})^T\,{\cal H}\,h^{-1}\;.
 \ee
For an element $S\in \text{Pin}(D,D)$ we
postulate
the following $\fancys$ transformation:
 \be\label{vmvg}
  \fancys^{\prime}(X^{\prime}) \ = \ (S^{-1})^{\dagger}\, \fancys(X) \,S^{-1}\;.
 \ee
Here $X^{\prime}= hX$,  where $h = \rho(S)$.   The compatibility
with (\ref{Htransform}) is verified by taking $\rho$ on both sides.
Suppressing  the
coordinate arguments, we indeed find
 \be
\begin{split}
{\cal H}' & \ = \ \rho(\fancys^{\prime}) \ = \  \rho \bigl(  (S^{-1})^\dagger  \, \fancys  \,S^{-1} \bigr)
\ = \ \rho\bigl( (S^{-1})^\dagger\bigr) \, \rho(\fancys)\,\rho(S^{-1})  \\
& \ = \
(\rho(S)^{-1})^T  \, {\cal H} \, \rho(S)^{-1} \ = \ (h^{-1})^T  \, {\cal H} \, h^{-1}  \ = \ h\circ {\cal H} \;.
\end{split}
\ee
We infer that ${\cal H}^{\prime}$ satisfies  (\ref{Htransform}).

Independently of the postulated transformation rule (\ref{vmvg}), we can ask
how $S_{\cal H}$, defined in (\ref{definitionSH}) in terms
of ${\cal H}$,
transforms under
a duality transformation generated by an element $S\in$ Pin$(D,D)$.
This transformation is simply given by
\be
S:~~S_{\cal H}  ~\to ~  S_{{\cal H}'} \,, ~~\hbox{where}~~
{\cal H}' =  \rho(S)\circ {\cal H} \,.
\ee
It is of interest to compare
\be
(S^{-1})^\dagger  \, S_{\cal H} \, S^{-1}  ~~\longleftrightarrow  ~~
S_{{\cal H}'} \,.
\ee
Under $\rho$ they both map to ${\cal H}'$, thus the two can be equal
or can differ by a sign.  Perhaps surprisingly, there is a
sign factor  that depends nontrivially on $\rho(S)$ and on ${\cal H}$.
We will write
\be
\label{signdef}
(S^{-1})^\dagger  \, S_{\cal H} \, S^{-1}  \ = \
\sigma_{\rho(S)} ({\cal H}) \, \,S_{\rho(S) \circ {\cal H}}\,.
\ee
In the remainder of this section we determine this sign factor.

The case of zero $b$ field and flat Minkowski background,
${\cal H}_0= {\rm diag}(k,k)$, is readily analyzed.  We consider
a factorized T-duality $h_i$ with spin representative
$S_i = \psi^i + \psi_i = S_i^{-1} = S_i^\dagger$.  Under this transformation
${\cal H}_0$ remains invariant, since it corresponds to a diagonal metric
with entries of absolute value one.   We then have, using (\ref{shzero}),
\be
  (S_i^{-1})^{\dagger}\,S_{{\cal H}_0}\,S_i^{-1} \ = \
    (\psi^i + \psi_i )\,(\psi^{1}\psi_{1}-\psi_{1}\psi^{1})\,(\psi^i + \psi_i)\;.
\ee
It is manifest that the right-hand side is equal to $S_{{\cal H}_0}$
when $i\not=1$, and a small calculation shows that is equal to
$\,-S_{{\cal H}_0}$ when $i=1$:
  \be\label{signchange0}
   (S_i^{-1})^{\dagger}\,S_{{\cal H}_0}\,S_i^{-1} \ = \
   (-1)^{\delta_{i,1}} \, S_{{\cal H}_{0}} \,.
 \ee
We see that the sign is negative for a timelike T-duality, while the
sign is positive for spacelike T-dualities.

There is a large set of $O(D,D)$ transformations $h$ for which the sign
in (\ref{signdef}) is plus.  As we show in the Appendix
\be
(S_h^{-1})^\dagger  S_{\cal H}  \,S_h^{-1}  \ = \ + \,S_{h\circ \cal H}\,,
~~~~ \hbox{when} ~~~h\in\,GL(D)\ltimes \mathbb{R}^{\frac{1}{2}D(D-1)}\,.
\ee
The group $GL(D)\ltimes \mathbb{R}^{\frac{1}{2}D(D-1)}$ is that generated by
successive applications of $GL(D)$ transformations and $b$-shifts,
transformations $h_b$ of the form indicated in (\ref{PinEl}), which define
the abelian subgroup $\mathbb{R}^{\frac{1}{2}D(D-1)}$.

It is the T-dualities that produce sign changes. We  therefore consider
the sign factor~in
\be
\label{kcvgsmllsg}
(S_i^{-1})^\dagger  \, S_{\cal H}  \,  S_i^{-1}  \ = \ \sigma_i ({\cal H})\, S_{h_i \circ {\cal H}} \,.
\ee
As we can see, the sign factor depends on the
particular ${\cal H}$
appearing on the left-hand side above. Our strategy in Appendix \ref{A2app}  is to determine $O(D,D)$ transformations $h$ that acting on 
${\cal H}$ do not change the sign factor.  We will show that  
 if $h\in GL(D)\ltimes \mathbb{R}^{\frac{1}{2}D(D-1)}$ and
$h_i \, h\, h_i \in GL(D)\ltimes \mathbb{R}^{\frac{1}{2}D(D-1)}$, then
\be
\label{jnvsvg99}
~\sigma_i (h\circ {\cal H}) = \sigma_i ( {\cal H})\,.
\ee
It turns out that $b$-shifts satisfy the above conditions.  Since
at any point $X$
the $b$-field of an ${\cal H}$ can be removed completely by a $b$-shift, we
learn that the sign factor depends only on the metric $g$ in ${\cal H}$:
\be
\sigma_i ({\cal H}) =  \sigma_i (g)\,.
\ee
We then find a restricted class of $GL(D)$ transformations that also
satisfy the conditions for invariance of the sign factor.  With these we
are able to show that the metric $g$ can be put in diagonal form, with entries
$\pm 1$.  The sign factor then becomes calculable, just like we had
for the case of ${\cal H}_0$.  Our final result is:
\be
\label{sgnresult}
\sigma_i ({\cal H}) \ = \ \hbox{sgn} (g_{ii}) \,.
\ee
It follows from this equation that for the flat Minskowski metric
 the duality transformation $J$
about all of the spacetime coordinates gives the sign factor:
$\sigma_{J} ({\cal H}_0) = -1$.  At the end of Appendix~\ref{A2app}
we prove that this result holds for a general background ${\cal H}$
whose metric has Lorentzian signature:
\be
\label{tdualJonS}
\sigma_{J} ({\cal H}) = -1\,.
\ee

It is possible to give some intuition for the appearance of the
minus signs under duality transformations. For more details
see Appendix~\ref{appendixexamples},
where an example is worked out as well.
Since a sign cannot change continuously,  $\sigma_i ({\cal H}+ \delta {\cal H}) = \sigma_i ({\cal H})$ as long as the variation $\delta {\cal H}$
does not generate
singularities in the fields $(g,b)$ or
their T-dual versions $h_i\circ (g,b)$
in equation (\ref{kcvgsmllsg}).  Consider now a continuous
family
${\cal H}(\alpha)$ parameterized
by $\alpha$ in which the metric component $g_{ii}(\alpha)$ changes sign
at some point $\alpha^*$.   Consider also the related family 
$h_i \circ {\cal H}(\alpha)$ obtained by T-duality about
the $i$-th direction.  Under this duality the new metric, indicated by
primes, is
\be
 g_{ii}' (\alpha) \ = \ \frac{1}{g_{ii} (\alpha) } \; .
\ee
It follows that $g_{ii}'$ diverges and is discontinuous at $\alpha=\alpha^*$.
Note, however, that the generalized metric $h_i \circ {\cal H}(\alpha)$ is regular throughout, since it is obtained from the regular ${\cal H}(\alpha)$ by the
action of the regular matrix $h_i$.
The discontinuity of $g'_{ii}$ implies a discontinuity in the vielbein
 $e'$ and a discontinuity in $S_{g'} = S_{e'} S_{k} S_{e'}^{\dagger}$.
This results in a discontinuity of
$S_{h_i\circ {\cal H} (\alpha)}$.  Since $h_i \circ {\cal H}(\alpha)$ is
continuous,
the only discontinuity in $S_{h_i\circ {\cal H} (\alpha)}$
consistent with the homomorphism $\rho$ is
a change of sign.  The right-hand side of (\ref{kcvgsmllsg}) changes sign at the point where the original metric component changes sign. This is consistent with our result (\ref{sgnresult}).

The issues of signs are not an artifact of our definition of $S_{\cal H}$.
In  Appendix \ref{appendixexamples} we construct
 a continuous family of regular generalized metrics ${\cal H}(\alpha)$
 for which ${\cal H}(0) = {\cal H}(\pi/2)$, so that  ${\cal H}(\alpha)$
 with $\alpha \in [0, \pi/2]$ is a closed path in the
space of generalized metrics.  If we define the representative
$S_{{\cal H}(0)}$ and then continuously deform this representative
 along the path we find that at the end of the path the representative is $\,-S_{{\cal H}(0)}$.  The lift to the spin group cannot be done continuously over the
space of generalized metrics.  If we do the lift using our definition
of $S_{\cal H}$ from ${\cal H}$ we find that for some intermediate
$\alpha$ the metric $g(\alpha)$ and the $b$-field $b(\alpha)$ become singular,
while ${\cal H}(\alpha)$ remains regular.  At this point the definition
of $S_{\cal H}$ gives a discontinuity.

There seems to be some tension between the defined
duality transformation of $\fancys$ in (\ref{vmvg}), 
which has no signs, and the duality transformation 
(\ref{signdef}) of its particular parametrization $S_{\cal H}$,
which shows some signs.
The sign-free transformation of $\fancys$ implies that the double field theory action
is fully invariant under all duality transformations, including those, like
timelike T-dualities, that
give a sign in (\ref{signdef}).  
Once we choose a parametrization by setting~\hskip2pt$\fancys=S_{\cal H}$, 
the sign factors in (\ref{signdef}) have two consequences. 
First, it follows that the Spin$(D,D)$ invariance of the action cannot be fully 
realized through transformations of the conventional fields $g$ and~$b$. 
More precisely, it can only be realized for $SO(D,D)$ transformations that do 
not involve a timelike T-duality.  
This means that if we  take
timelike T-dualities  seriously, we inevitably have to view $\fancys$ as the 
fundamental field. Second, when comparing the double field theory evaluated in one 
T-duality frame (as $\tilde{\partial}^{i}=0$) to the same theory evaluated in another T-duality 
frame obtained by a timelike T-duality transformation (as $\partial_{i}=0$),  
the conventional effective RR action changes sign.
This sign change corresponds precisely to the transition from 
type II to type II${}^*$ theories expected for timelike T-dualities.
Correspondingly, the freedom in the choice of parametrization for $\fancys$, 
namely $\pm S_{\cal H}$, has no physical significance in that
it merely fixes for which coordinates ($x$ or $\tilde{x}$) we obtain the type II and
for which we obtain the type II${}^*$ theory. Similarly, the actual sign of the RR term
in the double field theory action (\ref{totaction}) has no physical significance. 
Therefore, we find a consistent picture, though certain invariances of
the action cannot be fully realized on the conventional gravitational fields.

\subsection{Gauge transformations}\label{gaugtransfancys}

In this section we determine the gauge transformation of the spinor representative
$\fancys$ in such a way that it us consistent with the
known gauge variation of
the generalized metric ${\cal H}_{MN}$. This variation, given in (\ref{manifestH}), can be rewritten as:
\be\label{manifestH99}
  \delta_{\xi}{\cal H}^M{}_{P}  \ = \
  \xi^{L}\partial_{L}  {\cal H}^M{}_{P}
  +\big(\partial^{M}\xi_{K} -\partial_{K}\xi^{M}\big)\,{\cal H}^K{}_{P}
  +
  \big(\partial_{P}\xi^{K} -\partial^{K}\xi_{P}\big)\,{\cal H}^M{}_{K}\;,
 \ee
where we used
that the metric $\eta_{MN}$ that lowers
indices is gauge invariant.  We have positioned the indices of the generalized
metric as in ${\cal H}^{\bullet}{}_{\bullet}$ to emphasize its role as an
$O(D,D)$ group element.   We also recall that ${\cal H}^{M}{}_{K}{\cal H}^{K}{}_{N} = \delta^{M}{}_{N}$.  The matrix
${\cal H}$ used so far represents ${\cal H}_{\bullet\,\bullet}$.

It turns out to be convenient to write the gauge variation in terms of
the spin variable ${\cal K}$ defined by
\be
\label{clwltlckhrss}
{\cal K}\ \equiv \ C^{-1}\, \fancys \,.
\ee
This combination will be used to prove the gauge invariance
 of the action in section~\ref{gauinvoftheact}.
While $\fancys$ is a spin representative of ${\cal H}_{\bullet\,\bullet}$, we
now check that ${\cal K}$ is the spin representative of
 ${\cal H}^\bullet_{~\,\bullet}$.  Indeed recalling that $\rho(C^{-1}) = J$
 with $J$ defined in (\ref{Jdef}), we have
 \be
 \rho({\cal K}) = \rho(C^{-1}) \rho( \fancys) =  J {\cal H}_{\bullet\,\bullet}
 = {\cal H}^\bullet_{~\, \bullet}\,,
 \ee
 since $J$ is identical to the matrix $\eta^{-1}$ that raises indices.  We write this conclusion as
 \be\label{Krepres}
S_{{\cal H}^{\bullet}{}_{\bullet}}=\pm\, {\cal K}  \,.
 \ee

\medskip
\noindent  We will show that the gauge transformation of ${\cal K}$ compatible with that of $ {\cal H}^\bullet_{~\, \bullet}$ takes the form
\be \label{gtS}
\delta_{\xi} {\cal K}  \ = \ \xi^M \partial_M  {\cal K}
+ {1 \over 2} \big[ \Gamma^{PQ}  , \,  {\cal K} \, \big]  \partial_{P} \xi_{Q} \; ,
\ee
where $\Gamma^{PQ} \equiv \frac{1}{2}  [\Gamma^P , \Gamma^Q ]$.
We will prove the above in a different but equivalent form, which reads
\be\label{gtSextra2}
\delta_{\xi}{\cal K}
\ = \   \ \xi^M \partial_M  {\cal K}
+ \frac{1}{2}  \left( \Gamma^{PQ}
 -\Gamma^{RS}\, {\cal H}^P{}_{R}\,{\cal H}^Q{}_{S}\right) {\cal K} ~\partial_{P} \xi_{Q}\;.
\ee
This, in turn, can be written more suggestively as
\be\label{gtSextra22}
(\delta_{\xi}{\cal K})\,{\cal K}^{-1}
\ = \   \ \xi^M (\partial_M  {\cal K}){\cal K}^{-1}
+ \frac{1}{2}  \left( \Gamma^{PQ}
-\Gamma^{RS}\, {\cal H}^P{}_{R}\,{\cal H}^Q{}_{S}\right) \partial_{P} \xi_{Q}\;.
\ee
To see that (\ref{gtSextra2}) is equivalent to (\ref{gtS})
we use that (\ref{invgamma2}) implies for any $h\in O(D,D)$
 \be
 \label{clclt}
  S\,\Gamma_{PQ}\,S^{-1} \ = \ \Gamma_{RS}\,h^{R}{}_{P}\,h^{S}{}_{Q}  \;,
~~~\rho(S) \ = \ h\,.  \ee
Specialized to the $O(D,D)$ element ${\cal H}^{M}{}_{N}$ this yields
 \be\label{STEp123}
  S_{{\cal H}^{\bullet}{}_{\bullet}}\,\Gamma_{PQ}\,(S_{{\cal H}^{\bullet}{}_{\bullet}})^{-1} \ = \
  \Gamma_{RS}\,{\cal H}^{R}{}_{P}\,{\cal H}^{S}{}_{Q}\;,
  \ee
  and with use of (\ref{Krepres}) 
we find
 \be\label{STEp1239}
{\cal K}\,\Gamma_{PQ}\,{\cal K}^{-1} \ = \
  \Gamma_{RS}\,{\cal H}^{R}{}_{P}\,{\cal H}^{S}{}_{Q}\ \quad \to \quad
{\cal K}\,\Gamma^{PQ} \ = \
 \Gamma^{RS}\, {\cal H}^P{}_{R}{\cal H}^Q{}_{S}\,{\cal K}\,.
  \ee
This final identity demonstrates the equivalence of (\ref{gtSextra2})
and (\ref{gtS}).

The strategy in our construction will be to express the gauge transformations as Lie algebra identities that can be realized both
in the fundamental and spin representations of $O(D,D)$.
To begin, we consider the transport term $\delta_\xi^t$ in the transformation (\ref{manifestH99}) of the generalized metric, written
as follows
\be
  (\delta^t_{\xi}{\cal H}^M{}_{P}) ({\cal H}^{-1})^{P}{}_{N} \ = \
  \xi^{L}\partial_{L}  {\cal H}^M{}_{P}({\cal H}^{-1})^{P}{}_{N}\,.
   \ee
This equality of Lie-algebra elements is here realized in the fundamental
representation.  In the spin representation, where the group element
${\cal H}^\bullet{}_{\bullet}$ is represented by ${\cal K}$ we would have
\be
(\delta^t_{\xi}{\cal K}) {\cal K}^{-1}  \ = \
  \xi^{L} (\partial_{L}  {\cal K}) \, {\cal K}^{-1} \,.
\ee
This proves that the transport term in (\ref{gtSextra2}) is required by consistency.  Calling $\Delta_{\xi}{\cal H}^{M}{}_{P}$ the non-transport
terms in the transformation, we now have
 \be\label{step0099}
 \begin{split}
  \Delta_{\xi}{\cal H}^{M}{}_{P}\,({\cal H}^{-1})^{P}{}_{N} \ &= \
  \Delta_{\xi}{\cal H}^{M}{}_{P}\,{\cal H}^{P}{}_{N} \\
  \ &= \
  (\partial^{M}\xi_{N}-\partial_{N}\xi^{M})
  +(\partial_{P}\xi^{K}-\partial^{K}\xi_{P}){\cal H}^{M}{}_{K}{\cal H}^{P}{}_{N}    \\
  \ &= \ \bigl(\partial_{P}\xi_{Q}-\partial_{R}\xi_{S}\,{\cal H}^{R}{}_{P}\,{\cal H}^{S}{}_{Q}\bigr) \, \bigl(\eta^{MP} \delta^Q{}_{N}  - \eta^{MQ} \delta^P{}_{N} \bigr)  \;,
 \end{split}
 \ee
where the last equality is readily checked by expansion of the product.
We now recognize the last factor in the last line of the above equation
as  $(T^{PQ})^M{}_{N}$, the Lie algebra
generator in the fundamental representation, as introduced in~(\ref{fund}).
We thus have
 \be
   \Delta_{\xi}{\cal H}^{M}{}_{P}\,({\cal H}^{-1})^{P}{}_{N} \ = \
 \left(\partial_{P}\xi_{Q}-\partial_{R}\xi_{S}\,{\cal H}^{R}{}_{P}\,{\cal H}^{S}{}_{Q}\right)
   (T^{PQ})^{M}{}_{N}\;.
 \ee
Passing to the spin representation with matrices (\ref{spinmat}) we find
 \be
 \begin{split}
 ( \Delta_{\xi}{\cal K}) \, {\cal K} ^{-1} & \ = \
  \frac{1}{2}\left(\partial_{P}\xi_{Q}-\partial_{R}\xi_{S}\,{\cal H}^{R}{}_{P}\,{\cal H}^{S}{}_{Q}\right)
  \Gamma^{PQ} \\
  & \ = \ \frac{1}{2}\left(\Gamma^{PQ}-\Gamma^{RS}\,{\cal H}^{P}{}_{R}\,{\cal H}^{Q}{}_{S}\right)
  \partial_{P}\xi_{Q}\,.
  \end{split}
 \ee
This coincides exactly with the non-transport term in (\ref{gtSextra22})
and concludes our proof that the postulated gauge transformation
(\ref{gtS}) of $\cal K$ is consistent with that of the generalized metric.

\section{Action, duality relations, and gauge symmetries}
In this section we introduce the $O(D,D)$ covariant double field theory
formulation of the RR action and the duality relations. We prove T-duality invariance and
gauge invariance, and we determine the $O(D,D)$ covariant form of the field equations.

\subsection{Action, duality relations, and $O(D,D)$ invariance}
\label{actduarelandodd}

The dynamical field we will use to write an action is a spinor of
Pin$(D,D)$
written as in (\ref{genstate}):
 \be
 \label{gen_spinor_bz_new}
 \chi \ \equiv \ \ket{\chi} \ = \ \sum_{p=0}^{D}\frac{1}{p!}\,C_{i_1\ldots i_p}\,\psi^{i_1}\ldots\psi^{i_p}\ket{0}\;.
 \ee
Here the component forms $C_{i_1\ldots i_p} (x, \tilde x)$ are the dynamical fields and, as is usual in double field theory, they are
real functions of the full collection of $2D$ coordinates $x$ and $\tilde x$.
We will assume $\chi$ to have a definite chirality. Thus, as discussed in sec.~\ref{chiralspinors},
 it consists
either of only odd forms or even forms.
The bra associated with this ket is called $\chi^\dagger$ 
and is defined by 
 \be
 \label{gen_spinor_bra}
\chi^\dagger \ \equiv \ \bra{\chi} \ = \ \sum_{p=0}^{D}\frac{1}{p!}\,C_{i_1\ldots i_p}\,
 \bra{0} \psi_{i_p}\ldots\psi_{i_1}\;.
 \ee
We conventionally define the conjugate spinor using the $C$ matrix
defined in section~\ref{csection}:
 \be\label{conjSpinor}
  \bar{\chi} \ \equiv \ \chi^{\dagger}C\;.
 \ee
We will make use of a Dirac operator on spinors that behaves just
as an exterior derivative on the associated forms:
\be
\label{def-dir-op}
\slashed\partial  \equiv   {1\over \sqrt{2}} \, \Gamma^M \partial_M
=   \ \psi^i\partial_i+\psi_i\tilde{\partial}^{i} \,,
\ee
where we used (\ref{defgamma}).  The $\slashed{\partial}$
operator behaves like the exterior derivative $d$ in that its repeated action gives zero:
 \be
  \slashed{\partial}^2 \ = \ \frac{1}{2}\Gamma^{M}\Gamma^{N}\partial_{M}\partial_{N} \ = \  \frac{1}{4}\{ \Gamma^{M},\Gamma^{N}\}
  \partial_{M}\partial_{N} \ = \
  \frac{1}{2}\eta^{MN}\partial_{M}\partial_{N} \ = \ 0\,,
 \ee
by the constraint (\ref{ODDconstr}).
The $\slashed\partial$ operator will be used to define field strengths in
a Pin$(D,D)$  covariant way.
It is clear that
acting on forms that do not depend on $\tilde x$, the only term that
survives, $\psi^i \partial_i$, both differentiates with respect to $x$ and
increases the degree of the form by one.  More details will be given
in section~\ref{sss5}.

\medskip
We turn now to a discussion of the double field theory action.
We claim that the RR action is
$S = \int dx d \tilde{x} {\cal L}$, where the Lagrangian density ${\cal L}$
is simply given by
\be
\label{action-first-form-var}
{\cal L} \ = \    \frac{1}{4}(\slashed{\partial}{\chi})^\dagger \;
 \fancys \;  \slashed\partial\chi \,.
 \ee
The above Lagrangian is manifestly real:
${\cal L}^\dagger =  {\cal L}$ because the spinor
$\chi$ is Grassmann even and $\fancys$ is Hermitian.
The Lagrangian can be written using conjugate spinors
and the kinetic operator ${\cal K } = C^{-1} \fancys$.
We claim that the above Lagrangian is equal to
\be
\label{action-first-form}
{\cal L} \ = \ \frac{1}{8} \,\partial_{M}\bar{\chi}\,\Gamma^{M} \, {\cal K} \,  \Gamma^{N}\partial_{N}\chi \; .
 \ee
Indeed, using the conjugate spinor (\ref{conjSpinor}) and
(\ref{GammaunderC})
 this second version is written as
\be
\begin{split}
{\cal L} \ &= \ \frac{1}{8} \,\partial_{M}\chi^\dagger C \,\Gamma^{M} \, C^{-1}\, \fancys \,  \sqrt{2} \slashed{\partial}\chi
\ = \ \frac{1}{8} \,\partial_{M}\chi^\dagger  \,(\Gamma^{M})^\dagger \, \fancys \,  \sqrt{2} \slashed{\partial}\chi \\
\ &= \  \frac{1}{8} \,\sqrt{2} (\slashed{\partial}\chi)^\dagger   \, \fancys \,  \sqrt{2} \slashed{\partial}\chi \  =  \
\frac{1}{4}(\slashed{\partial}{\chi})^\dagger \;
 \fancys \;  \slashed\partial\chi\,.
\end{split}
\ee
The properties of bar conjugation allow us to recognize that
\be
\overline{\slashed\partial \chi} =  {1\over \sqrt{2}}  \bigl( \Gamma^M
\partial_M \chi\bigr)^\dagger  C =
 {1\over \sqrt{2}}  \,\partial_M \chi^\dagger  (\Gamma^{M})^\dagger C
=    {1\over \sqrt{2}}  \,\partial_M \bar\chi  C^{-1}
(\Gamma^{M})^\dagger C =  {1\over \sqrt{2}}  \,\partial_M \bar\chi
{\Gamma^{M}}\,,
\ee
and therefore we can write the action more compactly as
\be
\label{action-sec-form}
{\cal L} \ = \ \frac{1}{4}\,\overline{\slashed\partial \chi} \,\, {\cal K} \,\,  \slashed\partial \chi \; .
\ee

Our first task now  is to establish the global Spin$(D,D)$ invariance of this Lagrangian (the $dxd\tilde x$ measure is $O(D,D)$ invariant). This is the maximal invariance group that is consistent
with the fixed chirality of $\chi$.
Under the action of a Spin$(D,D)$ element $S$, whose associated
$O(D,D)$ element is $h = \rho (S)$, the spinor field $\chi$ transforms
as follows:
\be
\label{field-spinor-trans}
  \chi \,\rightarrow\, \chi^{\prime} \ = \ S\,\chi\;.
 \ee
 Implicit in here is that the coordinates the fields depend on are also
 transformed: primed fields depend on primed coordinates $X'^M
 = h^M{}_{N} X^N$.   Note also that the daggered state transforms~as
 \be
 \label{dagchi}
 \chi^\dagger \,\rightarrow ~\chi^\dagger\,  S^\dagger\;.
 \ee
 We also have that
 \be
 \label{vmbv}
 \begin{split}
  \slashed{\partial}\chi =
  {1\over \sqrt{2}} \Gamma^M \partial_M \chi \,\rightarrow\,
  & \frac{1}{\sqrt{2}}\,\Gamma^{M}(h^{-1})^{N}{}_{M}\partial_{N}S\chi
  \ = \, \frac{1}{\sqrt{2}}\,S \, [S^{-1} \Gamma^{M} S]
  (h^{-1})^{N}{}_{M}\partial_{N}\chi\,.
 \end{split}
\ee
We now use (\ref{invgamma}) to find
\be
\label{vmbvg}
  \slashed{\partial}\chi
 \,\rightarrow\,
   \, \frac{1}{\sqrt{2}}\,S \, h^M{}_{P} \Gamma^P
  (h^{-1})^{N}{}_{M}\partial_{N}\chi =   \frac{1}{\sqrt{2}} \,S \, \Gamma^{N}\partial_{N}\chi\;,
\ee
and therefore we have
\be\label{covslash}
 \begin{split}
  \slashed{\partial}\chi \, \rightarrow \,
   \ S\, \slashed{\partial}\chi\;.
\end{split}
\ee
We have thus leaned that $\slashed\partial \chi$ transforms just like
$\chi$.  In other words, the Dirac operator $\slashed\partial$ is
Spin$(D,D)$ invariant.   Recalling the transformation of $\fancys$
in (\ref{vmvg}) :  $\fancys \to   (S^{-1})^\dagger  \, \fancys \, S^{-1}$,
the invariance of the Lagrangian (\ref{action-first-form-var})  is essentially manifest:
\be
\label{action-first-form-var-99}
{\cal L} \ = \    \frac{1}{4}(\slashed{\partial}{\chi})^\dagger \;
 \fancys \;  \slashed\partial\chi  \;~\to~\;
  \frac{1}{4}(\slashed{\partial}{\chi})^\dagger  S^\dagger  \,  (S^{-1})^\dagger  \, \fancys \, S^{-1}  \, S  \slashed{\partial}{\chi} \ = \ {\cal L}\,.
 \ee

\medskip

The action must be supplemented by duality constraints
 among the field strengths.  We can write
 Spin${}^+(D,D)$
 covariant versions of the duality relations that relate all
RR field strengths:\footnote{For the special case
of type IIA, 
a similar $O(D,D)$-covariant form of the duality relations has also been proposed in the second reference of \cite{West:2010ev}.}
 \be\label{ODDdual}
  \slashed{\partial}\chi \ = \   -{\cal K}\, \slashed{\partial}\chi\;.
 \ee
According to (\ref{covslash}), the left-hand side transforms covariantly with $S\in{\rm Spin}(D,D)$.
The right-hand side transforms in the same way, since
 \be
  -{\cal K}\, \slashed{\partial}\chi \;\;\rightarrow \;\;
   -C^{-1}(S^{-1})^{\dagger}\,\fancys\,S^{-1}S\slashed{\partial}\chi
  \ = \ -S\, C^{-1}\, \fancys\,\slashed{\partial}\chi \ = \ -S\,{\cal K}\, \slashed{\partial}\chi\;,
 \ee
where we used that (\ref{finalSDagger}) implies $C^{-1}(S^{-1})^{\dagger}=SC^{-1}$
for $S\in{\rm Spin}^{+}(D,D)$. Thus, the duality relations are actually only invariant
under Spin$^{+}(D,D)$. This is to
be expected
since already for conventional
duality relations the presence of an epsilon tensor breaks the symmetry to
the group $GL^{+}(D)$ of parity-preserving transformations.

The relations (\ref{ODDdual}) require a consistency condition.
Acting on both sides of (\ref{ODDdual}) with ${\cal K}$, we see that
consistency requires ${\cal K}^2=1$, which in turn implies
 \be
 \label{ksqcon}
  {\cal K}^2 \ = \ C^{-1}\,\fancys\,C^{-1}\,\fancys \ = \
  C(\fancys\,C\,\fancys) \ = \ C(-C)
   \ = \ -(-1)^{\frac{1}{2}D(D-1)} \ = \ 1\;,
 \ee
where we used (\ref{Ctrans}) and (\ref{Cinverse}). Thus, the duality relations
are self-consistent in dimensions for which $\tfrac{1}{2}D(D-1)$ is odd.  For $D\leq 10$, these are
 \be\label{validD}
  D \, = \, \big\{\,10\,,7\,,6\,, 3\,,2 \,\big\}\;.
 \ee
We note that the
even dimensions above are precisely those for which
conventional self-duality relations can be imposed
consistently.
Indeed, the middle degree forms
corresponding to the self-dual field strengths
are then odd, and for them
$\star^2=1$
in Lorentzian signature.
 As we will show  in sec.~\ref{513}
the component form of (\ref{ODDdual}) contains one self-duality
relation in even dimensions, so this result is to be expected. In the following we will focus on $D=10$,
but we note that
$D=2,6$
can be seen as type II  toy models.
The possible significance of theories with odd $D$ will not be discussed here.

We close by giving the equations of motion of $\chi$, which are
readily derived from (\ref{action-first-form}),
 \be
 \slashed\partial \bigl( {\cal K}\,\slashed{\partial}\chi\bigr) \ = \ 0\;.
 \ee
As it should be, the equation of motion is the integrability condition
for the duality relations:  acting
with a $\slashed{\partial}$ on both sides of (\ref{ODDdual}), and using $\slashed{\partial}^2=0$,
we recover the field equation.

\subsection{Gauge invariance}

In this subsection we give the gauge transformation of the RR fields.
The $p$-form gauge transformations are manifestly invariances of
the Lagrangian and of the duality constraints.  For the gauge transformations parameterized by $\xi^M$ the transformation of
$\chi$ is nontrivial and so are the checks of gauge invariance of
the Lagrangian and the duality constraints.

\subsubsection{Gauge transformations}
We start by introducing the double field theory version of the abelian gauge symmetries
of the $p$-form gauge fields.
These are parameterized by a
spacetime dependent spinor $\lambda$:
\be\label{lambdagauge}
\delta_\lambda  \chi  =  \slashed{\partial} \lambda  \,.
\ee
Since $\lambda$ encodes a set of forms and $\slashed \partial$
acts as an exterior derivative, the above transformations are the
familiar ones.  It follows that
\be
\delta_\lambda  \slashed{\partial}\chi  =  \slashed{\partial}\slashed{\partial} \lambda = 0\,,
\ee
and this implies the gauge invariance of the
Lagrangian density~(\ref{action-first-form-var}) and of the duality constraint (\ref{ODDdual}).

For the gauge parameter
$\xi^M$ that encodes
the diffeomorphism and Kalb-Ramond gauge symmetries,
we postulate the gauge transformation
\be \label{gtchi}
~\begin{split}
\delta_\xi \chi \ = \ \widehat{\cal L}_\xi \chi  \ & \equiv
\ \xi^M \partial_M  \chi  \ + \   {1\over \sqrt{2}}\, \slashed{\partial} \xi^M
\Gamma_{\hskip-1pt M}\, \chi  \, \\
&= \ \xi^M \partial_M  \chi  \ + \ {1\over 2} \,  \partial_N \xi_M \Gamma^N
\Gamma^M \chi\;. ~~
\end{split} 
\ee
In the second form it is simple to verify that a gauge parameter of
the form $\xi_M = \partial_M \Theta$ is trivial in that it generates no gauge transformations:
\be
\delta_{\partial \Theta} \chi  \ = \  \partial^M \Theta\, \partial_M  \chi  \ +  \   {1\over 2} \partial_N
\partial_M \Theta\, \Gamma^N
\Gamma^M \chi \ = \  \frac{1}{2}\partial_N \partial_M \Theta\, \eta^{MN} \chi \ = \ 0\,.
\ee
A short calculation gives the gauge transformation of the conjugate
spinor $\bar \chi$:
\be
\label{conjgt}
\delta_\xi \bar\chi
\ = \ \xi^M \partial_M \bar \chi  \ + \ {1\over 2} \,  \partial_N \xi_M \,
\bar\chi\,\Gamma^M
\Gamma^N \,.
\ee

Let us now turn to the gauge algebra. 
We claim that
the gauge transformations parametrized by $\lambda$
and $\xi^{M}$ close as follows
\be
\big[\delta_\lambda \,, \delta_\xi \big]   =  \delta_{\widehat{\cal L}_\xi \lambda}  \;.
\ee
To check this we consider the left-hand side acting on $\chi$:
\be
\label{lsvm}
\begin{split}
\big[\delta_\lambda \,, \delta_\xi \big]  \chi =  \delta_\lambda \delta_\xi  \chi
\ = \ &  \delta_\lambda \bigl( \xi^M \partial_M  \chi  \ + \  {1\over 2} \partial_N \xi_M \Gamma^N
\Gamma^M \chi   \bigr) \\
 \ =\  &
\xi^M \partial_M \slashed{\partial} \lambda   \ + \  {1\over 2}  \partial_N \xi_M \Gamma^N
\Gamma^M \slashed{\partial} \lambda \;.
\end{split}
\ee
The right-hand side of the expected algebra is:
\be
\label{vm}
\begin{split}
\delta_{\widehat{\cal L}_\xi \lambda} \chi \ = \ &\slashed{\partial}
\widehat{\cal L}_\xi \lambda =   \slashed{\partial}
\bigl( \xi^M \partial_M  \lambda  \ + \  {1\over 2}  \partial_N \xi_M \Gamma^N
\Gamma^M \lambda \bigr)\\
\ = \   & \frac{1}{\sqrt{2}}  \partial_P
\bigl(  \xi^M \partial_M \Gamma^P \lambda  \ + \ {1\over 2}   \partial_N \xi_M \Gamma^P\Gamma^N
\Gamma^M \lambda \bigr)\\
\ = \  & \xi^M \partial_M  \slashed{\partial} \lambda  \
+\frac{1}{\sqrt{2}}  \partial_P
 \xi^M  \Gamma^P \partial_M\lambda
+  \   {1\over 2 \sqrt{2}}  \partial_N \xi_M \Gamma^P\Gamma^N
\Gamma^M \partial_P\lambda\;, \\
\end{split}
\ee
where the term with two derivatives on $\xi$ vanishes by the use
of $\slashed{\partial}^2 = 0$.
Using the commutator
\be
\label{2gs}
[\,  \Gamma^P\,, \Gamma^N
\Gamma^M\,]
=  2\eta^{PN}  \Gamma^M  - 2 \eta^{PM} \Gamma^N\,,
\ee
one can readily show that
\be
{1\over 2 \sqrt{2}}\, \partial_N \xi_M[\,  \Gamma^P\,, \Gamma^N
\Gamma^M\,] \partial_P\lambda \
=  -  {1 \over \sqrt{2}} \,\partial_P \xi^M  \, \Gamma^P \partial_M\lambda\;,
\ee
where we used the constraint and relabeled the indices.   Then,
returning to (\ref{vm}),
\be
\delta_{\widehat{\cal L}_\xi \lambda}   \chi
\ = \
\xi^M \partial_M  \slashed{\partial}
  \lambda  \ + \    {1 \over 2}  \,  \partial_N \xi_M\Gamma^N
\Gamma^M \slashed{\partial}\lambda\;.
\ee
This agrees with (\ref{lsvm}) confirming the closure of the gauge
algebra.
We have also verified that, as expected,
 $[\delta_{\xi_1}, \delta_{\xi_2}]
=-\delta_{[\xi_1, \xi_2]_{\rm C}}$, where $[\,\cdot \,, \, \cdot ]_{\rm C}$
is the C-bracket
discussed in~\cite{Hohm:2010pp}.

\subsubsection{Gauge invariance of the action
and the duality constraints}\label{gauinvoftheact}

The action is manifestly invariant under $p$-form gauge transformations. Here
we check the invariance under $\delta_\xi$.
We use the Lagrangian in (\ref{action-first-form}):
\be
\label{action-sec-form}
{\cal L} \ = \ \frac{1}{4}\,\overline{\slashed\partial \chi} \,\, {\cal K} \,\,  \slashed\partial \chi \; .
\ee
As usual, when we vary the Lagrangian, which has the index structure
of a scalar, we obtain a transport term and a `non-covariant' term
\be
\label{orgcalcgiL}
\delta_{\xi} {\cal L} \ = \  \xi^M \partial_{M}  {\cal L} + \Delta_{\xi}  {\cal L} \; .
\ee
Since $\Delta_\xi$ acts as a derivation and commutes with bar-conjugation,
\be
\label{Deltaxiaction}
\Delta_{\xi}  {\cal L}  = \frac{1}{4}\, \Bigl(
(\overline{\Delta_\xi\slashed\partial \chi}) \,\, {\cal K} \,\,  \slashed\partial \chi
 \ +\ \overline{\slashed\partial \chi} \,\,(\Delta_\xi {\cal K}) \,\,
\slashed\partial \chi
\ +\ \overline{\slashed\partial \chi} \,\, {\cal K} \,\, \Delta_\xi \slashed\partial \chi\Bigr)\,.
\ee
For the action to be gauge invariant, $\Delta_\xi{\cal L}$ must be such
that $\delta_\xi {\cal L}$ in (\ref{orgcalcgiL}) is a total derivative.
Since $\Delta_\xi {\cal K}$ can be read from (\ref{gtS}), we only have to
calculate $\Delta_\xi  \slashed{\partial} \chi$.  We begin my noting that
\be
\delta_{\xi} \big( \slashed{\partial} \chi \big) \ = \ \frac{1}{\sqrt{2}}
\Gamma^M \partial_{M} \Big( \xi^P \partial_P  \chi  \ + \ {1\over 2}   \partial_P \xi_Q \Gamma^P  \Gamma^Q \chi \Big) \; .
\ee
The noncovariant piece in this transformation includes all terms in the right-hand side except for
$\xi^P \partial_P \slashed{\partial} \chi$. Therefore we have
\be
\Delta_{\xi} \big( \slashed{\partial} \chi \big) \ = \ \frac{1}{\sqrt{2}}\Big(
 \partial_M \xi^P  \Gamma^M \partial_P\chi
+ \ {1\over 2} \, \partial_P \xi_Q  \Gamma^M\Gamma^P  \Gamma^Q \partial_{M}\chi\Big)  \,,
\ee
since the term with two derivatives on $\xi$ vanishes.  A short computation
using (\ref{2gs}) to bring $\Gamma^M$ next to the spinor gives the final
answer
\be
\Delta_{\xi} \big( \slashed{\partial} \chi \big) \ = \
  {1\over 2} \, \partial_P \xi_Q \, \Gamma^P\Gamma^Q \slashed{\partial} \chi  \,.
\ee
Bar conjugation immediately yields,
\be
\overline{\Delta_{\xi} \big( \slashed{\partial} \chi \big)} \ = \
  {1\over 2} \, \partial_P \xi_Q ~  \overline{ \slashed{\partial} \chi}\,\Gamma^Q\Gamma^P
   \,.
\ee
Using the above variations and (\ref{gtS}) we find that (\ref{Deltaxiaction})
gives
\be
\label{Deltaxiaction99}
\Delta_{\xi}  {\cal L}  = \frac{1}{8}\, \partial_P\xi_Q \, \,
\overline{ \slashed{\partial} \chi}\,
\Bigl( \Gamma^Q  \Gamma^P  \, {\cal K} +  \big[ \Gamma^{PQ}  ,   \,{\cal K}  \big]
+ \ {\cal K} \,\Gamma^P  \Gamma^Q \Bigr) \, \slashed\partial \chi  \,.
\ee
A short calculation shows that the factor in parenthesis equals $2\eta^{PQ}
{\cal K}$.  As a result we find
\be
\label{Deltaxiaction999}
\Delta_{\xi}  {\cal L}  = \frac{1}{4}\, \partial_M\xi^M \, \,
\overline{ \slashed{\partial} \chi}\,
 \, {\cal K}  \, \slashed\partial \chi  \ = \  \partial_M\xi^M \, {\cal L} \,.
\ee
Back in (\ref{orgcalcgiL})  we get
$\delta_{\xi} {\cal L}  =  \xi^M \partial_{M}  {\cal L} + ( \partial_M \xi^{M} ) {\cal L} = \partial_{M} (\xi^{M} {\cal L} )$, which confirms the gauge invariance of the action.

\medskip
Finally, we have to prove gauge
covariance
of the duality
constraints  $\slashed{\partial}\chi = - {\cal K}\slashed{\partial}\chi $.
We now take the gauge variation $\delta_\xi$ of both sides of the duality constraint.  The transport terms on both sides are identical, using the
duality constraint.  So only the non-covariant terms matter, and we can evaluate $\Delta_\xi$ on both sides of the constraint, finding
\be
\Delta_\xi\slashed{\partial}\chi \ =\  -(\Delta_\xi{\cal K})
\slashed{\partial}\chi \
 -\ {\cal K}\Delta_\xi\slashed{\partial}\chi \,.
\ee
Our task is to verify that this holds, using the duality constraint.  Bringing
all terms to one side we must check that
\be
\Delta_\xi\slashed{\partial}\chi \ + \ (\Delta_\xi{\cal K})
\slashed{\partial}\chi
 \ + \  {\cal K}\Delta_\xi\slashed{\partial}\chi   \ = \ 0 \,.
\ee
Using our earlier results we find that the left-hand side is equal to
\be
 {1\over 2} \, \partial_P \xi_Q \,
 \Bigl(  \Gamma^P\Gamma^Q  + \, [\Gamma^{PQ}, {\cal K} \, ]
 +  \ {\cal K} \, \Gamma^P\Gamma^Q \Bigr)   \slashed{\partial} \chi\,.
\ee
Expanding the commutator and using the duality constraint we find
that the above becomes
\be
 {1\over 2} \, \partial_P \xi_Q \,
 \Bigl( ( \Gamma^P\Gamma^Q- \Gamma^{PQ})    +  \ {\cal K} \, (\Gamma^P\Gamma^Q- \Gamma^{PQ}) \Bigr)   \slashed{\partial} \chi
 \ = \  {1\over 2} \, \partial_P \xi_Q \,  \eta^{PQ}
( 1   +  \ {\cal K}  )   \slashed{\partial} \chi  \ = \ 0\,.
\ee
This concludes our proof.

\subsection{General variation of $\fancys$ and gravitational
equations of motion}\label{genvarSgraveans}

In this section we determine the general variation of the action under a variation of $\fancys$
in order to determine the contribution of the new action to the field
equations. This is non-trivial since $\fancys$ is a \textit{constrained} field in that it takes
values in Spin$(D,D)$. The corresponding problem for the constrained variable given
by the generalized metric ${\cal H}$ has been discussed in \cite{Hohm:2010pp}, and the method
employed there can be elevated to $\fancys$, as we discuss next.

In \cite{Hohm:2010pp}, sec.~4,  it was
shown that a general variation of the
constrained variable ${\cal H}$ can be parametrized in terms of a symmetric but
otherwise unconstrained matrix ${\cal M}^{MN}$ as follows
 \be
 \begin{split}
  \delta{\cal H}^{MN} \ &= \ \frac{1}{4}\left[ \big(\delta^{M}{}_{P}+{\cal H}^{M}{}_{P}\big)
  \big(\delta^{N}{}_{Q}-{\cal H}^{N}{}_{Q}\big)+\big(\delta^{M}{}_{P}-{\cal H}^{M}{}_{P}\big)
  \big(\delta^{N}{}_{Q}+{\cal H}^{N}{}_{Q}\big)\right]{\cal M}^{PQ}\\
  &= \ {1\over 2}  \Bigl[ ~ {\cal M}^{MN} - {\cal H}^M{}_{P}\, {\cal M}^{PQ}\,
  {\cal H}^N{}_{Q}~ \Bigr]\,.
 \end{split}
 \ee
Lowering the $N$ index,
 \be
 \label{skns}
   \delta{\cal H}^M{}_{N} \ = \ {1\over 2}  \Bigl[ ~ {\cal M}^M{}_{N} - {\cal H}^M{}_{P}\, {\cal M}^{PQ}\,
  {\cal H}_{NQ}~ \Bigr]\,.
 \ee
 As in section~\ref{gaugtransfancys} we now form the Lie-algebra element
\be
\begin{split}
(\delta {\cal H}^M{}_{P}) {\cal H}^P{}_{N} \ & = \
{1\over 2}  \Bigl(  {\cal M}^M{}_{R}  {\cal H}^R{}_{N}
- {\cal H}^M{}_{R}  {\cal M}^R{}_{N}~\Bigr)\\
& = \   {1\over 2}  \, {\cal M}_{PR}  {\cal H}^R{}_{Q} \Bigl( \,
\eta^{MP} \delta^Q{}_{N}  - \eta^{MQ} \delta^P{}_{N} \Bigr) \\
& = \   {1\over 2}  \, {\cal M}_{PR}  {\cal H}^R{}_{Q} \bigl( T^{PQ}
\bigr)^M{}_{N} \,,
\end{split}
\ee
where we made repeated use of the symmetry properties of
${\cal H}$ and ${\cal M}$ and used (\ref{fund}).  In the spin
representation this equation yields
\be
(\delta {\cal K}) \,{\cal K}^{-1}  \ = \ {1\over 4}
 {\cal M}_{PR}  {\cal H}^R{}_{Q} \, \Gamma^{PQ}
 \ = \    {1\over 4}  \, {\cal M}_{MN}  {\cal H}^M{}_{P}  ~\Gamma^{NP} \,,
\ee
after some index relabeling.  Our final result for the variation is therefore
\be
\label{uncKvar}
\delta {\cal K}
 \ = \    {1\over 4}  \, {\cal M}_{MN}  {\cal H}^M{}_{P}  ~\Gamma^{NP}
 \, {\cal K} \,.
\ee
This, with ${\cal H}^\bullet_{~\bullet} = \rho({\cal K})$,
 is the general variation of ${\cal K}$ consistent
with its group property ${\cal K} \in \text{Spin}(D,D)$.  It is
consistent with the variation (\ref{skns}), and thus
 the variation of the NS-NS action is unmodified as compared to
the discussion in \cite{Hohm:2010pp}.

Next, we apply (\ref{uncKvar}) in order to compute the variation of the RR action
 \be
  \delta {\cal L} \ = \ \frac{1}{4}~\overline{\slashed{\partial}\chi} \;\delta\,
  {\cal K} \; \slashed{\partial}\chi
  \ = \ \frac{1}{16}{\cal M}_{MN}\, {\cal H}^{M}{}_{P}\,
  \, \overline{\slashed{\partial}\chi} \,
 \,\Gamma^{NP} {\cal K} \,\slashed{\partial}\chi   \;.
 \ee
Since ${\cal M}$ is an arbitrary symmetric matrix, we read off that the contribution to the
field equations is given by the symmetric `stress-tensor'
 \be
 \label{rrstress-tensor}
  {\cal E}^{MN} \ = \  \frac{1}{16} {\cal H}^{(M}{}_{P}\,
  \overline{\slashed{\partial}\chi} \,
 \,\Gamma^{N)P}\,{\cal K}~\slashed{\partial}\chi\;.
 \ee
 It is possible to verify that, as required, the above symmetric tensor
 is real $ ({\cal E}^{MN})^\dagger =   {\cal E}^{MN}$.  This
 calculation makes use of  $C^\dagger = C^{-1}$ and (\ref{STEp1239}).
 It is also important to note that ${\cal E}^{MN}$ transforms covariantly
 under duality:
 \be
 {\cal E}'^{MN}(X')  \ = \ h^M{}_{P}\, h^N{}_{Q} \, {\cal E}^{PQ}(X)\,.
 \ee
The explicit check makes use of (\ref{clclt}) and the duality properties
of ${\cal H}$.

 Taking the variation
of the NS-NS action into account, which leads to the tensor ${\cal R}_{MN}$  defined
in eq.~(4.58) of \cite{Hohm:2010pp}, this leads to the $O(D,D)$ covariant form of the
type II field equations,
 \be
  {\cal R}_{MN}+{\cal E}_{MN} \ = \ 0\;,
 \ee
supplemented by the duality constraint (\ref{ODDdual}).  In fact, the
duality constraint allows us to simplify ${\cal E}^{MN}$ considerably:
 \be
 \label{rrstress-tensor+duality}
  {\cal E}^{MN} \ = \  -\, \frac{1}{16} \,{\cal H}^{(M}{}_{P}\,
  \overline{\slashed{\partial}\chi} \,
 \,\Gamma^{N)P}\,\slashed{\partial}\chi\;.
 \ee
One may try to verify again the reality of this stress-tensor.  A short
calculation shows that it is only real whenever $C C = -1$.  This
is precisely the constraint for consistent duality constraints, as discussed
at the end of section~\ref{actduarelandodd}.  Since we work with real numbers
throughout, a non-real stress-tensor can only be equal to zero.

\section{Action and duality relations in the standard frame}\label{sss5}

In this section we examine the form of the action and duality relations when choosing
the `standard' duality frame $\tilde{\partial}^{i}=0$, and we show
that
they reduce
to the conventional democratic formulation of type II theories. For this we have to
assume that we are in a region with a
well-defined  metric,
so that we can
choose the parametrization $\fancys=S_{\cal H}$.
The physical significance of this particular parametrization will be discussed in
the next section.

\subsection{Action and duality relations  in  $\tilde\partial=0$ frame}
In this section we evaluate the action and duality relations in the standard frame $\tilde\partial=0$.
We begin by
deriving
some relations which will turn out to be useful for this analysis.

\subsubsection{Preliminaries}
Let us
derive some useful identities for the action of $S_{g}$ on general spinor states.
To this end we need to determine the action of an exponential of fermionic oscillators.
We find
 \be\label{expAction}
 \begin{split}
  e^{\psi^{i}R_{i}{}^{j}\psi_{j}}\psi^{k}\ket{0} \ &= \ \Big(1+R_{i}{}^{j} \psi^{i}\psi_{j}
  +\frac{1}{2}R_{i}{}^{j} R_{p}{}^{q}\psi^{i}\psi_{j}\psi^{p}\psi_{q}+\cdots\Big)\psi^{k}\ket{0} \\
  \ &= \ \Big(\psi^{k}+\psi^{i}R_{i}{}^{j}\big\{\psi_{j},\psi^{k}\big\}+\frac{1}{2}R_{i}{}^{j} R_{p}{}^{q}\psi^{i}
  \big\{\psi_{j},\psi^{p}\big\}\big\{\psi_{q},\psi^{k}\big\}+\cdots\Big)\ket{0} \\
  \ &= \ \Big(\delta_{l}{}^{k}+R_{l}{}^{k}+\frac{1}{2}R_{l}{}^{j} R_{j}{}^{k}+\cdots\Big)\psi^{l}\ket{0}
  \ = \
   \big(\exp R\big)_l{}^{k}\,\psi^{l}\ket{0}\;.
 \end{split}
 \ee

In order to determine now the action of $S_{g}=S_{e}S_{k}S_{e}^{\dagger}$ on general states,
we compute the action of the respective factors.
For $S_{e}$, we introduce $e = \exp(E)$
and we have
 \be\label{ActioN1}
  S_{e}\,\psi^{i}\,\ket{0} \
  = \  {1 \over \sqrt{\det{e}}}\, e^{ \psi^{j}E_{j}{}^{k}\psi_{k}}\,\psi^{i}\,\ket{0} \ = \
  {1 \over \sqrt{\det{e}}}\,
  \big(\exp E\big)_j{}^{i}
  \,\psi^{j}\,\ket{0} \ = \
  {1 \over \sqrt{\det{e}}}\,  e_j{}^{i}
  \psi^{j}\,\ket{0}\;.
 \ee
For $S_{e}^{\dagger}$  we find an expression
with unusual index position
 \be\label{ActioN2}
  S_{e}^{\dagger}\,\psi^{i}\,\ket{0} \ = \ {1\over \sqrt{\det{e}}}\, e_{i}{}^{j}\,\psi^{j}\,\ket{0}\;.
 \ee
The action of $S_{k}$ can be easily computed,
 \be\label{ActioN3}
  S_{k}\,\psi^{p}\, \ket{0} \ = \ \big(\psi^{1}\psi_{1}-\psi_{1}\psi^{1}\big)\,\psi^{p}\,\ket{0}
  \ = \ -  k_{pq}
  \,\psi^{q}\ket{0}\;,
 \ee
using the flat Lorentz metric $k=
\text{diag}(-1,1,\ldots,1)$
defined in (\ref{define-k-matrix}).
Using (\ref{ActioN1}), (\ref{ActioN2}) and (\ref{ActioN3}),  the action of $S_{g}$ is then given by
 \be\label{finalSg1}
 \begin{split}
  S_{g} \,\psi^{i}\,\ket{0} \ &= \ S_{e}S_{k}S_{e}^{\dagger}\,\psi^{i}\,\ket{0}
  \ =
  \ {1 \over \sqrt{\det e}}\,S_{e}S_{k}\, e_{i}{}^{j}\,\psi^{j}\,\ket{0}
  \ =
  \ -{1 \over \sqrt{\det e}}\,S_{e}\, e_{i}{}^{j}\,k_{jp}\,\psi^{p}\,\ket{0}\\
  \ &  =
  \ -{1 \over \det e}\, (e_{i}{}^{j}\,k_{jp}\,e_q{}^p) \psi^{q}\,\ket{0}
  =  \ -{1 \over \det e}\, (e_{i}{}^{j}\,e_q{}^p\, k_{jp}\,) \psi^{q}\,\ket{0} \\
  & =
  \ -{1 \over \det e}\, g_{iq} \,\psi^{q}\,\ket{0}
  \ =
   \ -{1 \over \sqrt{| \det g|}}\,g_{iq}\,\psi^{q}\,\ket{0}\;,
 \end{split}
 \ee
where we used the definition of the metric in  (\ref{define-k-matrix})
and wrote $\det e = \sqrt{| \det g |}$. Similarly, for $S_{g}^{-1}$ one finds
 \be\label{finalSg2}
  S_{g}^{-1} \,\psi^{i}\,\ket{0} \ = \ -\sqrt{| \det g|}\,\, g^{ij}
  \, \psi^{j}\,\ket{0}\;\,,
 \ee
where $g^{ij}$ is, as usual, the inverse of the metric $g_{ij}$.

All of the above relations straightforwardly extend to the case where $S_{g}$ acts
on multiple fermionic oscillators, for which eqs.~(\ref{finalSg1}) and (\ref{finalSg2})
are generalized to
 \be\label{FINALSg}
  \begin{split}
   S_{g}^{-1}\, \psi^{i_1}\cdots \psi^{i_p}\ket{0} \ &= \ -\sqrt{| \det g|}\, g^{i_1j_1}\cdots
   g^{i_pj_p}\,\,\psi^{j_1}\cdots \psi^{j_p}\,\ket{0}\;, \\
   S_{g}\, \psi^{i_1}\cdots \psi^{i_p}\ket{0} \ &= \ -\frac{1}{\sqrt{| \det g|}}\,
   g_{i_1j_1}\cdots g_{i_p j_p}\,\psi^{j_1}\cdots \psi^{j_p}\,\ket{0}\;.
  \end{split}
 \ee
With these ingredients we are now ready to evaluate the action.

\subsubsection{The action}
We start by writing the action in the duality frame $\tilde{\partial}=0$.  For this choice,
the field strength
\be\label{FieldStrength123}
\ket{F}\equiv \slashed\partial \ket{\chi}\,,
\ee
reduces to
 \be\label{ODDfields}
 \begin{split}
   \ket{F} \Big|_{\tilde{\partial}=0} \ &= \
   \sum_{p=0}^{D}\frac{1}{p!}\partial_{j}C_{i_1\ldots i_{p}}\,\psi^{j}\psi^{i_1}\cdots \psi^{i_{p}}\ket{0}
   \ = \
   \sum_{p=1}^{D}\frac{1}{(p-1)!}\partial_{[i_1}C_{i_2\ldots i_{p}]}\,\psi^{i_1}\cdots \psi^{i_{p}}\ket{0}\\
   \ &= \
   \sum_{p=1}^{D}\frac{1}{p!}F_{i_1\ldots i_p}\,\psi^{i_1}\cdots \psi^{i_{p}}\ket{0}\;,
 \end{split}
 \ee
where we performed an index shift and relabeled the indices. Thus, the components are given by the conventional field strengths
 \be\label{fieldstr}
  F_{i_1\ldots i_p} \ = \ p\,\partial_{[i_1}C_{i_2\ldots i_p]}\;.
 \ee
It is sometimes useful to avoid explicit indices and combinatorial
factors by using the language of differential forms.  In general, we identify
a spinor state $\ket{G_p}$ with a $p$-form $G^{(p)}$ as follows
\be
\ket{G_p}  = {1\over p!} \, G_{i_1 \cdots i_p}  \, \psi^{i_1}
 \cdots  \psi^{i_p} \ket{0}  ~\longleftrightarrow ~
G^{(p)} =  {1\over p!}\,  G_{i_1 \cdots i_p}  \, dx^{i_1} \wedge
\cdots  \wedge dx^{i_p}\;.
\ee
Whenever we speak of a $p$-form $G^{(p)}$ and its components
$G_{i_1 \ldots i_p}$, we will assume a normalization that includes the
$p!$ coefficient shown above.
It is now straightforward to translate (\ref{fieldstr}) to form language:
\be
\label{fsp}
F^{(p)} =  d C^{(p-1)} \,.
\ee
We  now collect all field strengths of different degrees into a
single form $F = \sum_p F^{(p)}$ and do the same for the potentials
$C = \sum_p C^{(p)}$.  We then have that (\ref{fsp}), or for that matter
(\ref{fieldstr}),  for all relevant
$p$ is summarized by
\be
F =  d C \,.
\ee

In order to evaluate the action we need to choose a parameterization for $\fancys$,
which we take to be $S_{\cal H}$,
 \be\label{spinH}
 \fancys \ = \ S_{\cal H} \ = \ e^{\frac{1}{2}b_{ij}\psi_{i}\psi_{j}}\, S_{g}^{-1}\,
  e^{-\frac{1}{2}b_{ij}\psi^{i}\psi^{j}}\;.
 \ee
The $b$-dependent terms in $S_{\cal H}$ suggest the
definition of
modified field strengths, related to the original field strengths
$\ket{F} = \slashed{\partial} \ket{\chi}$ by the addition of Chern-Simons
like terms:
\be\label{Fhatdef}
\ket{\widehat{F}} \ \equiv \ e^{- \frac{1}{2}b_{ij}\psi^{i}\psi^{j}}
\ket{F}
\ = \ \sum_{p=1}^{D}\frac{1}{p!}\widehat{F}_{i_1\ldots i_p}\,\psi^{i_1}\cdots \psi^{i_{p}}\ket{0}\; .
\ee
This relation is summarized in form language by
\be
\label{ayvg}
\widehat F \ = \ e^{-b^{(2)}} \wedge F \ = \
  e^{-b^{(2)}} \wedge dC  \,,~~~~\hbox{with} ~~~~
b^{(2)} \equiv~ {1\over 2} \, b_{ij} \,dx^i \wedge dx^j\,.
\ee
Explicitly, for example,
 \be
 \begin{split}
  \widehat{F}^{(3)} \ &= \  F^{(3)} -b^{(2)} \wedge F^{(1)}  \\
  \widehat{F}^{(5)}  \ &= \ F^{(5)} -b^{(2)} \wedge F^{(3)}\, + \,{1\over 2}
  b^{(2)} \wedge b^{(2)} \wedge F^{(1)} \;,\qquad \text{etc.}
 \end{split}
 \ee
The bra corresponding to $\ket{\widehat{F}}$ is given by
 \be\label{braF}
  \bra{\widehat{F}} \ = \
  \sum_{p=1}^{D}\frac{1}{p!} \bra{0}\psi_{i_p}\cdots \psi_{i_{1}}\,\widehat{F}_{i_1\ldots i_p}\; .
 \ee

Next, we can evaluate the Lagrangian (\ref{action-first-form-var})
using (\ref{spinH}), (\ref{Fhatdef}) and (\ref{braF}), which yields
 \be
  {\cal L} \ = \ \frac{1}{4}\bra{\widehat{F}}\,S_{g}^{-1}\,\ket{\widehat{F}} \ = \ \frac{1}{4}
  \sum_{p,q=1}^{D}\frac{1}{q!p!}\widehat{F}_{i_1\ldots i_p}\widehat{F}_{j_1\ldots j_q}
  \bra{0}\psi_{i_p}\cdots \psi_{i_1}\,S_{g}^{-1}\,\psi^{j_1}\cdots\psi^{j_q}\ket{0}\;.
 \ee
Using now (\ref{FINALSg}) for the action of $S_{g}^{-1}$ and the normalization
 \be
  \bra{0}\psi_{i_p}\cdots\psi_{i_1}\psi^{m_1}\dots\psi^{m_q}\ket{0}  \ = \
  \delta_{pq}\,p!\,\delta_{i_1}{}^{[m_1}\cdots\delta_{i_p}{}^{m_p]}\;,
 \ee
following from $\langle 0|0\rangle=1$,
the action reduces to
 \be
 \label{action-reduction}
  {\cal L} \ = \ -\frac{1}{4}\sqrt{g}\,\sum_{p=1}^{D}\,\frac{1}{p!}\,g^{i_1 j_1}\cdots g^{i_p j_p}\,
  \widehat{F}_{i_1\ldots i_p}\widehat{F}_{j_1\ldots j_p}\;,
 \ee
where we used the short-hand notation $\sqrt{g}=\sqrt{| \det g|}$.
This can also be written as
 \be
 \label{action-reduction99}
  {\cal L} \ = \ -\frac{1}{4}\sqrt{g}\,\sum_{p=1}^{D}\,
  | \widehat{F}^{(p)} |^2 ~~,
  \ee
  where we define for any $p$-form $\omega^{(p)}$:
  \be
  \label{defsquare}
   | \omega^{(p)} |^2 \equiv
  \frac{1}{p!}\,g^{i_1 j_1}\cdots g^{i_p j_p}\,
  \omega_{i_1\ldots i_p}\omega_{j_1\ldots j_p}\;.
 \ee
The result in (\ref{action-reduction99})
is the required sum of kinetic terms for all $p$-form gauge fields (of odd or even degree,
depending on the chirality of $\chi$), which appear in the democratic formulation.
This action needs to be supplemented by the duality relations, ensuring that we propagate
only the physical degrees of freedom of type II.
We consider these next.

\subsubsection{Self-duality relations in terms of field strengths}\label{513}
Here we show that for  $\tilde{\partial} = 0$
the self-duality conditions  $\slashed{\partial} \chi=-
{\cal K} \slashed{\partial} \chi$, c.f.~eq.~(\ref{ODDdual}),  reduce to
\be\label{dualityRel}
\widehat{F}^{(p)} \ = \ (-1)^{\frac{(D- p)(D- p-1)}{2}} *\widehat{F}^{(D-p)} \; .
\ee
These are conventional duality relations for $p$-form field strengths. In here we use the following definition of the Hodge-dual form:
\be
\label{definehodge}
(*A)_{i_1 \cdots i_p} \equiv \frac{1}{(D-p)!} 
\, g_{i_1 j_1} \cdots g_{i_p j_p} \varepsilon^{k_{p+1} \cdots k_{D} \, j_1 \cdots j_{p} } A_{k_{p+1} \cdots k_D}  \; .  
\ee
Our conventions for the epsilon symbols are as follows:
 \be\label{EpsConventions}
  \begin{split}
  \epsilon^{1\,2\ldots D} \ &= \ +1\;, \qquad
  \varepsilon^{i_1\ldots i_{D}} \ = \ \frac{1}{\sqrt{g}}\,\epsilon^{i_1\ldots i_D}\;, \\
  \epsilon_{1\,2\ldots D} \ &= \ -1 \;, \qquad
  \varepsilon_{i_1\ldots i_{D}} \ = \ \sqrt{g}\,\epsilon_{i_1\ldots i_D}\;,
  \end{split}
 \ee
i.e., $\epsilon$ is a tensor density, while $\varepsilon$ is a (pseudo-)tensor. As usual, lowering
the indices on $\varepsilon^{i_1\ldots i_{D}}$ with $g_{ij}$ yields $\varepsilon_{i_1\ldots i_{D}}$,
and $\varepsilon$ and $\epsilon$ coincide on flat space.
We note the familiar relation for the square of the Hodge star on
forms of degree $p$ in a $D$-dimensional spacetime with signature $s$:
\be
* * \omega^{(p)}  = (-1)^{p (D-p)} s\, \omega^{(p)} \,.
\ee
We can ask when is (\ref{dualityRel}) consistent with repeated application
of the Hodge star operation.  A calculation gives the condition
\be
s \, (-1)^{ {1\over 2} D (D-1)}  = \, 1\,.
\ee
Not surprisingly, in Lorentzian signature this agrees with the
result in (\ref{ksqcon}).  Finally, for $D=10$, the
duality constraints (\ref{dualityRel}) take the form
\be
\label{d-rel10}
\widehat{F}^{(p)} \ = \ - (-1)^{{1\over 2} p(p+1)} *\widehat{F}^{(D-p)} \; .
\ee

We can now begin our calculation.  Let us first introduce
the short-hand notation  
\be
{\bf B}\ = \ \tfrac{1}{2}b_{ij}\psi^{i}\psi^{j}\,, ~~~~
{\bf B}^\dagger \ = \ -\, \tfrac{1}{2}b_{ij}\,\psi_{i}\psi_{j}\,, 
\ee
which allows us to write $S_{\cal H}$ in (\ref{definitionSH}) as follows
\be
S_{\cal H} \ = \ e^{- {\bf B}^\dagger} S_g^{-1}
\, e^{- {\bf B}} \,.
\ee
The self-duality conditions $\slashed{\partial} \chi=-
{\cal K} \slashed{\partial} \chi$
can now be written~as
\be
e^{- {\bf B}} \ket{\slashed\partial\chi}\ = \  -e^{- {\bf B}} \, C^{-1} \, e^{- {\bf B}^\dagger} S_g^{-1}
\, e^{- {\bf B}} \ket{\slashed\partial\chi} \; ,
\ee
where we multiplied the factor $e^{- {\bf B}}$ from the left 
to form  the modified field strengths $\ket{\widehat{F}}$
defined in (\ref{Fhatdef}):
\be
\ket{\widehat{F}} \ = \   -e^{- {\bf B}} \, C^{-1} \, e^{- {\bf B}^\dagger}  \, S_g^{-1}\, \ket{\widehat{F}}\;.
\ee
Using (\ref{conpsiaction}) we readily verify that
\be
C e^{- {\bf B}}  C^{-1} =   e^{- C {\bf B} C^{-1}} = e^{- \tfrac{1}{2} b_{ij} \psi_i \psi_j} = e^{{\bf B}^\dagger} \,,
\ee
and, as a result,
\be
\ket{\widehat{F}}
 \ = \  -C^{-1} \, S_g^{-1} \ket{\widehat{F}} \,.
\ee
A further simplification is possible using (\ref{SginverseC}) in the
form $S_g = - C^{-1} S_g^{-1} C$, giving 
\be\label{dualitystep99}
\ket{\widehat{F}} \ = \   S_g \, C^{-1} \ket{\widehat{F}} \; .
\ee
Finally, we recall that in the dimensions with self-consistent
duality constraints (\ref{ksqcon}) we have $C^{-1} = - C$ and therefore
\be\label{dualitystep}
\ket{\widehat{F}} \ = \  -\, S_g \, C \ket{\widehat{F}} \; .
\ee
This is the simplest possible form of the duality constraints. 

We can now examine (\ref{dualitystep}) in terms of 
component fields, as defined in (\ref{Fhatdef}).  We find
 \be\label{sTEp}
  \sum_{p=1}^{D}\frac{1}{p!}\widehat{F}_{i_1\ldots i_p}\,\psi^{i_1}\cdots \psi^{i_{p}}\ket{0} \ = \
- \,   \sum_{p=1}^{D}\,\frac{1}{p!}\,
  \widehat{F}_{i_1\ldots i_p}~  S_g~\psi_{i_1}\cdots \psi_{i_{p}} \, C \, \ket{0}   \; ,
 \ee
where we used (\ref{conpsiaction}).  Next, we show that the charge conjugation matrix in (\ref{sTEp}) effectively acts like an epsilon symbol. In fact,
by multiple application of the oscillator algebra one can verify that
\be\label{Cduality}
\begin{split}
\psi_{i_1}\cdots \psi_{i_{p}} \, C \, \ket{0} \ &= \  \psi_{i_1}\cdots \psi_{i_{p}} \psi^1 \psi^2 \cdots \psi^D \ket{0} \\
 &= \ \frac{1}{(D-p)!}  \epsilon^{i_p i_{p-1} \cdots i_1 \, j_{p+1} \cdots j_{D}} \, \psi^{j_{p+1} }  \cdots \psi^{j_{D}}  \ket{0} \\
 &= \ \frac{1}{(D-p)!} (-1)^{\frac{p(p-1)}{2}} \epsilon^{i_1 i_{2} \cdots i_p \, j_{p+1} \cdots j_{D}} \, \psi^{j_{p+1} }  \cdots \psi^{j_{D}}  \ket{0}  \; .
\end{split}
\ee
Back in  (\ref{sTEp}) and defining $ \tilde p =  D- p$ we have
\be
\begin{split}
& \sum_{p=1}^{D}\frac{1}{p!}\widehat{F}_{i_1\ldots i_p}\,\psi^{i_1}\cdots \psi^{i_{p}}\ket{0} \\
\ & = - \,   \sum_{p=1}^{D}\,(-1)^{\frac{p(p-1)}{2}}\frac{1}{p! \,\tilde p !}\,
  \widehat{F}_{i_1\ldots i_p}~    \epsilon^{i_1 i_{2} \cdots i_p \, j_{1} \cdots j_{\tilde p}} \,S_g ~ \psi^{j_{1} }  \cdots \psi^{j_{\tilde p}}  \ket{0} \\ 
 \ & = \,   \sum_{p=1}^{D}\,(-1)^{\frac{p(p-1)}{2}}\frac{1}{p! \,\tilde p !}\,
    {1\over \sqrt{g}} \,\epsilon^{i_1 i_{2} \cdots i_p \, j_{1} \cdots j_{\tilde p}}
     \widehat{F}_{i_1 \ldots i_p}\, \, g_{j_1 k_1} \cdots g_{j_{\tilde p} 
     k_{\tilde p}} ~  \psi^{k_{1} }  \cdots \psi^{k_{\tilde p}} \ket{0} \\  
   \ & = \,   \sum_{p=1}^{D}\,(-1)^{\frac{p(p-1)}{2}}\frac{1}{\,\tilde p !} ~ \frac{1}{p!}\, \,g_{k_1 j_1} \cdots g_{ k_{\tilde p}j_{\tilde p}} ~ \varepsilon^{i_1 i_{2} \cdots i_p \, j_{p} \cdots j_{\tilde p}}  \,\widehat{F}_{i_1\ldots i_p}~   \psi^{k_{1} }  \cdots \psi^{k_{\tilde p}} \ket{0} \\  
   \ & = \,   \sum_{p=1}^{D}\,(-1)^{\frac{p(p-1)}{2}}\frac{1}{\,\tilde p !} ~(*
   \widehat F)_{k_1 \cdots k_{\tilde p} }    \psi^{k_{1} }  \cdots \psi^{k_{\tilde p}} \ket{0} \\  
\ &= \ \sum_{p=1}^{D} (-1)^{\frac{(D- p)(D- p-1)}{2}}\frac{1}{p!}  \, (*\widehat F)_{i_1 \cdots i_{p}} \, \psi^{i_1 }  \cdots \psi^{i_p}  \ket{0} \; .
\end{split}
\ee
In obtaining this result we made use of (\ref{FINALSg}), the definition
(\ref{definehodge}) and some simple manipulations.  
Thus, we have shown that the duality constraint implies the claimed duality relations (\ref{dualityRel}).

\subsection{Conventional gauge symmetries}
Let us now verify that the gauge transformations parameterized by $\xi^{M}$ and $\lambda$
reduce to the conventional gauge symmetries of type II theories in the frame $\tilde{\partial}^{i}=0$.
We start with the $p$-form gauge symmetries (\ref{lambdagauge})
whose parameter we write in components as
 \be
  \ket{\lambda} \ = \ \sum_{p=0}^{D}\frac{1}{p!}\lambda_{i_1\ldots i_p}\,\psi^{i_1}\cdots\psi^{i_p}\ket{0}\;.
 \ee
For $\tilde{\partial}=0$ this implies
 \be
  \delta_{\lambda}\ket{\chi} \ = \ \slashed{\partial}\ket{\lambda} \ = \ \psi^{j}\partial_{j}\ket{\lambda} \ = \
  \sum_{p=1}^{D}\frac{1}{(p-1)!}\partial_{[i_1}\lambda_{i_2\ldots i_p]}\,\psi^{i_1}\cdots\psi^{i_p}\ket{0}\;,
 \ee
from which we read off
 \be
  \delta_{\lambda}C_{i_1\ldots i_p} \ = \ p\,\partial_{[i_1}\lambda_{i_2\ldots i_p]}\;.
 \ee
These are the conventional $p$-form gauge transformations.
In form language they read
\be
\label{pformlangCtr}
\delta_\lambda\, C \ = \ d \lambda \,.
\ee

Let us now discuss the gauge transformations parameterized by $\xi^{M}=(\tilde{\xi}_{i},\xi^{i})$.
We first claim that the $C$ forms
transform as $p$-forms under diffeomorphisms
parameterized by $\xi^{i}$.
To see this, we compute
 \be
  \delta_{\xi}\ket{\chi} \ = \ \left(\xi^{j}\partial_{j}+\partial_{j}\xi^{k}\,\psi^{j}\psi_{k}\right)
  \sum_{p=0}^{D}\,\frac{1}{p!}\,C_{i_1\ldots i_p}\,\psi^{i_1}\cdots \psi^{i_p}\ket{0}\;.
 \ee
The transport term just gives rise to the transport term of the component fields.
The second term can be evaluated using (\ref{spinident}),
which then implies for the components
 \be\label{formdiff}
  \delta_{\xi}C_{i_1\ldots i_p} \ = \ \xi^{j}\partial_{j}C_{i_1\ldots i_p}
  +p\, \partial_{[i_1}\xi^{j}\,C_{|j|i_2\ldots i_p]} \ \equiv \ {\cal L}_{\xi}C_{i_1\ldots i_p}\;.
 \ee
This is the usual diffeomorphism symmetry which infinitesimally acts via
the Lie derivative.

We now consider the $\tilde \xi_i$ parameters, which are parameters
for the $b$-field gauge transformations.
It turns out that the $C$ forms transform non-trivially under this symmetry. In order to see this
we compute for $\tilde{\partial}=0$
 \be
 \begin{split}
  \delta_{\tilde{\xi}}\ket{\chi} \ &= \ \partial_{k}\tilde{\xi}_{l}\,\psi^{k}\psi^{l}\ket{\chi} \ = \
  \sum_{p=0}^{D}\frac{1}{p!}\partial_{[i_1}\tilde{\xi}^{}_{i_2}\,C_{i_3\ldots i_{p+2}]}
  \psi^{i_1}\cdots \psi^{i_{p+2}}\ket{0}\\
   \ &= \  \sum_{p=2}^{D}\frac{1}{(p-2)!}\partial_{[i_1}\tilde{\xi}^{}_{i_2}\,C_{i_3\ldots i_{p}]}
  \psi^{i_1}\cdots \psi^{i_{p}}\ket{0} \;,
 \end{split}
 \ee
where we performed an index shift $p\rightarrow p+2$ in the last equation. We thus read off
 \be \label{RRformbgauge}
  \delta_{\tilde{\xi}}C_{i_1\ldots i_p}  \ = \ p(p-1)\partial_{[i_1}\tilde{\xi}^{}_{i_2}\,C_{i_3\ldots i_p]}\;.
 \ee
 In the language of forms the above equation reads
 \be
 \label{RRformbgauge2}
 \delta_{\tilde{\xi}} C \ = \  d\tilde \xi \wedge C \,.
 \ee
 Note that this implies that
 \be
  \delta_{\tilde\xi} C^{(0)}  =   \delta_{\tilde\xi} C^{(1)} = 0 \,,~~~\delta_{\tilde\xi}C^{(2)}=
  d\tilde \xi \cdot C^{(0)} \,, ~~ \ldots ~,~~  \delta_{\tilde\xi} C^{(p)} = d\tilde \xi \wedge C^{(p-2)}\,.
 \ee
Recalling that
\be
\delta_{\tilde \xi} \, b^{(2)}  =  d\tilde{\xi}\,,
\ee
it is straightforward to define new potentials
$\hat{A}$ that do not transform under $\tilde \xi$:
\be
\label{ahatpot}
\hat{A}  \ \equiv \ e^{-b^{(2)}} \wedge C \,.
\ee
Indeed it is simple to verify that
\be
\delta_{\tilde \xi}\,  \hat A = 0 \,.
\ee
One can also write the $C$ fields in terms of the $\hat A$ fields easily:
\be\label{CANDA}
C  \ = \ e^{b^{(2)}} \wedge \hat{A} \,.
\ee
The $\hat A$ potentials are hatted
to distinguish them from conventional type II potentials to be discussed below.

\subsection{Democratic formulation}

The democratic formulation of type II theories introduces an action for all even and odd forms,
which is then supplemented by duality relations between the corresponding field strengths.
The resulting equations of motion are equivalent to the standard equations of motion by virtue of
the Bianchi identities of the field strengths \cite{Fukuma:1999jt,Bergshoeff:2001pv}. 
Here we briefly introduce this formulation and show the equivalence with the conventional formulation.

\subsubsection{Review and comments on the standard formulation}

The standard 10-dimensional low energy action for type II theories is given by
\be \label{origaction}
S \ = \  S_{\rm NS-NS} + S_{\rm RR}
\; ,
\ee
where $S_{\rm NS-NS}$ is the same for both type IIA and type IIB and written as
\be \label{10dimnsaction}
S_{\rm NS-NS} =  \int d^{10} x \sqrt{-g} e^{-2 \phi}  \left[ R + 4 (\partial \phi)^2  - \frac{1}{2} | H^{(3)} |^2 \right] \; .
\ee
The RR actions $S_{\rm RR}$
for  type IIA and type IIB are given by, respectively,
\bea
\label{10dimIIARaction}  S_{\rm RR}^{\rm (IIA)}
&=& \hskip-4pt- \frac{1}{ 2} \int d^{10} x \sqrt{- g} \, \Bigl(
|\widehat{F}^{(2)} |^2 + |\widehat{F}^{(4)} |^2\Bigr)    +   \frac{1}{ 2} \int b^{(2)} \wedge d {A}^{(3)} \wedge d{A}^{(3)} \; ,
\\[0.7ex]
 S_{\rm RR}^{\rm (IIB)}
 &=& \hskip-4pt- \frac{1}{ 2 } \int \hskip-2pt
 d^{10} x \sqrt{- g} \,\Bigl(|\widehat{F}^{(1)} |^2 + |\widehat{F}^{(3)} |^2 + \frac{1}{2} |\widehat{F}^{(5)}|^2\Bigr)  + \  \frac{1}{ 2} \int b^{(2)} \wedge d {A}^{(4)} \wedge d{A}^{(2)}\,, ~~~~
\eea
with the additional self-duality condition $\widehat{F}^{(5)} = * \widehat{F}^{(5)} $ for type IIB, which has to
be imposed on the field equations after varying the action.
We also note   
that the type II$^{\star}$ theories take the same 
form,  
with the overall sign of the kinetic terms for the RR fields
(but not of the Chern-Simons terms)  reversed. 
The field strengths $\widehat{F}^{(n)}$ are defined in terms of the original RR potentials ${A}^{(n)}$ as
\be
\begin{array}{ll}
\widehat{F}^{(1)} \equiv dA^{(0)} 
& \widehat{F}^{(2)} \equiv dA^{(1)}  
\\[0.2ex]
\widehat{F}^{(3)} \equiv  dA^{(2)}  
 +H^{(3)} \wedge {A}^{(0)} &
 \widehat{F}^{(4)} \equiv  dA^{(3)}  
 + H^{(3)} \wedge {A}^{(1)} \\[0.2ex]
\widehat{F}^{(5)} \equiv dA^{(4)} 
+ \frac{1}{2} H^{(3)} \wedge {A}^{(2)} - \frac{1}{2} b^{(2)} \wedge
dA^{(2)} 
\,.
\end{array}
\ee
The field strengths above must be invariant under $p$-form gauge
transformations of the potentials.  But the presence of $A$-forms without
an exterior derivative acting on them implies that the $p$-form gauge
transformations of $A$'s are a bit nontrivial:
\be
\label{pfgtA}
\begin{array}{ll}
\delta_\lambda A^{(0)}  = ~0
&\delta_\lambda A^{(1)} = ~d\lambda^{(0)}
\\[0.2ex]
\delta_\lambda A^{(2)} =  ~d\lambda^{(1)}
  &
\delta_\lambda A^{(3)} =  ~d\lambda^{(2)}
 - b^{(2)} \wedge d\lambda^{(0)} \\[0.2ex]
\delta_\lambda A^{(4)}  = ~d\lambda^{(3)}
- \frac{1}{2} \,b^{(2)} \wedge d\lambda^{(1)} \,.
\end{array}
\ee
One can readily verify that $\delta_\lambda \widehat{F}^{(p)} = 0$ and
that the Chern-Simons terms are invariant because
the integrands change by a $d$-exact form.  Since the $\widehat{F}$'s involve the field $b^{(2)}$, the $A$ potentials are not invariant
under the $b^{(2)}$ gauge transformations, $\delta_{\tilde \xi} b^{(2)}
= d\xi$.  The invariance of the $\widehat{F}$'s requires
\be
\label{axigt}
\begin{array}{ll}
\delta_{\tilde \xi} A^{(0)}  = ~0
&\delta_{\tilde \xi}  A^{(1)} = ~0
\\[0.2ex]
\delta_{\tilde \xi}  A^{(2)} =  ~0
  &
\delta_{\tilde \xi}  A^{(3)} =  ~0 \\[0.2ex]
\delta_{\tilde \xi}  A^{(4)}  =
~ \frac{1}{2} \,d\tilde\xi \wedge A^{(2)} \,.~~
\end{array}
\ee
One can readily verify that $\delta_{\tilde \xi} \widehat{F}^{(p)} = 0$ and
that the Chern-Simons terms are invariant because
the integrands change by a $d$-exact form (use $dA^{(2)} \wedge
dA^{(2)} = 0$, for the IIB case).

A set of
modified RR potentials $C^{(n)}$ are constructed by combining the
NS-NS 2-form $b^{(2)}$ and the original RR potentials ${A}^{(n)}$:
\be \label{CandAprime}
\begin{array}{ll}
C^{(0)} \equiv {A}^{(0)} & C^{(1)} \equiv {A}^{(1)} \\
C^{(2)} \equiv {A}^{(2)} +b^{(2)} \wedge {A}^{(0)} & C^{(3)} \equiv {A}^{(3)} + b^{(2)} \wedge {A}^{(1)} \\
C^{(4)} \equiv {A}^{(4)} + \frac{1}{2} b^{(2)} \wedge {A}^{(2)} + \frac{1}{2} b^{(2)} \wedge b^{(2)} \wedge {A}^{(0)}\,.
\end{array} \;
\ee
These transformations have one peculiar feature.  The field $C$ fails
to be equal to
 $e^{b^{(2)}} \wedge A$ because of the terms in $C^{(4)}$.
As we will argue below, this is because matters can be simplified by
using a different $A^{(4)}$ field.  The inverse relations are
\be \label{CandAprime-inv}
\begin{array}{ll}
A^{(0)} = C^{(0)} & A^{(1)} = {C}^{(1)} \\
A^{(2)} = {C}^{(2)} -b^{(2)} \wedge {C}^{(0)} & A^{(3)} \equiv {C}^{(3)} - b^{(2)} \wedge {C}^{(1)} \\
A^{(4)} \equiv {C}^{(4)} - \frac{1}{2} b^{(2)} \wedge {C}^{(2)} \,.~~
\end{array} \;
\ee

We now claim that the $C$ fields
defined above coincide with the $C$ fields we have been using in
this paper; the fields that transform naturally under T-duality
and have conventional $p$-form gauge transformations.  Indeed, a short calculation shows that the $p$-form
gauge transformations in (\ref{pfgtA}) imply
\be
\delta_\lambda C =  d \lambda \,,
\ee
in agreement with (\ref{pformlangCtr}).  Moreover, the $\tilde{\xi}$
gauge transformations in (\ref{axigt}) imply that the $\tilde{\xi}$
gauge transformations of the $C$ fields are summarized by
\be
\delta_{\tilde\xi} C =  d \tilde \xi \wedge C \,,
\ee
in agreement with (\ref{RRformbgauge2}).
Finally, the field strengths $\widehat{F}$ take a simple form in terms of the $C$ forms
\be \label{simplerelationFandC}
\widehat{F} = e^{- b^{(2)} } \wedge d C \; ,
\ee
in agreement with (\ref{ayvg}).  The desired properties
$\delta_\lambda \widehat {F}=0$ and $\delta_{\tilde \xi} \widehat{F}=0$
are now manifest.

We noted earlier in (\ref{ahatpot}) that the potentials
\be
\label{gljdf}
\hat A
=e^{- b^{(2)}} \wedge C\,,
\ee are invariant under $\delta_{\tilde \xi}$.
Comparing with (\ref{CandAprime-inv}) we see that
\be
\hat{A}^{(p)} =  A^{(p)}  \,,  ~~~  p \not= 4 \,,
\ee
and for the case $p=4$ a short calculation shows that
\be
{A^{(4)}} = \hat{A}^{(4)} + \frac{1}{2}\, {b^{(2)}} \wedge \hat{A}^{(2)} \; .
\ee
This equation is consistent with (\ref{axigt}) and $\delta_{\tilde\xi} \hat{A}^{(4)} = 0$.
Moreover, using (\ref{CANDA}) and (\ref{simplerelationFandC}) we find
\be
\widehat{F} =  e^{-b^{(2)}} \wedge d \Bigl(  e^{-b^{(2)}} \wedge A\Bigr)\,,
\ee
which quickly yields
\be
\widehat{F}  =  d \hat{A}  + H^{(3)} \wedge \hat {A} \,.
\ee

The above means that a formulation with potentials $\hat{A}$ is
somewhat more efficient than the conventional formulation.  Indeed, a small calculation shows that
the type IIB  Chern-Simons term, expressed in terms of $\hat{A}^{(4)}$
and $\hat{A}^{(2)}$ takes exactly the same form as before, thus
\bea
\label{10dimIIARaction-improved}  S_{\rm RR}^{\rm (IIA)}
&=& \hskip-4pt- \frac{1}{ 2} \int d^{10} x \sqrt{- g} \, \Bigl(
|\widehat{F}^{(2)} |^2 + |\widehat{F}^{(4)} |^2\Bigr)    +   \frac{1}{ 2} \int b^{(2)} \wedge d \hat{A}^{(3)} \wedge d\hat{A}^{(3)} \; ,
\\[0.7ex]
 S_{\rm RR}^{\rm (IIB)}
 &=& \hskip-4pt- \frac{1}{ 2 } \int \hskip-2pt
 d^{10} x \sqrt{- g} \,\Bigl(|\widehat{F}^{(1)} |^2 + |\widehat{F}^{(3)} |^2 + \frac{1}{2} |\widehat{F}^{(5)}|^2\Bigr)  + \  \frac{1}{ 2} \int b^{(2)} \wedge d \hat{A}^{(4)} \wedge d\hat{A}^{(2)}\,. ~~~~
\eea
Here, collecting information,
\be
\widehat{F}  =  d \hat{A}  + H^{(3)} \wedge \hat {A} \,, ~~~
\delta_\lambda \hat{A} =  e^{-b^{(2)}} \wedge d\lambda\,, ~~
\delta_{\tilde \xi} \hat{A} = 0 \,.
\ee
The advantage of this formulation is that the haphazard $\delta_\lambda$
transformations of the $A$'s now takes a closed form expression and
the $\delta_{\tilde\xi}$ symmetry is manifest.

\medskip
The formulation in terms of $C$ potentials is also elegant as it brings
out the duality properties more clearly.  This time the Chern-Simons terms
are a bit more complex, however.  One finds
\be
\begin{split}
\label{10dimIIARaction-improved-dual}  S_{\rm RR}^{\rm (IIA)}
& \ = \ -\, \frac{1}{ 2} \ \int d^{10} x \sqrt{- g} \, \Bigl(
|\widehat{F}^{(2)} |^2 + |\widehat{F}^{(4)} |^2\Bigr)   \\
& ~~~~~ +   \frac{1}{ 2} \int b^{(2)} \wedge d\Bigl(C^{(3)} - b^{(2)}\wedge C^{(1)}\Bigr) \wedge  d\Bigl(C^{(3)} - b^{(2)}\wedge C^{(1)}\Bigr) \; ,
\\[0.7ex]
 S_{\rm RR}^{\rm (IIB)}
 &\ =\ - \,\frac{1}{ 2 } \,\int
 d^{10} x \sqrt{- g} \,\Bigl(|\widehat{F}^{(1)} |^2 + |\widehat{F}^{(3)} |^2 + \frac{1}{2} |\widehat{F}^{(5)}|^2\Bigr) \\
 & ~~~~~+ \  \frac{1}{ 2} \int b^{(2)} \wedge
 d \Bigl( C^{(4)} - {1\over 2} b^{(2)} \wedge C^{(2)}\Bigr)
  \wedge d\Bigl( C^{(2)}- b^{(2)} \wedge C^{(0)}\Bigr) \,,
  \\
& \widehat{F} = e^{- b^{(2)} } \wedge d C \,, ~~~
\delta_\lambda C = d \lambda \,, ~~~ \delta_{\tilde\xi} C = d\tilde\xi \wedge C\,.
\end{split}
\ee

\subsubsection{Democratic formulation and equivalence}

Let us now turn to the democratic formulation.
The democratic action features kinetic terms for all forms, but no explicit Chern-Simons terms,
\bea
\label{democratIIAaction} S_{\rm RR}^{\rm (IIA),dem} &=& - \frac{1}{ 4} \int d^{10} x \sqrt{- g}  \sum_{n = 2, 4, 6, 8}|\widehat{F}^{(n)} |^2  \ = \ \frac{1}{ 4 } \int   \sum_{n = 2, 4, 6, 8}
\widehat{F}^{(n)} \wedge * \widehat{F}^{(n)} \; , \\
 S_{\rm RR}^{\rm (IIB),dem} 
 &=& - \frac{1}{ 4} \int d^{10} x \sqrt{- g}  \sum_{n = 1, 3, 5, 7, 9}|
 \widehat{F}^{(n)} |^2 \ = \  \frac{1}{ 4} \int   \sum_{n = 1, 3, 5, 7, 9}
 \widehat{F}^{(n)} \wedge * \widehat{F}^{(n)} \; . \quad \
\eea
Note that the normalization of the action has a factor of $1/2$ relative
to similar (non Chern-Simons) terms  in the standard formulation (\ref{10dimIIARaction}). The above actions are supplemented by the duality relations
\be \label{japanduality}
\begin{array}{lll}
* \widehat{F}^{(1)} \ = \ \widehat{F}^{(9)} \;, & \quad & * \widehat{F}^{(2)} \ = \ -  \widehat{F}^{(8)}\;, \\[0.3ex]
* \widehat{F}^{(3)} \ = \ - \widehat{F}^{(7)} \;,& \quad & * \widehat{F}^{(4)} \ = \   \widehat{F}^{(6)} \;,\\[0.3ex]
* \widehat{F}^{(5)} \ = \ \widehat{F}^{(5)} \;,& \quad & * \widehat{F}^{(6)} \ = \ -  \widehat{F}^{(4)} \;,\\[0.3ex]
* \widehat{F}^{(7)} \ = \ - \widehat{F}^{(3)} \;,& \quad & * \widehat{F}^{(8)} \ = \   \widehat{F}^{(2)} \;,\\[0.3ex]
* \widehat{F}^{(9)} \ = \ \widehat{F}^{(1)} \;.& \quad &
\end{array}
\ee
The above are indeed the duality constraints we obtained before
at the component level, as can be readily checked using~(\ref{d-rel10}).
Moreover, our action, evaluated for $\tilde \partial =0$, is also identical to the above democratic action.
This can be seen in (\ref{action-reduction99}), where we also recall
in (\ref{defsquare}) the definition of $|\ldots|^2$ on forms. By showing
that the democratic formulation agrees with the standard formulation
we will have shown that our double field theory type II action, for $\tilde \partial=0$, agrees with the type II theories.
We note that the democratic formulation of the type II$^{\star}$ theories
is completely analogous; it introduces a RR action with a kinetic term for each potential
together with duality constraints, both with a reversed overall sign.

 The claim is that the field equations of the standard action are the same as those of the democratic action after imposing these duality relations. We present this equivalence for type IIA. The story for type IIB is completely analogous.

The field equation for $g_{ij}$ is relatively straightforward. Since the NS-NS action is the same for the standard formulation and the democratic formulation, it is sufficient to examine the energy-momentum tensor for both formulations. The energy-momentum tensor $T_{ij}$ in the standard formulation does not receive contributions from the Chern-Simons terms. We simply have
\be \label{original_em}
T_{ i j } =
{\cal E}_{ij} (\widehat{F}^{(2)}) + {\cal E}_{i j} (\widehat{F}^{(4)}) \; ,
\ee
where
\be
{\cal E}_{i j} (\widehat{F}^{(n)}) \ \equiv \ \frac{1}{(n-1)!}  \widehat{F}_{i k_1 k_2 \cdots k_{n-1}} \widehat{F}_{j}{}^{ k_1 k_2 \cdots k_{n-1}} - \frac{1}{2} g_{i j} |\widehat{F}^{(n)}|^2 \; .
\ee
The energy-momentum tensor resulting from the democratic action is given by
\be \label{democrat_em}
T_{i j} = \frac{1}{2} \sum_{n = 2, 4, 6, 8} {\cal E}_{i j} (\widehat{F}^{(n)}) \; ,
\ee
where the $1/2$ factor is due to the normalization, as mentioned
above (\ref{japanduality}).  From the identity ${\cal E}_{i j} (\widehat{F}^{(n)} ) = {\cal E}_{i j} (*
\widehat{F}^{(n)})$ and the duality relations,
we infer that the energy-momentum tensor in the democratic action is equal
to that in the standard action.
Both formulations give the same Einstein equations.

In the standard formulation (c.f.~(\ref{10dimnsaction}) and (\ref{10dimIIARaction}))  the field equation for $b^{(2)}$ is
\be
\label{jkjk}
d ( e^{-2 \phi} * H^{(3)}) + \widehat{F}^{(2)} \wedge *\widehat{F}^{(4)} - \frac{1}{2} \widehat{F}^{(4)} \wedge \widehat{F}^{(4)} \ = \ 0 \; ,
\ee
and in the democratic formulation (\ref{democratIIAaction}) with (\ref{japanduality}) the field equation for $b^{(2)}$ reads
\be
d ( e^{-2 \phi} * H^{(3)}) + \frac{1}{2} \widehat{F}^{(2)} \wedge * \widehat{F}^{(4)} + \frac{1}{2} \widehat{F}^{(4)} \wedge * \widehat{F}^{(6)} + \frac{1}{2} \widehat{F}^{(6)} \wedge * \widehat{F}^{(8)} \ = \ 0 \; ,
\ee
which is equivalent to (\ref{jkjk})
after imposing the duality relations (\ref{japanduality}).

The most nontrivial checks in the equivalence of
the two formulations are the field equations for
 $C^{(n)}$. In the standard formulation we have nontrivial Bianchi identities from (\ref{simplerelationFandC}):
\be
\label{bident}
d \widehat{F}^{(n)} = - H^{(3)} \wedge \widehat{F}^{(n-2)} \; .
\ee
The field equations for $C^{(1)}$ and $C^{(3)}$ in the standard formulation read, respectively,
\be
\label{c1c3}
\begin{split}
0 \ = \  & d ~\Bigl( - *\widehat{F}^{(2)} + b^{(2)} \wedge *\widehat{F}^{(4)}  + \frac{1}{2} b^{(2)} \wedge b^{(2)} \wedge \widehat{F}^{(4)} + \frac{1}{6} b^{(2)} \wedge b^{(2)} \wedge b^{(2)} \wedge \widehat{F}^{(2)}  \Bigr)  \;,  \\
0 \ = \ & d~ \Bigl( * \widehat{F}^{(4)} + b^{(2)}\wedge \widehat{F}^{(4)} + \frac{1}{2} b^{(2)} \wedge b^{(2)} \wedge \widehat{F}^{(2)} \Bigr) \; .
\end{split}
\ee
In the democratic formulation the field equations for all odd forms $C^{(1)}, C^{(3)}, C^{(5)}, C^{(7)}$ are respectively given by
\be
\begin{split}
0 &\ = \ d \,\left( - * \widehat{F}^{(2)} + b^{(2)} \wedge * \widehat{F}^{(4)} - \frac{1}{2} b^{(2)} \wedge b^{(2)} \wedge * \widehat{F}^{(6)} + \frac{1}{6} b^{(2)} \wedge b^{(2)} \wedge b^{(2)} \wedge * \widehat{F}^{(8)} \right) \; , \quad   \\
0 & \ = \ d \, \left( * \widehat{F}^{(4)} - b^{(2)} \wedge * \widehat{F}^{(6)} + \frac{1}{2} b^{(2)} \wedge b^{(2)} \wedge * \widehat{F}^{(8)} \right) \; , \\
0 & \ = \ d\, \left( - * \widehat{F}^{(6)} + b^{(2)} \wedge * \widehat{F}^{(8)} \right)  \; , \\
0 & \ = \  d\, \left( * \widehat{F}^{(8)} \right) \;  .
\end{split}
\ee
By imposing the duality relations (\ref{japanduality}) the last two equations become the Bianchi identities for $\widehat{F}^{(4)}$
and $\widehat{F}^{(2)}$
in (\ref{bident})
and the first two equations are equivalent to
 the field equations (\ref{c1c3})
for $C^{(1)}$ and $C^{(3)}$.
In summary, for the common potentials the equations of motion agree
after use of duality relations.  For the potentials in the democratic formulation that are absent in the standard formulation, the democratic
equations of motion arise from the Bianchi identities of the potentials
in the standard formulation.
The analysis given above explicitly shows that in the democratic formulation the field equations are
equivalent to those of the standard formulation.

\section{IIA versus IIB}  
Here we consider  double field theory  evaluated in frames with $\tilde{\partial}^{i}\neq 0$.
In the first part, we review the results of \cite{Hohm:2010jy} for the NS-NS sector and give
an intuitive picture of how this generalizes to the RR sector. In the second part, we give
a more explicit treatment of the RR action when evaluated in different T-duality frames.

\subsection{Review of NS-NS sector and motivation for RR fields}

In the previous section we have seen that for fields with no $\tilde x$ dependence
or, equivalently, setting $\tilde{\partial}^{i}=0$, the proposed double field theory reduces to the type IIA or type IIB theory in the
democratic formulation, depending on
the chosen chirality of $\chi$.
It is equally consistent with the strong constraint, however, to keep
the $\tilde x$ dependence of fields while dropping the $x$ dependence
by setting $\partial_{i}=0$.
We will see that if the theory reduces to type IIA when setting
 $\tilde{\partial}^{i}=0$,
the same theory reduces  to type IIA$^{\star}$ when setting
$\partial_{i}=0$,  and vice versa. Similarly, for the opposite chirality
of $\chi$, in one frame the theory reduces to type IIB and
in the other frame to type IIB$^{\star}$.

More generally, we can consider intermediate frames that originate from the $\tilde{x}_{i}=0$ frame
by an arbitrary $O(D,D)$ transformation.
Specifically, with the subgroup
$ O(n-1,1)\times O(d,d)\subset O(D,D)$ acting on coordinates
$(x^\mu, x^a, \tilde x_a)$,
with $\mu=0, \ldots, n-1$ and $a = 1, \ldots , d$,
we can consider the $O(d,d)$ transformation that maps the
$\tilde{x}_a=0$ frame to the $x^a=0$ frame.
Here we find that
the resulting theory is equivalent to the original one
if $d$ is even
or to the theory with opposite chirality if $d$ is odd.
In other words, for $d$ odd, if we start with a chirality such that the theory reduces to IIA
for $\tilde{x}_{a}=0$, the same theory reduces to type IIB
 for  $x^{a}=0$, and vice versa.

In order to set the stage to discuss the above claims, let us first review the
transition from the $\tilde x=0$ frame to the $x=0$
frame
for the pure NS-NS sector.
This matter was analyzed in sec.~3.2 of \cite{Hohm:2010jy}. The two T-duality
frames $\tilde{\partial}^{i}=0$ and $\partial_{i}=0$ are mapped into each other by
the $O(D,D)$ transformation $J$ that exchanges $x$ and $\tilde{x}$,
 \be\label{Itrans}
  J^{M}{}_{N} \ = \   \begin{pmatrix} 0 &  1 \\ 1 & 0 \end{pmatrix} \;.
 \ee
The action evaluated in one duality frame is
equivalent to the action evaluated in
the other duality frame, but written in terms of field variables that are redefined according
to the $O(D,D)$ transformation (\ref{Itrans}). To make this more 
explicit, we introduce
 \be\label{Hdual}
  \tilde{\cal H} \ \equiv \
  J\,{\cal H}\,J \ = \ {\cal H}^{-1} \;.
 \ee
In components, we obtain
 \be\label{dualfields99}
  \tilde{{\cal H}} \ = \
  \begin{pmatrix} g_{ij}-b_{ik}g^{kl}b_{lj} &  b_{ik}g^{kj} \\[0.3ex] -g^{ik}b_{kj} & g^{ij} \end{pmatrix}\,.
 \ee
 If we view $\tilde{\cal H}$ as the generalized metric associated
 with a new metric $g'$ and a new
 antisymmetric field $b'$, following (\ref{firstH})
 we would write
  \be
  \label{firstHvar}
  \tilde{\cal H} \ = \  \begin{pmatrix}    g'^{ij} & -g'^{ik}b'_{kj}\\[0.5ex]
  b'_{ik}g'^{kj} & g'_{ij}-b'_{ik}g'^{kl}b'_{lj}\end{pmatrix}
  =  \begin{pmatrix}  \tilde{g}_{ij} &  -\tilde{g}_{ik}\tilde{b}^{kj}
  \\[0.5ex] \tilde{b}^{ik}\tilde{g}_{kj}  & \tilde{g}^{ij}-\tilde{b}^{ik}\tilde{g}_{kl}\tilde{b}^{lj} \end{pmatrix}\;,
 \ee
where in the second step we defined the tilde fields by
\be
\label{tfdef}
 \tilde g^{ij} \equiv g'_{ij}   \quad \to \quad \tilde g_{ij} =  g'^{\,ij} \,, ~~~~
\hbox{and}~~~\tilde b^{ij}  \equiv  b'_{ij} \,.
\ee
Note that the change of index position in passing from primed to tilde
variables makes the right-hand sides of (\ref{dualfields99}) and (\ref{firstHvar}) have consistent index positions:
\be\label{dualfields}
  \tilde{\cal H} \ = \
  \begin{pmatrix} g_{ij}-b_{ik}g^{kl}b_{lj} &  b_{ik}g^{kj} \\ -g^{ik}b_{kj} & g^{ij} \end{pmatrix}
  \ \equiv \
   \begin{pmatrix}  \tilde{g}_{ij} &  -\tilde{g}_{ik}\tilde{b}^{kj}
  \\ \tilde{b}^{ik}\tilde{g}_{kj}  & \tilde{g}^{ij}-\tilde{b}^{ik}\tilde{g}_{kl}\tilde{b}^{lj} \end{pmatrix}\;.
 \ee
The dilaton is invariant under this inversion duality: $\tilde{d}=d$.

Let us verify directly
 that the field redefinition in (\ref{dualfields}) is equivalent to
the change of variable induced by T-duality, following eqs.~(3.20)--(3.22) of \cite{Hohm:2010jy}. In there, we considered
the fundamental field ${\cal E}_{ij}=g_{ij}+b_{ij}$ represented by the matrix
${\cal E}$  and the T-dual field $\tilde {\cal E} = {\cal E}^{-1}$,
writing
 \be
  \tilde{\cal E}^{ij} \ \equiv \ \big({\cal E}^{-1}\big)^{ij} \ \equiv \ \
   \tilde{g}^{ij}+\tilde{b}^{ij}
   \qquad \Rightarrow \qquad
  {\cal E}_{ik} \tilde{\cal E}^{kj} \ = \ \delta_{i}{}^{j}\;,
 \ee
where $\tilde{g}^{ij}$  and $\tilde{b}^{ij}$ are the symmetric and antisymmetric parts of
$\tilde{\cal E}^{ij}$, respectively. Consequently,
$\tilde{g}^{ij}$ is interpreted as the metric and $\tilde{g}_{ij}$
denotes the inverse metric.  The duality transformations of the metric
imply that they satisfy~\cite{Hohm:2010jy}:
 \be
  \tilde{g}_{ij} \ = \ {\cal E}_{ki}\,g^{kl}\,{\cal E}_{lj} \;, \qquad
  g^{ij} \ = \ \tilde{\cal E}^{ik}\,\tilde{g}_{kl}\,\tilde{\cal E}^{jl}\;.
 \ee
Writing these equations in terms of $g$ and $b$ (or their dual variables $\tilde{g}$ and $\tilde{b}$),
we recover (\ref{dualfields}) for the diagonal matrix entries. For the off-diagonal entries
we compute, for instance,
 \be
 \begin{split}
  -\tilde{g}_{ik}\tilde{b}^{kj} \ &= \ -\tilde{g}_{ik}\big(\tilde{\cal E}^{kj}-\tilde{g}^{kj}\big) \ = \
  -\tilde{g}_{ik}\tilde{\cal E}^{kj}+\delta_{i}{}^{j}
   \ = \
  -{\cal E}_{pi}g^{pq}{\cal E}_{qk}\tilde{\cal E}^{kj}+\delta_{i}{}^{j} \\
  \ &= \ -{\cal E}_{pi}g^{pj}+\delta_{i}{}^{j} \ = \ -(g_{pi}+b_{pi})g^{pj}+\delta_{i}{}^{j}
  \ = \ b_{ip}g^{pj}\;,
 \end{split}
 \ee
confirming the equality of the off-diagonal entries in (\ref{dualfields}).

We note that the
field redefinitions (\ref{tfdef})
 interchange upper with lower indices in order to
 work consistently with
the \textit{lower} indices of the dual coordinates $\tilde{x}_{i}$.
In particular, the diffeomorphisms in the dual coordinates are generated by $\tilde{\xi}_{i}$
in that the gauge transformations (see (2.37) and (2.38) of~\cite{Hohm:2010jy})
reduce for $\partial_{i}=0$ to
 \be\label{dualdiff}
   \delta_{\tilde{\xi}}\tilde{\cal E}^{ij} \ = \ \tilde{\xi}_{k}\tilde{\partial}^{k}\tilde{\cal E}^{ij}
   +\tilde{\partial}^{i}\tilde{\xi}_{k}\,\tilde{\cal E}^{kj}
   +\tilde{\partial}^{j}\tilde{\xi}_{k}\,\tilde{\cal E}^{ik}\; .
 \ee
Viewing  
$\tilde{\cal E}^{ij}$ with upper indices as a \textit{covariant} rather than a contravariant tensor,
this is the conventional transformation of such a tensor under
infinitesimal diffeomorphisms.

The double field theory action $S_{\rm NS-NS}$ for the
NS-NS fields is, of course, the same as the
 double field theory action $S_{\rm DFT}$ for the low energy
 bosonic string.
 We thus write
 \be
 \label{clcl99}
  S_{\rm NS-NS} \Big|_{\tilde{\partial}=0}  \ = \   S_{\rm DFT} \Big|_{\tilde{\partial}=0}
    \ = \ S\Big[g,b,d,\partial\Big]\;,
 \ee
with $S$ a function of the four arguments written above.
In the dual frame $\partial =0$, our discussion above implies
that we have
\be
 \label{clcl}
 S_{\rm NS-NS} \Big|_{\partial=0}  \ = \
    S_{\rm DFT} \Big|_{\partial=0}
    \ = \ S\Big[\tilde{g},\tilde{b},\tilde{d},\tilde{\partial}\Big]\;.
 \ee
 The replacements in the arguments of $S$ are, explicitly,
\be
g_{ij} \to \tilde g^{ij}  \,,  ~~ g^{ij} \to \tilde g_{ij}  \,, ~~ b_{ij} \to \tilde b^{ij} \,,
~~\partial_i \to \tilde \partial^i \,.
\ee

\medskip
Let us now see how this generalizes in presence of the RR fields.
Before we give a general discussion in the next section, it will be instructive
to first examine more explicitly, along the lines reviewed above, what happens in the
frame $\partial_i=0$ with $\tilde{\partial}^{i}\neq 0$.
Let us first evaluate the
field strength $\ket{F}$ in this frame,
 \be\label{newF}  
 \begin{split}
  \ket{F}\Big|_{\partial_i=0} \ &= \ \slashed{\partial}\ket{\chi} \ = \ \psi_j \tilde{\partial}^{j} \sum_{p=0}^D  {1\over p!}  C_{i_1\cdots i_p} \,\psi^{i_1} 
  \cdots \psi^{i_p} \ket{0}  \\
 \ &= \
  \sum_{p=1}^{D}\frac{1}{p!}\,\tilde{\partial}^{j}C_{ji_2\ldots i_p} \, p\, \psi^{i_2}\cdots\psi^{i_p}\ket{0}\;
\  = \
  \sum_{p=1}^{D}\frac{1}{(p-1)!}\,\tilde{\partial}^{j}C_{ji_1\ldots i_{p-1}} \, \psi^{i_1}\cdots\psi^{i_{p-1}}\ket{0}\;.\\
 \ &= \
  \sum_{p=0}^{D-1}\frac{1}{p!}\tilde{\partial}^{j}C_{ji_1\ldots i_p}\psi^{i_1}\cdots\psi^{i_p}\ket{0}\;.
 \end{split}
 \ee
At first sight this looks rather different from the conventional field strength of a $p$-form, but
it can actually be brought to the form of a `dual field strength' if we introduce a dual potential $\tilde{C}$
according to
 \be\label{dualform}
  C_{i_1\ldots i_p} \ = \ \alpha_p\, \epsilon_{i_1\ldots i_p j_1\ldots j_{D-p}}\,\tilde{C}
  ^{j_1\ldots j_{D-p}}\;,
 \ee
where the numerical coefficients $\alpha_p$ 
will be fixed below. We recall that the epsilon symbol is constant
and equal to $\pm 1$, i.e., it is not a tensor but rather a density.
In terms of this new variable, (\ref{newF}) reads
 \be\label{DUALfieldstrength}
 \begin{split}
   \ket{F}\Big|_{\partial_i=0}    \ &= \
   \sum_{p=0}^{D-1}\frac{\alpha_{p+1}}{p!}
   \epsilon_{j i_1\ldots i_p j_1\ldots j_{D-p-1}}\tilde{\partial}^{j}
   \tilde{C}^{j_1\ldots j_{D-p-1}}\psi^{i_1}\cdots\psi^{i_p}\ket{0}
     \\
    \ &\equiv \ \sum_{p=0}^{D-1}\frac{\alpha_{p+1}(-1)^p}{p!(D-p)}\,
   \epsilon_{ i_1\ldots i_p j_1\ldots j_{D-p}}\tilde{F}^{j_1\ldots j_{D-p}}\,
   \psi^{i_1}\cdots\psi^{i_p}\ket{0}\;,
 \end{split}
 \ee
where we introduced in analogy to (\ref{fieldstr})
 \be
  \tilde{F}^{j_1\ldots j_p} \ = \ p\,\tilde{\partial}^{[j_1}\tilde{C}^{j_2\ldots j_p]}\;.
 \ee

We should stress that (\ref{dualform}) does not involve any metric
and so this is \textit{not} the Hodge dual. Consequently, $\tilde{C}$ is not a covariant tensor
in the usual sense.
However, what we actually have to verify is that, as in (\ref{dualdiff}), this is a tensor in the T-dual sense that it
transforms under $\tilde{\xi}_{i}$ rather than $\xi^{i}$ with a Lie derivative.
To see this, we examine the gauge transformation (\ref{gtchi})   
 \be
  \delta_{\tilde{\xi}}\ket{\chi} \ = \ \tilde{\xi}_{j}\tilde{\partial}^{j}\ket{\chi}
  +\tilde{\partial}^{j}\tilde{\xi}_{k}\,\psi_{j}\psi^{k}\ket{\chi}\;.
 \ee
The transport term gives manifestly rise to the correct structure,
so we focus on the second term, denoted by $\bar{\delta}_{\tilde{\xi}}$, which yields
 \be\label{DEltaxi}
 \begin{split}
   \bar{\delta}_{\tilde{\xi}}\ket{\chi} \ &= \ \sum_{p=0}^{D}\frac{p+1}{p!}
   \tilde{\partial}^{j}\tilde{\xi}_{[j}\,C_{i_1\ldots i_p]}\psi^{i_1}\cdots\psi^{i_p}\ket{0} \\
   \ &= \ \sum_{p=0}^{D}\frac{\alpha_p (p+1)}{p!}
   \tilde{\partial}^{j}\tilde{\xi}_{[j}\, \epsilon_{i_1\ldots i_p]k_1\ldots k_{D-p}}
   \tilde{C}^{k_1\ldots k_{D-p}}\psi^{i_1}\cdots\psi^{i_p}\ket{0}\;.
 \end{split}
 \ee
To simplify this, we use that a fully antisymmetric tensor 
with 
$D+1$ indices in $D$ dimensions
vanishes identically,
 \be
  \begin{split}
   0 \ &= \ (D+1)\tilde{\partial}^{j}\tilde{\xi}_{[j}\,\epsilon_{i_1\ldots i_p k_1\ldots k_{D-p}]} \\
   \ &= \ (p+1)\tilde{\partial}^{j}\tilde{\xi}_{[j}\,\epsilon_{i_1\ldots i_p] k_1\ldots k_{D-p}}
   -(D-p)\tilde{\partial}^{j}\tilde{\xi}_{[k_1}\,\epsilon_{|i_1\ldots i_p j|k_2\ldots k_{D-p}]}\;.
  \end{split}
 \ee
Using this in (\ref{DEltaxi}), one obtains
 \be
   \bar{\delta}_{\tilde{\xi}}\ket{\chi} \ = \ \sum_{p=0}^{D} \frac{\alpha_p (D-p)}{p!}
   \epsilon_{i_1\ldots i_p k_1\ldots k_{D-p}}\tilde{\partial}^{k_1}\tilde{\xi}_{j}\,
   \tilde{C}^{j k_2\ldots k_{D-p}}\psi^{i_1}\cdots \psi^{i_p}\ket{0}\;,
 \ee
where we relabeled $k_1\leftrightarrow j$.  In total, we read off
 \be\label{dualLie}
  \delta_{\tilde{\xi}}\tilde{C}^{i_1\ldots i_{D-p}} \ = \
  \tilde{\xi}_{j}\tilde{\partial}^{j}  \tilde{C}^{i_1\ldots i_{D-p}}
  +(D-p)\tilde{\partial}^{[i_1}\tilde{\xi}_{k}\,\tilde{C}^{|k|i_2\ldots i_{D-p}]} \ \equiv \
  {\cal L}_{\tilde{\xi}}\tilde{C}^{i_1\ldots i_{D-p}}\;.
 \ee
This is the dual Lie derivative with respect to $\tilde{\xi}_{i}$ of a dual $p$-form, where we
note that upper indices are now covariant indices and so the signs in (\ref{dualLie})
are the conventional ones, c.f.~(\ref{formdiff}) and (\ref{dualdiff}).

\subsection{RR action in different T-duality frames}

So far we have seen explicitly that the field strengths in the dual frame $\partial_{i}=0$,
$\tilde{\partial}^{i}\neq 0$, take the conventional form when written in terms of the right `T-dual' variables $\tilde{C}^{i_1\cdots i_p}$.
We will now prove more generally that the action and duality relations in the frame $\partial_i=0$
yield the T-dual type II theory written in terms of the T-dual variables 
(\ref{dualfields}) for the NS-NS fields and $\tilde{C}$ for the RR fields. 
Since the $O(D,D)$ transformation
inverts all space-time dimensions, it contains a timelike T-duality and thus maps, say,
IIA and IIA$^{\star}$ into each other.

To proceed, we describe the field redefinition (\ref{dualform}) by introducing the following
tilde variable of the $O(D,D)$ spinor,
 \be\label{redefchi}
  \tilde\chi \ = \ S_{J}\,\chi \,, \qquad S_{J} \ = \ C   \;.
 \ee
This corresponds to the action of the spinor representative of the $O(D,D)$ transformation $J=J^{-1}$
that exchanges
$x^{i}$ and $\tilde{x}_{i}$, which for convenience we have chosen to be $C$, but we stress 
that this field redefinition does not affect the coordinate arguments. 

We can then verify that the field redefinition
$\chi\rightarrow\tilde{\chi}$ indeed amounts to the duality transformation
(\ref{dualform}).
In fact, with (\ref{conpsiaction}) and (\ref{Cduality}) we obtain
\be\label{CJtrans}
\begin{split}
 \tilde{\chi}  \ &\equiv \ \sum_{p}\frac{1}{p!}\,\tilde{C}^{i_1\ldots
i_p}\,\psi^{i_1}\cdots\psi^{i_p}\ket{0}
\ = \ C{\chi} \ = \ \sum_{p}\frac{1}{p!}\,C_{i_1\ldots
i_p}\,\psi_{i_1}\cdots
 \psi_{i_p}C\ket{0} \\
 \ &= \ \sum_{p}\frac{1}{p!(D-p)!}(-1)^{\frac{1}{2}p(p-1)}\,C_{i_1\ldots i_p}\,
 \epsilon^{i_1\ldots i_p j_{1}\ldots j_{D-p}}\,\psi^{j_{1}}\cdots
\psi^{j_{D-p}}\ket{0} \;.
 \end{split}
\ee
This equation determines  the tilde variables in terms of the original
ones:
\be
\tilde{C}^{i_1\ldots i_p} \ = \  (-1)^{\frac{1}{2}(D-p)(D+p-1)}\frac{1}{(D-p)!} 
\epsilon^{i_1 \ldots i_p j_1\ldots j_{D-p}}
C_{j_1\ldots j_{D-p}}\;,
\ee
where we performed an index shift. 
It can be checked with the standard identity
\be\label{epsilonID}
 \epsilon^{i_1\ldots i_p j_1\ldots j_{D-p}}\,\epsilon_{i_1\ldots i_p k_1\ldots
k_{D-p}} \ = \
 -p!(D-p)!\,\delta^{[j_1}{}_{k_1}\cdots \delta^{j_{D-p}]}{}_{k_{D-p}}\;,
\ee
following from (\ref{EpsConventions}),
that this coincides with (\ref{dualform}) for
$\alpha_{p}=(-1)^{\frac{1}{2}p(p-1)+1}/(D-p)!$.

In terms of the tilde variables (\ref{redefchi}) we have, using (\ref{invgamma}), 
 \be  
 \begin{split}
  \slashed{\partial}\chi
   \ & = \ \frac{1}{\sqrt{2}}\Gamma^{M}\partial_{M}(S_{J}^{-1}\tilde{\chi})
     \ = \ \frac{1}{\sqrt{2}}\Gamma^{M}S_{J}^{-1} \partial_{M}\tilde{\chi} \\
  \ & = \ \frac{1}{\sqrt{2}}J^{M}{}_{N}S_{J}^{-1}\Gamma^{N}\partial_{M}\tilde{\chi}
   \ = \ S_{J}^{-1}\frac{1}{\sqrt{2}}\Gamma^{N} (J^{M}{}_{N}\partial_{M})\tilde{\chi} \ = \   
   S_{J}^{-1}\hat{\slashed{\partial}}\tilde{\chi}\;,
\end{split}
 \ee
where we introduced a redefined derivative and Dirac operator, 
\be
\hat{\slashed{\partial}} \ \equiv\  {1\over \sqrt{2}} \,\Gamma^N \hat{\partial}_N \,, ~~~~~\hat{\partial}_{N}\ \equiv \ J^{M}{}_{N}\partial_{M}\, .
\ee
Recalling that the matrix $J^M{}_N$ has only the non-vanishing matrix elements
$J^{ij}$ and $J_{ij}$ that are equal to Kronecker
deltas we find that 
 \be
  \hat{\slashed{\partial}} \ = \ \psi^{i}\tilde{\partial}^{i}+\psi_{i}\partial_{i}\;.
 \ee
As expected, the $\partial_i$ and $\tilde\partial^i$ derivatives have
been exchanged. For the Lagrangian we now find
 \be\label{redefaction}
  \begin{split}
   {\cal L} \ &= \ {1\over 4}(\slashed{\partial}\chi)^{\dagger}\,S_{\cal H}\, \slashed{\partial}\chi
    \ = \ {1\over 4}(\hat{\slashed{\partial}}\tilde{\chi})^{\dagger}(S_{J}^{-1})^{\dagger}\,S_{\cal H}\,S_{J}^{-1}
    \hat{\slashed{\partial}}\tilde{\chi}
    \ = \ \, -{1\over 4}(\hat{\slashed{\partial}}\tilde{\chi})^{\dagger} \,S_{\tilde{\cal H}}\,
    \hat{\slashed{\partial}}\tilde{\chi}\;,
  \end{split}
 \ee
where we used the sign factor  in~(\ref{tdualJonS}).  
We see that in  tilde-variables the RR action takes the same form as in the original variables,
up to a sign. It can also be checked that the duality constraints in the dual frame take the form 
\be
 \hat{\slashed{\partial}}\tilde{\chi} \ = \   C^{-1} S_{\tilde{\cal H}}\,  \hat{\slashed{\partial}}\tilde{\chi} \,,
\ee
which differs from the constraints in the original frame by a sign factor.

It follows now that  setting $\partial_{i}=0$ in the evaluation of the
Lagrangian as written in the 
first form in (\ref{redefaction}) is
equivalent to setting  $\hat{\slashed{\partial}} \ = \ \psi^{i}\tilde{\partial}^{i}$ 
in the evaluation of the Lagrangian as written in the 
last form in (\ref{redefaction}).  But this latter evaluation is identical 
 to our original computation in sec.~5, with $\partial_i$ derivatives
 replaced by $\tilde\partial^i$ derivatives and $C_{i_1\ldots i_p}$
 replaced by $\tilde{C}^{i_1 \ldots i_p}$.  Of course, this time we get
 an extra minus sign.

Due to this sign change in the RR action
we conclude that if the theory reduces for $\tilde{\partial}^{i}=0$ to IIA, the
same theory reduces for $\partial_{i}=0$ to IIA$^{\star}$, but written in terms of the T-dual
variables.
We thus have, for instance,
 \be\label{frameaction}
  S_{{\rm DFT}_{\rm II}} \Big|_{\tilde{\partial}=0} \ = \  S_{\rm IIA}\Big[g,b,d,C,\partial\Big]\;, \qquad
  S_{{\rm DFT}_{\rm II}} \Big|_{\partial=0} \ = \
  S_{{\rm IIA}^{\star}}\Big[\tilde{g},\tilde{b},\tilde{d},\tilde{C},\tilde{\partial} \Big]\;,
 \ee
where we indicated by $S_{{\rm DFT}_{\rm II}}$ the full double field theory action of type II, while
$S_{\rm IIA}$ and $S_{{\rm IIA}^{\star}}$ are the low-energy actions of IIA and IIA$^{\star}$, respectively.  Moreover, the corresponding duality
constraints differ by a sign. 
 This is the expected sign given that the stress-tensor from the RR
sector in the dual frame must have a sign opposite to the one in the 
original frame. 

Similarly, if the chosen chirality is such that
the theory reduces in the $\tilde{\partial}^{i}=0$ frame to type IIB, the same theory
reduces in the $\partial_{i}=0$ frame to type IIB$^{\star}$.
We finally note that had we chosen the equally valid parametrization $\fancys=-S_{\cal H}$,
we would have obtained either IIA$^{\star}$ or IIB$^{\star}$ in the frame $\tilde{\partial}^{i}=0$
and the conventional IIA or IIB theories in the opposite frame.

It is instructive to reconsider the above analysis in somewhat more explicit terms by performing an 
expansion of the RR action in tilde derivatives $\tilde{\partial}$,
 \be\label{RRExpand}
  S^{}_{\rm RR} \ = \ S_{\rm RR}^{(0)}+S_{\rm RR}^{(1)}+S_{\rm RR}^{(2)}\;,
 \ee
where the superscript denotes the number of $\tilde{\partial}$. For simplicity, let us assume
that the $b$-field vanishes.
The first term $S_{\rm RR}^{(0)}$ is a conventional type II action
as discussed in sec.~5. The remaining terms can be straightforwardly computed using
that, by the linearity of the Dirac operator $\slashed{\partial}$, the full field strength
(\ref{FieldStrength123}) is simply the sum of (\ref{ODDfields}) and (\ref{DUALfieldstrength}),
 \be\label{Ffull}
  \ket{F} \ = \  \sum_{p=0}^{D}\frac{1}{p!}\,{\cal F}_{i_1\ldots i_p}\,\psi^{i_1}\cdots\psi^{i_p}\ket{0}\;,
\ee
where
\be\label{Ffull2}
  {\cal F}_{i_1\ldots i_p} \ \equiv \  F_{i_1\ldots i_p}+\beta_p\,\epsilon_{i_1\ldots i_p j_1\ldots j_{D-p}}
  \tilde{F}^{j_1\ldots j_{D-p}} \;,
 \ee
and $\beta_p=(-1)^p \alpha_{p+1}/(D-p)$. In here $F$ is the conventional field strength,
depending on derivatives $\partial$, and $\tilde{F}$ is the field strength in terms of the
dual variables, depending on the dual derivatives $\tilde{\partial}$. Precisely as in sec.~5, one then finds
for the full RR-action
 \be
   S_{\rm RR} \ = \ -\frac{1}{4}\sum_{p}\frac{1}{p!}\,\sqrt{g}\,
   g^{i_1j_1}\cdots g^{i_pj_p}\, {\cal F}_{i_1\ldots i_p}{\cal F}_{j_1\ldots j_p}\;.
 \ee
Insertion of (\ref{Ffull2}) then
gives
 \be\label{S0S2}
  \begin{split}
   S_{\rm RR}^{(0)} \ &= \ -\frac{1}{4}\sum_{p}\frac{1}{p!}\,\sqrt{g}\,g^{i_1j_1}\cdots g^{i_pj_p}\,
   F_{i_1\ldots i_p}F_{j_1\ldots j_p}\;,
   \\
   S_{\rm RR}^{(2)} \ &= \ +\frac{1}{4}\sum_{p}\frac{1}{p!}\,\frac{1}{\sqrt{g}}\,g_{i_1j_1}\cdots g_{i_pj_p}\,
   \tilde{F}^{i_1\ldots i_p}\tilde{F}^{j_1\ldots j_p}\;.
  \end{split}
 \ee
For the second equation we shifted the summation index $p$ and used the
identity
 \be\label{epsilonID2}
  \sqrt{g}\,g^{i_1j_1}\cdots g^{i_pj_p}\,\epsilon_{i_1\ldots i_pk_1\ldots k_{D-p}}\,
  \epsilon_{j_1\ldots j_p l_1\ldots l_{D-p}} \ = \ -\,\frac{1}{\sqrt{g}}\,p!(D-p)!\,
  g_{[k_1|l_1|}\cdots g_{k_{D-p}]l_{D-p}}\;,
 \ee
which follows from (\ref{epsilonID}) and (\ref{EpsConventions}).
 We stress that the  minus sign on the right-hand side of this identity is due to the Lorentzian signature.
It is this sign that is responsible for the relative sign  between $S^{(0)}$
and $S^{(2)}$ in (\ref{S0S2}). We have thus re-derived the sign change of (\ref{redefaction})
for the special case of vanishing $b$-field.  
Let us note that, as discussed in sec.~\ref{dualsubsection}, not all invariances of the original action are still present once we parametrize 
$\fancys$ in terms of the conventional fields. For instance, the transformation $J$ maps, using (\ref{CJtrans}),  
 \be
   \partial_i\,\rightarrow\,\tilde{\partial}^{i}\;, \quad
   C_{i_1\ldots i_p}\,\rightarrow\, \tilde{C}^{i_1\ldots i_p}\qquad\Rightarrow \qquad 
   F_{i_1\ldots i_p}\;\rightarrow\; \tilde{F}^{i_1\ldots i_p}\;.
\ee
Therefore, $S^{(0)}$ in (\ref{S0S2}) is transformed without a sign change, i.e.,  
$S^{(0)}$ is mapped to $-S^{(2)}$ and so (\ref{RRExpand}) is not invariant.

\medskip

\medskip

We close this section with a brief discussion of intermediate frames,
which we illustrate with the simplest case of one T-duality inversion. 
Thus, we split the indices as $x^{i}=(x^{1},x^{a})$ and assume that 
the non-trivial derivatives are $(\tilde{\partial}^{1},\partial_{a})$, 
where `1' denotes the special direction.  
As above, we consider a field redefinition that takes the form of the T-duality
inversion, 
 \be
 \begin{split}
  {\chi}^{\prime} \ = \ S_1\chi \ &= \ (\psi^{1}+\psi_1)
  \sum_p\frac{1}{p!}\left(C_{a_1\ldots a_p}\psi^{a_1}\cdots \psi^{a_p}
  +pC_{\,1a_1\ldots a_{p-1}}\psi^{1}\psi^{a_1}\cdots \psi^{a_{p-1}}\right)\ket{0} \\
  \ &= \ \sum_p \frac{1}{p!}\left(C_{a_1\ldots a_p}\psi^1\psi^{a_1}\dots \psi^{a_p}
  +pC_{1\,a_1\ldots a_{p-1}}\psi^{a_1}\cdots \psi^{a_{p-1}}\right)\ket{0} \\
  \ &\equiv \  \sum_p \frac{1}{p!} C^{\prime}_{i_1\ldots i_p}\psi^{i_1}\cdots \psi^{i_p}\ket{0}\;.
 \end{split}
 \ee
This implies that the redefined $C^{(p)}$ are given in terms of the original ones by 
 \be\label{fform}
   C^{\prime}_{i_1\ldots i_p} \ = \   \left\{
  \begin{array}{l l}
   C_{a_2\ldots a_p} & \quad \text{if\; $i_1=1$,\; $i_2=a_2\,,\,\ldots$ $\,,i_p=a_p$}\\
   C_{1 a_1\ldots a_p}  & \quad \text{if\, $i_1=a_1\,,\, \ldots$ $\,,i_p=a_p$\,.}\\
  \end{array} \right.   
\ee
Put differently, the new $p$-forms are obtained from the original ones by adding or 
deleting the special index. It follows that this redefinition interchanges even 
and odd forms and thus changes the chirality of $\chi$. The field strength then reads 
 \be 
  \slashed{\partial}\chi \ = \ \big(\psi^{a}\partial_{a}+\psi_1\tilde{\partial}^{1}\big)
  \big(\psi^1+\psi_1\big)\chi^{\prime} \ = \  \big(\psi^1+\psi_1\big)\big(\psi^{a}(-\partial_a)
  +\psi^{1}\tilde{\partial}^{1}\big)\chi^{\prime} \ = \ S_1\,\psi^{i}\partial_{i}^{\prime}
   \chi^{\prime} \;,
 \ee
where we recognized the transformed (primed) derivatives 
$\partial_{i}^{\prime}=(\tilde{\partial}^{1}, -\partial_a)$, recalling 
that the transformation $h_{i}$ in (\ref{PinEl33}) changes the overall sign of the 
coordinates  $x^{a}$. 

In precise analogy to (\ref{redefaction}), we can now conclude that the action 
in the frame with $\tilde{\partial}^{1},\,\partial_{a}\neq 0$ takes the same form 
as in the frame $\tilde{\partial}^{i}=0$, just with all field variables replaced by
primed variables. Since the primed variables have the opposite chirality, 
it follows that if the theory reduces for $\tilde{\partial}^{i}=0$ to, say, 
type IIA, in the new frame it reduces to type IIB if $g_{11}$ is positive 
and to type IIB$^{\star}$ if $g_{11}$ is negative. 
More generally,  if we evaluate the theory in any frame that results from 
the $\tilde{\partial}^{i}=0$ frame by an $O(d,d)$ transformation, we obtain 
the corresponding T-dual theory.

\section{Discussion and conclusions}

In this paper we introduced a double field theory formulation for the low-energy limit of type II strings.  
T-duality 
relates different type II theories, a feature that does not
occur in bosonic string theory.  
In the double field theory built here  
each of the type II theories can be obtained
by choosing different `slicings' within the doubled coordinates.
Consistent slicings are those allowed by the $O(D,D)$ covariant strong constraint $\partial^M\partial_M = 0$ that originates from the $L_0 - \bar L_0=0$ constraint of closed string theory.   
If we consider two slicings  
related by an odd number of spacelike T-duality
inversions and one yields  type IIA, the other must yield type IIB.    
The  double field theory necessarily features the
so-called type IIA$^{\star}$ and type IIB$^{\star}$ theories, which are related
to the conventional type II theories via T-dualities along \textit{timelike} directions.

Despite this unification, 
 the actual  invariance group 
of the theory is only Spin$^{+}(D,D)$
and therefore does not contain any of the 
T-duality transformations that relate different 
type II theories. This means that the 
Pin$(D,D)$ transformations that are not in Spin$^{+}(D,D)$
must be viewed as dualities rather than invariances. 
More precisely,  while we 
fix the chirality of the spinor $\chi$ from the outset, 
the opposite chirality is obtained 
by the field redefinition induced by the appropriate T-duality 
transformation. The situation is similar to theories that depend on a background
but which are nevertheless background-independent in the sense that 
any shift of the background can be absorbed into a
field redefinition. Just  as one may then ask for a manifestly background independent formulation,
we may now wonder if there is a formulation with 
full Pin$(D,D)$ invariance.  
This would presumably require the introduction of a spinor without a chirality condition, together with an additional gauge symmetry to remove the new unphysical degrees of freedom.

Further generalizations of this work are possible. 
It would be interesting to see if this type II  double field theory
allows for an enhancement of the global symmetry to a U-duality group,
such that the NS-NS and RR fields transform in an irreducible representation. 
Results on reformulations of 11-dimensional supergravity may 
be relevant, see \cite{Hillmann:2009ci} and \cite{Berman:2010is,Thompson:2011uw}. 
Moreover, exceptional groups are of particular interest since they naturally combine
fundamental and spinor representation, and in this context the Kac-Moody
algebras  $E_{11}$ \cite{West:2001as,West:2010ev}
and $E_{10}$ \cite{Damour:2002cu,Kleinschmidt:2004dy}
have been proposed.  Being infinite-dimensional, they easily accommodate
the massless fields of various string theories, but they also give rise to an infinite set of
further representations for which a physical interpretation has yet to be found.

The work here may also contain pointers for a yet to be constructed
string field theory of type II strings.  This is an outstanding problem
since these remain the only string theories for
which no string field action is known.  
Finally, there might be applications to generalized Kaluza-Klein type reductions
or to  the construction of T-duality invariant world-volume theories of branes.
We leave these and other questions for future research.

\section*{Acknowledgments}
For useful discussions we would like to thank Marco Gualtieri, Juan Maldacena,
Ashoke Sen and Paul Townsend. We thank Chris Hull for comments,
and understand from him that he has worked in
closely related directions.

This work is supported by the U.S. Department of Energy (DoE) under the cooperative
research agreement DE-FG02-05ER41360. The work of OH is supported by the DFG -- The German
Science Foundation, and the work of SK is supported in part by a Samsung Scholarship.

\appendix

\section{Duality transformations of $S_{\cal H}$}
\setcounter{equation}{0}
In this appendix we discuss the transformation behavior of $S_{\cal H}$ in some detail.
We first give the general proofs of the transformation rules stated in the main text, and
then give an example to illustrate these rules.

\subsection{$GL(D)$ and $b$-shifts}
Our goal is to determine the sign factor $\sigma$ appearing in the transformation of $S_{\cal H}$
under $O(D,D)$,
\be\label{signesoncemore}
(S^{-1})^\dagger  \, S_{\cal H} \, S^{-1}  \ = \
\sigma_{\rho(S)} ({\cal H}) \, \,S_{\rho(S) \circ {\cal H}}\,.
\ee
We start by considering the `geometric subgroup'. It consists of $GL(D)$ transformations
and the abelian subgroup $\mathbb{R}^{\frac{1}{2}D(D-1)}$ of $b$-shifts, which
together form the semi-direct product $GL(D)\ltimes \mathbb{R}^{\frac{1}{2}D(D-1)}$.
We show that for this subgroup no sign factor arises:\\
\noindent
{\em Theorem}: Given an arbitrary ${\cal H}$, for any $h_r \in GL(D)$ and
$h_b \in \mathbb{R}^{\frac{1}{2}D(D-1)}$
\be
\sigma_{h_r} ( {\cal H}) \ = \  \sigma_{h_b} ( {\cal H})  \ = \ 1 \; .
\ee
We can then immediately conclude that $\sigma_{h} ( {\cal H})=1$ for any
$h\in GL(D)\ltimes \mathbb{R}^{\frac{1}{2}D(D-1)}$.

In the remainder of this subsection we will prove this theorem. We first present the proof for $b$-shifts,
and then discuss $GL^+(D)$ and $GL^-(D)$, respectively.  \\[2ex]
\underline{$b_{}$-shifts:}\;
The $O(D,D)$ element which shifts $b \rightarrow b - \Delta b $ and its corresponding $\text{Spin} (D, D)$ element are given by, respectively,
\be
h_{\Delta b} \ = \ \begin{pmatrix}1 & - \Delta b \\ 0 & 1 \end{pmatrix} \;, \qquad
\  S_{\Delta b} = e^{- \frac{1}{2} \Delta b_{ij} \psi^i \psi^j} \; .
\ee
Then the duality transformation of $S_{\cal H}$ under $b$-shifts can be written as
\bea
(S_{\Delta b}^{-1})^{\dagger}\,S_{\cal H}\,S_{\Delta b}^{-1}    & = &  e^{- \frac{1}{2} \Delta b_{ij} \psi_i \psi_j}   S_{\cal H} \, e^{ \frac{1}{2} \Delta b_{ij} \psi^i \psi^j}  \ = \ e^{- \frac{1}{2} \Delta b_{ij} \psi_i \psi_j}    e^{\frac{1}{2}b_{ij}\psi_{i}\psi_{j}}\, S_{g}^{-1}\,
e^{ -\frac{1}{2}b_{ij}\psi^{i}\psi^{j}}  \, e^{ \frac{1}{2} \Delta b_{ij} \psi^i \psi^j} \nonumber  \\
&=& e^{\frac{1}{2} (b_{ij}- \Delta b_{ij})\psi_{i}\psi_{j} }  \, S_{g}^{-1}\, e^{ -\frac{1}{2} (b_{ij} - \Delta b_{ij} )\psi^{i}\psi^{j}}  \ = \ S_{\cal H '}  \;.
\eea
We conclude $\sigma_{h_b} ( {\cal H}) = 1$. \\[2ex]
\underline{$GL^{+}(D)_{}$:}\;
An arbitrary $O(D,D)$ element in $GL^{+}(D)$ and its corresponding $\text{Spin} (D, D)$ element can be written as
\be
h_r \ = \  \begin{pmatrix} r &  0 \\ 0 & (r^{-1})^{t} \end{pmatrix} \;, \qquad
S_{r} \ = \ \frac{1}{\sqrt{\det r} } e^{\psi^i R_{i}{}^{j} \psi_j} \; ,
\ee
with $\det r >0$.
Under this $O(D,D)$ transformation, $g$ and $b$ transform covariantly,
\be\label{gldgb}
g \,\rightarrow\, r\, g\, r^{t} \;, \qquad  b \,\rightarrow\, r\, b\, r^{t} \; .
\ee
This transformation of the metric $g$ is induced by the covariant transformation
$e \rightarrow re$ of the vielbein. The duality transformation of $S_{\cal H}$ under $GL^{+}(D)$ is then
\be\label{GLdstep90}
(S_{r}^{-1})^{\dagger}\,S_{\cal H}\,S_{r}^{-1}
\ = \   (S_{r}^{-1})^{\dagger} S_b^{\dagger}\,  S_g^{-1}\, S_{b} S_{r}^{-1}
\ = \ \left[ (S_{r}^{-1})^{\dagger} S_b^{\dagger}\,  S_{r}^{\dagger} \, \right]  (S_{r}^{-1})^{\dagger} S_g^{-1} S_{r}^{-1} \left[ \, S_{r} \, S_b  \, S_{r}^{-1} \right] \; .
\ee
We first evaluate the terms in the square parentheses. We only need to evaluate the second parenthesis since the term in the first parenthesis is just its hermitian conjugate,
\bea\label{hermiconjb}
S_{r} \, S_b \, S_{r}^{-1} &=&  S_{r}\, e^{ -\frac{1}{2}b_{ij}\psi^{i}\psi^{j}}  \,  S_{r}^{-1} \ = \
e^{-\frac{1}{2} b_{ij} (\psi^{k} r_{k}{}^{i} ) ( \psi^{l} r_{l}{}^{j} )}
\ = \  e^{- \frac{1}{2} (r b r^t)_{kl} \psi^{k} \psi^{l}} 
\ = \ S_{b^{\prime}} \; ,
\eea
where we used
\be
S_{r}  \psi^i S_{r}^{-1} \ = \ \psi^{k} r_{k}{}^{i} \; , \qquad S_{r}  \psi_i S_{r}^{-1} \ = \ \psi_{k} (r^{-1})_{i}{}^{k}  \; .
\ee
Thus we see that the $b$-field transforms exactly as required by (\ref{gldgb}). It remains
to inspect the following term in (\ref{GLdstep90})
\be\label{Srg234}
(S_{r}^{-1})^{\dagger} S_g^{-1} S_{r}^{-1} \ = \  (S_{r}^{-1})^{\dagger} (S_{e}^{-1})^{\dagger}\,S_{k}\,S_{e}^{-1} S_{r}^{-1} \; .
\ee
We write now $S_{e}$ in terms of the oscillators as
\be
S_e \ = \ \frac{1}{\sqrt{\det e }} e^{\psi^i E_{i}{}^{j} \psi_{j}} \; ,
\ee
where $\exp(E) = e$. To simplify the computation of (\ref{Srg234}), it is convenient to
note that with $\overline {A} \equiv \psi^{i} A_{i}{}^{j} \psi^{j}$ we have
\be
[ \overline{A} , \overline{B} ] = \overline{[A,B]} \quad  \Rightarrow \quad  e^{\overline{R}}  e^{\overline{E}} \ = \ \overline{ \left(e^{R}e^{E} \right)} \ = \ \overline{re} \ = \ \overline{e^{\log (re)}} \; .
\ee
Thus,
\be\label{sTEp34}
 S_{r} S_e \ = \ \frac{1}{\sqrt{\det r}} \, e^{\psi^i R_{i}{}^{j} \psi_j}  \frac{1}{\sqrt{\det e}} \, e^{\psi^i E_{i}{}^{j} \psi_j}  \ = \ \frac{1}{\sqrt{\det (re)}} e^{ \psi^i (\log (re))_{i}{}^{j} \psi_j } \ = \ S_{re} \ = \ S_{e^{\prime}}\; ,
\ee
where $\log (re)$ is defined by $e^{\log(re)} = re$.
Using this in (\ref{Srg234}) gives
 \be
(S_{r}^{-1})^{\dagger} S_g^{-1} S_{r}^{-1} \ = \ S_{g'}^{-1} \; .
\ee
In total, combining this and (\ref{hermiconjb}) we obtain
\be
(S_{r}^{-1})^{\dagger}\,S_{\cal H}\,S_{r}^{-1}  \ = \ S_{b'}^{\dagger}\,  S_{g'}^{-1}\, S_{b'}  \ = \ S_{\cal H '} \; ,
\ee
which proves $\sigma_{h_r}( {\cal H}) = 1$ for $h_r \in GL^{+}(D)$. \\[2ex]
\underline{$GL^{-}(D)_{}$:}\;
An arbitrary $GL^{-}(D) $ matrix and its corresponding $\text{Spin} (D, D)$ element
are given by, respectively,
\be
h_{r} \ = \  \begin{pmatrix} r &  0 \\ 0 & (r^{-1})^{t} \end{pmatrix} \;, \qquad
S_r \ = \ (\psi^i \psi_i - \psi_i \psi^i ) \frac{1}{\sqrt{|\det r|} } e^{\psi^k R_{k}{}^{l} \psi_l}  \; ,
\ee
with $\det r <0$. The index $i$ is fixed but arbitrary; in particular, there is no sum over $i$.
$R_{i}{}^{j}$ is defined by
\be\label{FAktorization}
 e^{R} \ = \ r_{+} \;, \qquad \text{s.t. \ }  r \ = \ k_i \, r_{+}   \; ,
\ee
where $k_i = \text{diag} (1, \cdots , -1, \cdots, 1)$ is the diagonal matrix that has a $-1$ in the diagonal entry $i$ and $r_{+} \in GL^{+}(D)$. Under this $O(D,D)$ transformation, $g$ and $b$
transform covariantly as in (\ref{gldgb}).
We have to keep in mind, however, that in writing the metric as $g = e k e^t$,
we require $e$ be positive definite, and thus we cannot write $e' = r e$.
One way to resolve this is to define a positive definite $e'$ as
\be
e' \ = \ r \, e \, k_i 
\; .
\ee
Since $k = k_i k  k_i$, this definition of $e'$ correctly gives $g' = r e  k  (re)^{t} = r g r^{t}$. The duality transformation of $S_{\cal H}$ under $GL^{-}(D)$ is then
\bea
(S_{r}^{-1})^{\dagger} S_{\cal H} S_{r}^{-1}  &=&  (S_{r}^{-1})^{\dagger} S_b^{\dagger}\,  S_g^{-1}\, S_{b} S_{r}^{-1}  \\ \nonumber
&=& \left[ (S_{r}^{-1})^{\dagger} S_b^{\dagger}\,  S_{r}^{\dagger} \, \right]  (S_{r}^{-1})^{\dagger} S_g^{-1} S_{r}^{-1} \left[ \, S_{r} \, S_b  \, S_{r}^{-1} \right] \; .
\eea
It is straightforward to see that, as in the case of $GL^{+}(D)$, $S_{r} \, S_b \, S_{r}^{-1} = S_{b'}$.
The remaining part is more subtle. We first compute
\bea
(S_{r}^{-1})^{\dagger} S_g^{-1} S_{r}^{-1}  &= &  (S_{r}^{-1})^{\dagger} (S_{e}^{-1})^{\dagger}\,S_{k}\,S_{e}^{-1} S_{r}^{-1} \\ \nonumber
&=& (S_{r}^{-1})^{\dagger} (S_{e}^{-1})^{\dagger} (\psi^i \psi_i - \psi_i \psi^i )\,S_{k}\, (\psi^i \psi_i - \psi_i \psi^i )S_{e}^{-1} S_{r}^{-1}  \; .
\eea
Here we used $S_{k} =(\psi^i \psi_i - \psi_i \psi^i )\,S_{k}\, (\psi^i \psi_i - \psi_i \psi^i )$, which
can be straightforwardly verified both for the case that $i$ is equal to the timelike direction
and for the case that it is different from the timelike direction.
This guarantees that the proof is independent of the particular factorization in (\ref{FAktorization}). Since
\be
(\psi^i \psi_i - \psi_i \psi^i )\,\psi^j\, (\psi^i \psi_i - \psi_i \psi^i ) \ = \ \psi^l (k_i)_{l}{}^{j} \;, \quad
 (\psi^i \psi_i - \psi_i \psi^i )\,\psi_j\, (\psi^i \psi_i - \psi_i \psi^i ) \ = \ \psi_l (k_i)_{j}{}^{l} \; ,
\ee
we obtain as in (\ref{sTEp34})
\be
\begin{split}
S_r S_{e} (\psi^i \psi_i - \psi_i \psi^i ) \ &= \ (\psi^i \psi_i - \psi_i \psi^i )\frac{1}{\sqrt{|\det r|} } e^{\psi^i R_{i}{}^{j} \psi_j}  \frac{1}{\sqrt{\det e}} \, e^{\psi^i E_{i}{}^{j} \psi_j}  (\psi^i \psi_i - \psi_i \psi^i ) \\
\ &= \ (\psi^i \psi_i - \psi_i \psi^i ) \frac{1}{\sqrt{|\det (re)|}} e^{ \psi^i (\log (r_{+}   e))_{i}{}^{j} \psi_j } (\psi^i \psi_i - \psi_i \psi^i ) \\
\ &= \  \frac{1}{\sqrt{|\det (re)|}}  e^{ \psi^l (k_i)_{l}{}^{p} (\log (r_{+}   e))_{p}{}^{q} \psi_m (k_i)_{q}{}^{m} }
\ = \ S_{e'} \; .
\end{split}
\ee
Summarizing, we have shown
\be
(S_{r}^{-1})^{\dagger} S_{\cal H} S_{r}^{-1} \ = \ S_{b'}^{\dagger}\,  S_{g'}^{-1}\, S_{b'}  \ = \ S_{\cal H '} \; ,
\ee
which proves $\sigma_{h_r} ({\cal H}) = 1$ for $h_r \in GL^{-}(D)$.

\subsection{T-dualities}\label{A2app}
We turn now to the sign factors in (\ref{signesoncemore}) for factorized T-dualities.
Using the shorthand notation $\sigma_{i} ({\cal H}) =\sigma_{h_i} ({\cal H})$, we will prove
that $\sigma_{i} ({\cal H}) =1$ if the $i$th direction is spacelike and $\sigma_{i} ({\cal H}) = - 1$ if this direction is timelike.

We start by establishing a simple lemma that allows to distinguish elements of the
geometric subgroup just discussed from genuine T-duality transformations. \\[1.5ex]
{\em Lemma 1}:  An $O(D,D)$ matrix of the
form
 \be\label{uppertri}
  h \ = \ \begin{pmatrix} \star & \star \\ 0 & \star \end{pmatrix}\;,
 \ee
where $\star$ stands for nonzero
blocks, is an element of $GL(D)\ltimes \mathbb{R}^{\frac{1}{2}D(D-1)}$
and can be written as the product of a $GL(D)$ element and an element of
$\mathbb{R}^{\frac{1}{2}D(D-1)}$.

\noindent{\em Proof:}  The group properties of $O(D,D)$ imply for a general matrix
of the form (\ref{uppertri})
 \be
  h \ = \ \begin{pmatrix} a & b \\ 0 & d \end{pmatrix}\; \qquad \Rightarrow \qquad
  d \ = \ (a^{-1})^T\;, \qquad a^{-1}b\;\;\, \text{antisymmetric}\;.
 \ee
Then  the matrix takes the form
\be
\begin{pmatrix} a & b \\ 0 & (a^{-1})^t \end{pmatrix} =
\begin{pmatrix} a & 0 \\ 0 & (a^{-1})^t \end{pmatrix}
\begin{pmatrix} 1 & a^{-1}b \\ 0 & 1 \end{pmatrix} \,,
\ee
proving the claim.
\medskip

Let us now
assume that
\be
(S_i^{-1})^\dagger  \, S_{\cal H}  \,  S_i^{-1}  \ = \ \sigma_i ({\cal H})\, S_{h_i \circ {\cal H}} \,.
\ee
Using $S_i = S_i^{-1} = S_i^\dagger$, this can also be written as
\be
\label{kcvg}
S_i  \, S_{\cal H}  \,  S_i  \ = \ \sigma_i ({\cal H})\, S_{h_i \circ {\cal H}} \,.
\ee
Letting $h$ denote an $O(D,D)$ transformation,  the above equation implies that
\be
\label{kcsvg}
S_i  \, S_{h\circ\cal H}  \,  S_i  \ = \ \sigma_i (h\circ {\cal H})\, S_{h_i h\circ {\cal H}} \,.
\ee
We want to determine the equivalence class of $h\in O(D,D)$ satisfying
$\sigma_i (h\circ {\cal H}) = \sigma_i ( {\cal H})$.  A sufficient condition is given by the following lemma:

\noindent
{\em Lemma 2:}~ If $h \in GL(D)\times \mathbb{R}^{\frac{1}{2}D(D-1)}$ and
$h_i \, h\, h_i \in GL(D)\ltimes \mathbb{R}^{\frac{1}{2}D(D-1)}$, then
\be
\label{jnvsvg}
~\sigma_i (h\circ {\cal H}) = \sigma_i ( {\cal H})\,.
\ee

\medskip
\noindent
{\em Proof:}  We write
\be
\label{vmvsvg}
 h_\star  \equiv h_i \, h \, h_i  \, \in GL(D)\ltimes \mathbb{R}^{\frac{1}{2}D(D-1)}\,,
\ee
and note that
\be
\label{vmsbvg9}
S_{h_\star} \ = \ \pm S_i \, S_h\, S_i  \,.
\ee
Since
$h\in GL(D)\ltimes \mathbb{R}^{\frac{1}{2}D(D-1)}$,
\be
\label{sgvg}
(S_h^{-1})^\dagger  S_{\cal H} \,S_h^{-1}
\ = \ +  \,S_{h\circ {\cal H}} \,.
\ee
We calculate the left-hand side of
(\ref{kcsvg}), using (\ref{sgvg})  in the first step,
\be
\label{jnbvg}
\begin{split}
S_i  \, S_{h\circ \cal H}  \,  S_i  & \ = \
S_i \,(S_h^{-1})^\dagger ~ S_{\cal H} ~S_h^{-1} \,  S_i \\
& \ = \ S_i \,(S_h^{-1})^\dagger S_i \, \bigl( S_i \, S_{\cal H} \,S_i \bigr)
~S_i  \,S_h^{-1} \, S_i \\
& \ = \ \bigl( \bigl( S_i \,S_h\, S_i)^{-1} \bigr)^\dagger \, \bigl( \sigma_i({\cal H})\, S_{h_i \circ \cal H} \bigr)
~\bigl(S_i  \,S_h \, S_i\bigr)^{-1}\,,
\end{split}
\ee
where we made use of (\ref{kcvg}).  Making use of (\ref{vmsbvg9}),
\be
S_i  \, S_{h\circ \cal H}  \,  S_i   \ = \  \sigma_i({\cal H})\,
 \bigl( S_{h_\star}^{-1}\bigr)^\dagger \, S_{h_i \circ \cal H}
~S_{h_\star}^{-1}\,\\
 \ = \  \sigma_i({\cal H})\,
\, S_{h_\star h_i \circ \cal H} \, ,
\ee
since $h_\star\in GL(D)\ltimes \mathbb{R}^{\frac{1}{2}D(D-1)}$.
 We now note that using (\ref{vmvsvg})
 \be
h_\star \, h_i \circ {\cal H}
 = h_i \, h \, h_i  \, h_i\circ {\cal H}
 =  h_i \, h \circ {\cal H}\,,
 \ee
and therefore we have obtained,
\be
S_i  \, S_{h \circ \cal H}  \,  S_i    \ = \
\sigma_i({\cal H})\, S_{h_i  h\circ {\cal H}}\,.
\ee
Comparing with (\ref{kcsvg}) we conclude that (\ref{jnvsvg}) is true, as we wanted to prove.

\medskip
As a first application we show that  $b$-shifts satisfy the
conditions of Lemma~2.   Indeed, taking $h= h_{b}$ for some arbitrary
$D\times D$ matrix $b$, a small computation confirms that
\be
\label{vmbvg99}
h_i \, h_{b} \, h_i \ = \  \begin{pmatrix}
1 +  b e_i &  - b + e_i b+ be_i\\ 0  &  1 + e_i b \end{pmatrix}
\,\in\, GL(D)\ltimes \mathbb{R}^{\frac{1}{2}D(D-1)}\,.
\ee
 It then follows that
\be
\sigma_i \bigl( h_b \circ {\cal H}\bigr)  =  \sigma_i ({\cal H})\,.
\ee
Since at any point $X$ the $b$ field in ${\cal H}$ can be removed completely by a
$b$-shift,  the sign $\sigma_i ({\cal H})$ is in
fact a function $\sigma_i (g)$ of the metric only:
\be
\sigma_i ({\cal H}) =  \sigma_i (g)  \,.
\ee

\medskip
In order to determine now $\sigma_i (g)$, we use $GL(D)$ transformations
that bring the metric into a simpler form.
There is an important complication, however: for arbitrary $r \in\,GL(D)$ it is
not generally true that $h_i h_r h_i$ is in $GL(D)\ltimes \mathbb{R}^{\frac{1}{2}D(D-1)}$,
and thus Lemma 2 cannot be generally applied.
For the Lemma to be applicable, the lower left block of the matrix $h_i h_r h_i$
must vanish.  A small calculation shows that this requires
\be
-e_i r (1- e_i)  - (1-e_i) (r^{-1})^T \, e_i  = 0 \,\,   ~~(i ~\hbox{not summed})\,.
\ee
Using $e_i A e_i = A_{ii} e_i$ for any matrix
$A$, and $e_i e_i = e_i$, we can rewrite the above condition as
\be
\label{ssvg}
-e_i (r  - r_{ii}  e_i)  - \Bigl( (r^{-1})^T  - ((r^{-1})^T)_{ii} e_i \Bigr)  e_i  = 0 \,.
\ee
Consider the condition that the first term vanishes:
\be
e_i (r  - r_{ii}  e_i) = 0 \,.
\ee
This requires the $i$-th row of $r$ to vanish, except for the diagonal
element $r_{ii}$ that can be arbitrary.  Without loss of generality, and to
display more easily the matrices, let us take $i=1$.
The condition then gives
\be
\label{jbvg}
e_1 (r  - r_{11}  e_1) = 0\quad \to \quad
r = r_\star \equiv \begin{pmatrix}  r_{11}  &  \vec{0}^{\,T} \\[0.5ex]  \vec{V} &  \hat r \end{pmatrix}\,,
\ee
letting $r_\star$ denote the solution of this constraint.
We decomposed the matrix $r_\star$ into a $1\times 1$ corner block
with element $r_{11}$,
a  $(D-1)$ column vector $\vec{V}$,  the vanishing
$(D-1)$ row vector, and the $(D-1)\times (D-1)$
invertible matrix $\hat r$.
A small calculation shows that
\be
r_\star^{-1} = \begin{pmatrix}  1/r_{11}  &  \vec{0}^{\,T} \\[0.5ex]
- \hat{r}^{-1} \vec{V}/r_{11} &  \hat{r}^{-1} \end{pmatrix}\,.
\ee
This shows that $(r_\star^{-1})^T$ has a vanishing first column, except
for its diagonal element, which implies that
\be
 \Bigl( (r_\star^{-1})^T  - ((r_\star^{-1})^T)_{11} e_1 \Bigr)  e_1  = 0   \,.
\ee
Thus for $r_\star$ the second term in (\ref{ssvg}) vanishes as well.  This
shows that for $r_\star$ as in (\ref{jbvg}), we have
$h_1 h_{r_\star} h_1 \in GL(D)\ltimes \mathbb{R}^{\frac{1}{2}D(D-1)}$, and the conditions
of Lemma~2 apply.  We thus have
\be
 \sigma_1 (g) = \sigma_1 \,(\,h_{r_\star}\hskip-2pt \circ g)
= \sigma_1 \,\bigl( g_\star \bigr) \,, ~~\hbox{with} ~~
 g_\star = r_\star\, g \, r_\star^T ~.
\ee
Let us compute the rotated metric assuming a block
decomposition for $g$:
\be
g= \begin{pmatrix}  g_{11}  &  \vec{A}^{\,T} \\[0.5ex]
\vec{A} &  \hat g\end{pmatrix}\,.
\ee
A small calculation gives
\be
g_\star = r_\star\, g \, r_\star^T = \begin{pmatrix}r_{11} g_{11} r_{11}
&  r_{11} (g_{11} \vec{V} + \hat{r} \vec{A}\, )^T  \\[1.5ex]
r_{11} (g_{11} \vec{V} + \hat{r} \vec{A}\, ) &
~~(g_{11} \vec{V} + \hat{r} \vec{A} \,) \vec{V}^{\,T}+ \vec{V} \vec{A}^{\,T} \hat{r}^T
+ \hat r \, \hat g\,  \hat{r}^T ~
\end{pmatrix}\,.
\ee
Choosing
\be
r_{11} = {1\over \sqrt{|g_{11}|}} \,, ~~~ \vec{V} =  -{1\over g_{11}}\,\hat r \vec{A} \,,
\ee
we find
\be
 g_\star = \begin{pmatrix} \hbox{sgn} (g_{11})
&  \vec{0}^{\,T} \\[1.5ex]
\vec{0}  &
\,\hat r \,\Bigl( \hat g - {1\over g_{11} }\vec{A} \,\vec{A}^{\,T}\Bigr)  \hat{r}^T ~
\end{pmatrix}\;.
\ee
By the general result on diagonalization of quadratic forms, we can choose
$\hat {r}$ in such a way that the lower-right block becomes a diagonal
matrix with entries equal to plus or minus ones,
\be
 g_\star \, = \begin{pmatrix} \hbox{sgn} (g_{11})
&  \vec{0}^{\,T} \\[1.5ex]
\vec{0}  &
\,\hat k ~
\end{pmatrix}\,, ~~~   \hat k ~\hbox{diagonal with~} \pm 1 ~\hbox{entries.}
\ee
By Sylvester's theorem of inertia, the matrix $g_\star$ has a single
$-1$ entry.  Thus either $g_{11}$ is negative and 
$g_\star = k$, with $k$ the Minkowski metric, 
or $g_{11}>0$ and $g_\star = k_i,$ for some $ i \not=1$. In either case we know
how to determine the sign factor:
\be
\sigma_1 (g) = \sigma_1 (g_\star) = \hbox{sgn}(g_{11})\,.
\ee
Since our choice of the first coordinate was just irrelevant, this holds
for a factorized T-duality about any coordinate. Our final result is therefore
\be
\label{jnsbtfl}
\sigma_i ({\cal H}) =  \hbox{sgn} (g_{ii}) \,.
\ee
Equivalently, $\sigma_i=1$ for a T-duality along a coordinate direction $x^i$
that is space-like, and $\sigma_i=-1$ for a T-duality along a coordinate direction that is timelike.

\medskip

In order to use (\ref{jnsbtfl}) 
for the successive application of several 
T-dualities, we have to keep in mind that each action of $h_{i}$ transforms
the full metric  $g_{ij}$ non-trivially and therefore the full 
sign factor cannot 
be inferred from the signs of the diagonal 
entries of the initial metric $g_{ij}$.
For the special case of  
the transformation $J$, i.e., T-dualities 
along all coordinates, however, we can show
that $\sigma_{J} ({\cal H}) = -1$ 
as follows.

Under the $O(D,D)$ transformation $J$, ${\cal H}$ transforms as 
\be
{\cal H}' \ = \ J {\cal H} J   \, .  
\ee
We define $h \equiv h_{e}^{-1} h_b$ and rewrite ${\cal H}$ and ${\cal H}'$ as
\be
{\cal H} \ = \ h^T {\cal H}_0 h\;, \qquad 
{\cal H}'  
\ = \  \ h^{-1}  {\cal H}_0 (h^{-1})^T \; .
\ee
With the corresponding spin representative $S_{h}=  S_e^{-1} S_b$ of $h$
we then have, by definition, 
\be
\label{popop}
S_{\cal H} \ = \  S_{h}^{\dagger}   S_{{\cal H}_0}  S_{h} \; , \qquad
S_{{\cal H}^{\prime}} \ = \  S_{h^{-1}}   S_{{\cal H}_0}  S_{h^{-1}}^{\dagger}\;.
\ee
Using that $S_J^{-1} =  S_J^\dagger$ we have 
\be
\label{clgnrstts}
(S_{J}^{-1})^{\dagger} S_{\cal H} \, S_{J}^{-1}   \ = \ S_{J} S_{h}^{\dagger}   S_{{\cal H}_0}  S_{h}  \, S_{J}^{\dagger} \ = \ \left[ (S_{h})^{-1} S_{h} S_{J} S_{h}^{\dagger}  \right] S_{{\cal H}_0} \left[ S_{h} S_{J}^{\dagger} S_{h}^{\dagger} (S_{h}^{\dagger})^{-1} \right] \; .
\ee
$J$ is an invariant matrix, $h J h^T = J$, and thus in Pin$(D,D)$ we have 
\be
S_h S_J S_h^{\dagger} \ = \ \pm S_J\;.
\ee
We can thus simplify (\ref{clgnrstts})  
\be
(S_{J}^{-1})^{\dagger} S_{\cal H} \, S_{J}^{-1}   \ = \ S_{h}^{-1} S_{J} S_{{\cal H}_0}  S_{J}^{\dagger} (S_{h}^{\dagger})^{-1} \ = \ -  S_{h}^{-1} S_{{\cal H}_0} (S_{h}^{-1})^{\dagger}  \ = \ - S_{{\cal H}'} \; ,
\ee
where we used $S_{J} S_{{\cal H}_0}  S_{J}^{\dagger} \ = \ - S_{{\cal H}_0}$ in the second equality and (\ref{popop}) for the last equality.   We have
thus shown that 
 $\sigma_{J} ({\cal H}) = -1$.

\subsection{Example} \label{appendixexamples}
Next, we present an instructive example concerning the above rules of sign factors.
We construct a closed loop in the space of ${\cal H}$ in $SO^{-}(D,D)$
that cannot be lifted to a closed loop in Spin$^{-}(D,D)$. 

Consider for $D=2$ the one-parameter family of $SO^{+}(D,D)$ transformations
parameterized by $\alpha$:
 \be\label{hfamily}
  h(\alpha) \ = \ \exp \big[ \alpha \,T  \big]\;, \qquad \alpha\in \big[\,0,\frac{\pi}{2}\big]\;,
 \ee
where $T$ is the Lie algebra generator
 \be\label{Genrator}
  T \ \equiv \ T^{14}+T^{12}+T^{32}+T^{34} \ = \
  \begin{pmatrix}  \phantom{-}0 & 1 &  \phantom{-}0 & 1 \\ -1 & 0 & -1 & 0 \\
   \phantom{-}0 & 1 &  \phantom{-}0 & 1 \\ -1 & 0 & -1 & 0  \end{pmatrix}\;,
 \ee
 and $T^{MN}$ are the standard fundamental generators (\ref{fund}).
 A computation gives:
 \be
 \label{vm2}
 h(\alpha) =
 \begin{pmatrix}
 \cos^2 \alpha  & {1\over 2} \sin 2\alpha&  - \sin^2 \alpha &
  {1\over 2} \sin 2\alpha \\[0.3ex]
 - {1\over 2} \sin 2\alpha &  \cos^2 \alpha  &
  -  {1\over 2} \sin 2\alpha & - \sin^2 \alpha \\[0.3ex]
   - \sin^2 \alpha &
  {1\over 2} \sin 2\alpha &  \cos^2 \alpha  & {1\over 2} \sin 2\alpha\\[0.3ex]
   - {1\over 2} \sin 2\alpha &  \sin^2 \alpha  &
    - {1\over 2} \sin 2\alpha &  \cos^2 \alpha
 \end{pmatrix}\,.
 \ee
Since $T^t = - T$, we have $h(\alpha)^t  = h(\alpha)^{-1}$. For later use we also note that (\ref{hfamily})
can be defined for arbitrary $\alpha$, which then has the periodicity
$h (\alpha) = h (\alpha + \pi)$.
This family of transformations was designed so that for $\alpha = \pi/2$ we get the product of the two
T-dualities $h_1$ and $h_2$:
 \be
 h(\tfrac{\pi}{2}) \ = \
  \begin{pmatrix}  \phantom{-}0 &\phantom{-} 0 & -1 & \phantom{-}0 \\ \phantom{-}0 &\phantom{-} 0 & \phantom{-}0 & -1 \\ -1 &\phantom{-} 0 & \phantom{-}0 &\phantom{-} 0 \\ \phantom{-}0 & -1 & \phantom{-}0 & \phantom{-}0  \end{pmatrix}
  \ = \ h_1\,h_2  \;.
 \ee

Consider now the `flat' generalized metric ${\cal H}_0=\text{diag}(k,k)\in SO^-(D,D)$.
The $SO^+(D,D)$ transformations $h(\alpha)$ acting on this
generalized metric give us a family of rotated metrics,
 \be
 \label{sgvg99}
  {\cal H}(\alpha) \ = \ (h(\alpha)^{-1})^{t}\,{\cal H}_0\,h(\alpha)^{-1}
   \ = \
   h(\alpha) \,{\cal H}_0\,h(\alpha)^{-1} \,.
 \ee
 A computation of the matrix product gives
 \be
 \label{vmvg2}
 {\cal H}(\alpha) =
 \begin{pmatrix}
 -\cos^2 2\alpha  & {1\over 2} \sin 4\alpha&   \sin^2 2\alpha &
  {1\over 2} \sin 4\alpha \\[0.4ex]
  {1\over 2} \sin 4\alpha &  \cos^2 2\alpha  &
    {1\over 2} \sin 4\alpha & - \sin^2 2\alpha \\[0.4ex]
    \sin^2 2\alpha &
  {1\over 2} \sin 4\alpha & - \cos^2 2\alpha  & {1\over 2} \sin 4\alpha
  \\[0.4ex]
   {1\over 2} \sin 4\alpha &  -\sin^2 2\alpha  &
     {1\over 2} \sin 4\alpha &  \cos^2 2\alpha
 \end{pmatrix}\,.
 \ee
As it turns out, the transformation $h_1 h_2$ leaves ${\cal H}_0$ invariant, thus ${\cal H}(\alpha)$ traces a closed curve as $\alpha\in [0, \pi/2]$:
\be
{\cal H}(0)={\cal H}(\tfrac{\pi}{2})={\cal H}_0\,.
\ee
  For general $\alpha$,
the metric and $b$ field read off from ${\cal H}= {\cal H}_{\bullet \bullet}$ are
 \be
 \label{clvg}
  g_{ij}(\alpha) \ = \  \begin{pmatrix} -1 & \tan 2\alpha \\ \tan 2\alpha & 1 \end{pmatrix}\;, \qquad
  b_{ij}(\alpha) \ = \  \begin{pmatrix} 0 & \tan 2\alpha \\ -\tan 2\alpha & 0 \end{pmatrix}\;.
 \ee
For $\alpha=\tfrac{\pi}{4}$ both the metric and the $b$ field components  become infinite, even though  the generalized metric
is still perfectly regular:
 \be
  {\cal H}(\tfrac{\pi}{4}) \ = \  \begin{pmatrix}
  0 & \phantom{-}0 & 1 & \phantom{-}0 \\ 0 & \phantom{-}0 & 0 & -1 \\ 1 &
  \phantom{-}0 & 0 & \phantom{-}0 \\ 0 & -1 & 0 & \phantom{-}0  \end{pmatrix}\;.
 \ee
At this singular point we expect that our explicit formula for $S_{\cal H}$ is affected by  some kind of
`phase transition'.  

 Next we turn to the study of the corresponding elements in
 Spin$(D,D)$.
Since $h(\alpha)$ is in the component of the group connected to the identity, its
spinor representative follows directly from (\ref{Genrator}),
 \be
  S(\alpha) \ \equiv \  S_{h(\alpha)} \ = \ \exp\big[\, \alpha\, \hat{\Gamma}\,\big]\;, \qquad
  \hat{\Gamma} \ \equiv \ {1\over 2} \Bigl(\Gamma^{14} +\Gamma^{12} +\Gamma^{32} +\Gamma^{34}\Bigr)\;.
 \ee
Recalling that $\Gamma^{MN} = {1\over 2} (\Gamma^M \Gamma^N - \Gamma^N \Gamma^M)$ and that $\Gamma^{M}=\sqrt{2}\,(\psi_1,\psi_2,\psi^1,\psi^2)$ we infer
 \be
  \hat{\Gamma} \ = \ \psi_1\psi^2  + \psi_1\psi_2+\psi^1\psi_2+\psi^1\psi^2 \ = \
  (\psi^1+\psi_1)(\psi^2+\psi_2)\;.
 \ee
As $\hat{\Gamma}^2=-{\bf 1}$, we get in closed form
 \be\label{Salphaclosed}
  S(\alpha) \ = \ \cos\alpha\cdot {\bf 1}+\sin\alpha \cdot\hat{\Gamma}\;.
 \ee
We can now investigate its action on the spinor representative for ${\cal H}_0$, which we choose
to be
 \be
  S_{{\cal H}_0}  \ = \ \psi^1 \psi_1-\psi_1\psi^1\;,
 \ee
where we denoted the timelike direction by $1$.
We then define
\be
\label{skvg}
 S_{\cal H}(\alpha) \ \equiv \ (S(\alpha)^{-1})^{\dagger}\,S_{{\cal H}_0}\,S(\alpha)^{-1} \,,
\ee
such that $S_{\cal H}(0) =  S_{{\cal H}_0}$. Taking the $\rho$ homomorphism of (\ref{skvg})
we conclude that  $S_{\cal H}(\alpha)$, so defined,
is
\be
\label{jnvg}
S_{\cal H}(\alpha) \ = \ \pm\,  S_{{\cal H}(\alpha)}\,.
\ee
Since the plus sign holds for $\alpha=0$ and both sides appear to
be defined by continuous deformations, it is puzzling that the sign
becomes minus at some point.  This is what we want to understand.

The explicit calculation of (\ref{skvg}) gives with (\ref{Salphaclosed})
 \be\label{Salpha99}
 \begin{split}
  S_{\cal H}(\alpha) 
  \ &= \  \cos(2\alpha)
  \Bigl[ \psi^1\psi_1-\psi_1\psi^1\,-\, \tan(2\alpha)(\psi^1-\psi_1)(\psi^2+\psi_2) \Bigr]\;.
 \end{split}
 \ee
Recalling that ${\cal H}({\pi\over 2}) = {\cal H}_0$, we observe that
 \be\label{SigNChange}
  S_{\cal H}(\tfrac{\pi}{2}) \ = \ -(\psi^1\psi_1-\psi_1\psi^1) \ = \ -S_{{\cal H}_0} = - S_{{\cal H}({\pi\over 2})} \,.
 \ee
We have gotten now a minus sign in (\ref{jnvg}). Alternatively, while
$\alpha \in [0, \pi/2]$ gives a closed loop for ${\cal H}(\alpha)$ it gives an open loop for $S_{\cal H} (\alpha)$.  The aim of the following discussion is
to see how this minus sign arises.

\medskip
We should compare $S_{\cal H} (\alpha)$ with the family $S_{{\cal H}(\alpha)}$, which can be
defined independently as:
\be
 S_{{\cal H}(\alpha)} \ \equiv \ S_{b(\alpha)}^{\dagger}\,S_{g(\alpha)}^{-1}\,S_{b(\alpha)} \,,
~~\hbox{with} ~~  S_{g(\alpha)}^{-1} \ = \ (S_{e(\alpha)}^{-1})^{\dagger}\,S_{k}\,S_{e(\alpha)}^{-1}\;.
\ee
In this definition one must extract $g(\alpha)$ and $b(\alpha)$ from ${\cal H}(\alpha)$ and use
$g(\alpha)$ to define a vielbein $e(\alpha)$ from $g(\alpha) = e(\alpha) k e(\alpha)^t$.  The potential difficulty here is
the possibility that divergent $g$'s can lead to discontinuous $e$'s and thus a discontinuous definition of $S_{\cal H}$.

We begin the calculation by computing the vielbein using
$g=e\,k\,e^{T}$ and the metric from (\ref{clvg}).
The vielbein is not unique, but one representative is  \be
  e(\alpha) \ \ = \ \begin{pmatrix} |\sec 2\alpha | & \tan2\alpha \\
  0 & 1 \end{pmatrix}\;.
 \ee
 Next, we need a matrix $E$ such that $e=\exp(E)$:
 \be
 \label{kcvg}
  E \ = \ \begin{pmatrix} u & v \\ 0 & 0 \end{pmatrix} \quad \to \quad
  \exp(E) = \begin{pmatrix} e^u & {v\over u} (e^u-1) \\ 0 & 1 \end{pmatrix}
= \begin{pmatrix} |\sec 2\alpha | & \tan2\alpha \\
  0 & 1 \end{pmatrix}\,. \ee
We then compute
 \be
  S_{e}^{-1} \ = \ \sqrt{\det{e}}\,e^{-\psi^1 E_{1}{}^{1}\psi_1-\psi^1 E_{1}{}^{2}\psi_2}   \ = \ \sqrt{\det{e}}\,e^{-u \psi^1\psi_1-v\psi^1\psi_2}\;.
  \ee
The exponential can be worked out explicitly, giving
  \be
 \begin{split}
  S_{e}^{-1} 
    \ &= \ \sqrt{\det{e}}\,\big(1-e^{-u}(e^u-1))(\psi^1\psi_1+vu^{-1}\psi^1\psi_2)\big) \;.
 \end{split}
 \ee
Therefore, using (\ref{kcvg}) we find
\be
 S_{e}^{-1} = |\sec 2\alpha|^{1/2} \Bigl( 1 - |\cos 2\alpha| \Bigl[
 (|\sec 2\alpha| -1) \psi^1 \psi_1  + \tan 2\alpha\, \psi^1 \psi_2 \Bigr] \Bigr)\;.
\ee
We then obtain for the metric
 \be
 \begin{split}
  S_{g}^{-1} 
  \ & = \  |\sec 2\alpha|  \Bigl(  \cos^2 2\alpha\, \psi^1 \psi_1 - \psi_1 \psi^1
- \sin 2\alpha \cos 2\alpha (\psi^1 \psi_2 + \psi^2 \psi_1)
+ \sin^2 2\alpha \, \psi^2 \psi_1 \psi^1 \psi_2 \Bigr)\;.
 \end{split}
 \ee
The $b$-field contributions are given by
 \be
  S_{b} \ = \ e^{-\frac{1}{2}b_{ij}\psi^{i}\psi^{j}} \ = \ 1-b_{12}\psi^{1}\psi^{2}\;,
 \ee
while all higher terms vanish in $D=2$.  Using (\ref{clvg}),
 \be
  S_{b} \ = \ 1-\tan(2\alpha)\,\psi^{1}\,\psi^2\;, \qquad
  S_{b}^{\dagger} \ = \ 1-\tan(2\alpha)\,\psi_2\,\psi_1\;.
 \ee
After some further calculation we get
 \be
 \begin{split}
  S_{{\cal H}(\alpha)} 
  \ &= \ |\cos(2\alpha)|(\psi^1\psi_1-\psi_1\psi^1) -\frac{|\cos(2\alpha)|}{\cos(2\alpha)}\sin(2\alpha)
  (\psi^1-\psi_1)(\psi^2+\psi_2) \\
  \ &= \ {\rm sgn}(\cos(2\alpha))\,S_{\cal H}(\alpha)\;,
 \end{split}
 \ee
where ${\rm sgn}$ denotes the sign of its argument, and we compared with
(\ref{Salpha99}).
This result is perfectly consistent with the sign change found in (\ref{SigNChange}). For small values of
$\alpha>0$, the sign is positive and
so this agrees with (\ref{jnvg}) using the $+$ sign. At $\alpha=\tfrac{\pi}{4}$, $S_{{\cal H}(\alpha)}$ is discontinuous. 
For $\alpha>\tfrac{\pi}{4}$, (\ref{jnvg}) holds for the minus sign,
 as it should be in order to be
consistent with the final relative sign at $\alpha=\tfrac{\pi}{2}$.

\medskip

Let us finally 
reconsider the above analysis in a different approach. Specifically, since we saw above that
the sign change occurs at a singular point for which $g$ degenerates and $S_{\cal H}$
becomes ill-defined, it is natural to inquire what happens if one employs a definition that only
requires ${\cal H}$ to be regular, but not necessarily decomposable into $h_g$ and $h_b$.
Such a definition can indeed be given by separating ${\cal H}(\alpha)$ into two pieces:
\be\label{HonceMore}
{\cal H}(\alpha) = {\cal H}_0 ( {\cal H}_0 {\cal H}(\alpha)) \equiv {\cal H}_0 \hat{\cal H}(\alpha) \; ,
\ee
where ${\cal H}_0$ is an $O(D,D)$ element disconnected from the identity while $\hat{\cal H}$ is an
$O(D,D)$ element connected to the identity.
Thus, $\hat{\cal H}$ can be written in exponential form.
Indeed, with the explicit forms (\ref{vm2}) and (\ref{vmvg2})
one can verify
\be \label{hhat}
\hat{\cal H} (\alpha) \ = \ \exp \big[ - 2 \alpha \,T  \big]\; .
\ee
Therefore, its spinor representative can also be defined as an exponential which, after choosing
$S_{{\cal H}_0}$, gives a spinor representative for ${\cal H}(\alpha)$
according to (\ref{HonceMore}).

At first sight, this leads to a well-defined and smooth spinor representative for all $\alpha$.
There is a subtlety, however, which is due to the following periodicity of $\hat{\cal H}$,
\be
\hat{\cal H} (\alpha) 
\ = \ \hat{\cal H} (-\tfrac{\pi}{2} + \alpha)  \; ,
\ee
following analogously to the periodicity of $h(\alpha)$ noted after (\ref{vm2}).
Consequently, given an ${\cal H}(\alpha)$, there is no unique parameter value
$\alpha$ that reproduces this generalized metric according to (\ref{hhat}), and
therefore there is no unique spinor representative of $\hat{\cal H}$. More precisely,
if we attempt to define the exponential form of $S_{\hat{\cal H} }$ by replacing $T$ by $\hat{\Gamma}$
in (\ref{hhat}), there are actually two choices,
\be\label{TwoChoice}
S_{\hat{\cal H}(\alpha)} \ = \
\left\{  \begin{array}{l} \exp \big[ - 2 \alpha \, \hat{\Gamma} \big] \\\exp \big[  (- 2 \alpha + \pi) \, \hat{\Gamma} \big]  \end{array} \right. \; .
\ee
Since, using (\ref{Salphaclosed}), $\exp [ \pi  \hat{\Gamma} ] = -1$,
these two choices differ precisely by a sign.
This has the consequence that there is no continuos and single-valued way to choose the
spin representative over the complete path of ${\cal H}(\alpha)$. In fact, since the path is closed,
single-valuedness requires $S_{\hat{\cal H}(0)} = S_{\hat{\cal H}(\frac{\pi}{2})} = {\bf 1}$.
This, in turn, can only be achieved if we choose in (\ref{TwoChoice})
the first parametrization of $S_{\hat{\cal H}}$
for $\alpha = 0$ and the second parametrization for $\alpha = \tfrac{\pi}{2}$.
Thus, at some point in the interval $(0,\tfrac{\pi}{2})$ we need to change the parametrization,
leading to a non-continuous $S_{\hat{\cal H}}$ and $S_{\cal H}$.
(In the previous approach, this point was at $\alpha = \tfrac{\pi}{4}$.) Thus, we conclude
that while in this approach
the `point of discontinuity' can be chosen arbitrarily in the interval $(0,\tfrac{\pi}{2})$,
the associated sign change as in (\ref{SigNChange}) is unavoidable.

\end{document}